%% file: long.tex
\documentstyle[preprint,aps,eqsecnum]{revtex}
\newcommand{\spone}{0.9}
\newcommand{\singlespace}{\edef\baselinestretch{\spone}\Large\normalsize}
\catcode`@=11
\font\fivemib = cmmib10  \@ptscale5 
\font\sevenmib = cmmib10  \@ptscale7 
\font\tenmib = cmmib10   

\newfam\biffam \def\bif{\fam\biffam\tenmib}
\textfont\biffam=\tenmib \scriptfont\biffam=\sevenmib
\scriptscriptfont\biffam=\fivemib

\catcode`@=12
\def\O{{\cal O}}

\begin{document}


\input{beginning-aol.tex} 

\section{Introduction and Summary}

	There is considerable theoretical and experimental
interest at present in the question of whether neutrinos have mass
\cite{NEUTRINO94,BV,K,MOR93}.  On the theoretical side the existence of massive
neutrinos would require a modification of the standard SU(2)
$\times$ U(1) model of the electroweak interaction, and could also
provide some insight into the question of how other particles
acquire their masses.  These and other related theoretical
questions, such as the missing mass problem in the Universe
\cite{NEUTRINO94,C93}, have served to stimulate continuing experimental
searches for evidence of a nonzero neutrino mass.  To date there
is no compelling direct evidence for massive neutrinos, and the
existing upper limits for $\nu_{e}$, $\nu_{\mu}$, and $\nu_{\tau}$
(or for the corresponding antiparticles) are \cite{PDG94}
\begin{equation}
	m(\nu_{e}) \lesssim 7.0\,\,\mbox{eV};
	\phantom{space}
	m(\nu_{\mu}) \lesssim 0.27\,\,\mbox{MeV};
	\phantom{space}
	m(\nu_{\tau}) \lesssim 31\,\,\mbox{MeV}.
\label{m nu limits}
\end{equation}
Notwithstanding the bounds in Eq.~(\ref{m nu limits}), there is
mounting indirect evidence for massive neutrinos from a number of
different experiments as discussed in 
Refs.~\cite{BER94,RAG95,SAR95,BAH95,GT95}.  
We will return to Eq.~(\ref{m nu limits}) in Sec.~VIII
below.

	The object of the present paper is to develop a new
technique for studying neutrino mass which exploits the fact that
massless neutrinos give rise to long-range forces.  Recent
interest in the question of weak long-range forces \cite{Nature
Review,Adelberger,Fujii} has
focussed on the possible existence of new ultra-light bosonic
fields as the mediators of such interactions.  It is well known,
however, that long-range forces can also arise from the exchange
of neutrino-antineutrino ($\nu\bar{\nu}$) pairs, and the history
of various attempts to calculate the $\nu\bar{\nu}$-exchange
force is summarized in Ref.~\cite{FS68}.  The first correct
calculation of the 2-body $\nu\bar{\nu}$-exchange potential was
carried out by Feinberg and Sucher (FS) \cite{FS68} who used an
effective low-energy 4-fermion interaction involving only charged
currents.  They found for the potential energy $V_{ee}^{(2)}(r)$
describing the interaction of two electrons via the
$\nu\bar{\nu}$-exchange diagram shown in Fig.~1,
\begin{equation}
	V_{ee}^{(2)}(r) = G_{F}^{2}/4\pi^{3}r^{5},
\label{Vee}
\end{equation}
where $r = |\vec{r}_{1} - \vec{r}_{2}|$ is the separation of the
electrons, and $G_{F} = 1.16639(2) \times 10^{-5}$ GeV$^{-2}
(\hbar c)^{3}$ is the Fermi decay constant.  Subsequently
Feinberg, Sucher, and Au (FSA) \cite{FSA89} recalculated
$V_{ee}^{(2)}(r)$ in the framework of the Standard Model and
obtained
\begin{equation}
	V_{ee}^{(2)}(r) = G_{F}^{2}(2\sin^{2}\theta_{W} +
			1/2)^{2}/4\pi^{3}r^{5},
\label{SM Vee}
\end{equation}
where $\theta_{W}$ is the weak mixing
angle, with $\sin^{2}\theta_{W} = 0.2319(5)$. The result in
Eq.~(\ref{SM Vee}) has been rederived recently by Hsu and Sikivie
\cite{HS94} using a different formalism from that of FS.  
Numerically,
$V_{ee}^{(2)}(r) = 3\times 10^{-82}r^{-5}$ eV where $r$ is in meters,
from which one can see that the interaction energy of two
electrons arising from neutrino-exchange is smaller than their
mutual gravitational interaction for $r \gtrsim r_{0} = 5 \times
10^{-8}$m.  However, for $r \lesssim r_{0}$, the electroweak
interaction arising from $\gamma$- and $Z^{0}$-exchange would
dominate, and hence there appears to be no distance scale over
which $V_{ee}^{(2)}(r)$ leads to detectable effects.   As we note in
Appendix~A, the neutrino-neutron and neutrino-proton coupling
constants, $a_{n}$ and $a_{p}$ respectively, are even smaller
than the neutrino-electron coupling constant $a_{e}$, so the same
conclusions would hold for the corresponding 2-body potentials
$V_{nn}^{(2)}(r)$ and $V_{pp}^{(2)}(r)$.  In what follows it will
be shown that although the 2-body potential energy is indeed
extremely small, the many-body interaction energy arising from
neutrino-exchange can be extremely large in neutron stars and
white dwarfs.  This observation eventually leads to the
conclusion that there is a lower bound on the mass of any
neutrino or antineutrino, as given in Eq.~(\ref{final m limit}) below.

	Many-body effects arising from neutrino-exchange have
been considered previously by a number of authors
\cite{HAR70,IS36,FEYN71}, often in connection with attempts to
understand the gravitational interaction.  Feynman \cite{FEYN71}
considered the many-body interaction that would describe the
effective 2-body force that arises when two test masses interact
via neutrino-exchange in the presence of distant matter.
Although Feynman's attempt to explain the inverse-square law of
Newtonian gravity in terms of neutrino-exchange ultimately proved
unsuccessful, he made an observation about such interactions
which forms part of the basis of the present work.  Feynman noted
that higher order (in $G_{F}$) many-body interactions could be
important because they depended on higher powers of the masses of
the interacting objects.  This is a special case of an
observation made earlier by Primakoff and Holstein (PH)
\cite{PH39} in their classic analysis of many-body effects in
atomic and nuclear systems.  PH noted that in a nucleus
containing $N$ particles the magnitude of the total $k$-body
interaction ($k = 2,3,\cdots$). grows as the binomial coefficient
${N\choose k}$,
\begin{equation}
	{N \choose k} = \frac{N!}{k!(N - k)!},
\label{bin coeff}
\end{equation}
which counts the number of distinct $k$-body amplitudes that can
be formed from $N$ particles.  Subsequently Chanmugam and
Schweber \cite{CS70} suggested that many-body electromagnetic
effects could be important in white dwarfs due to the presence of
the binomial coefficient.  We will return shortly to discuss the
combinatorics of many-body diagrams using Eq.~(\ref{bin coeff}).
Subsequently Hartle \cite{HAR70} used the same phenomenological
weak coupling employed by FS to evaluate the $\O(G_{F}^{4})$
4-body contribution corresponding to the diagram in 
Fig.~\ref{First 4-body figure}a
and obtained the result quoted in Eq.~(\ref{V4}) below.  Using
this result Hartle demonstrated that the energy for two electrons
interacting with each other and with a spherical shell of matter
varied with the separation $r$ of the electrons approximately as
$1/r$.  For the systems that Hartle considered the effects
arising from the 4-body potential were too small to be of
interest.  Moreover, even though electrons can interact via
long-range neutrino-exchange forces, these cannot be used to
measure the electron-number of a black hole (``a black hole has
no neutrino hair'' \cite{HAR72}).  Although these results appear
to support the conclusion derived from the 2-body potential that
neutrino-exchange forces are unimportant, we show in what follows
that this is not necessarily the case when computing the
self-energy of a compact object such as a neutron star or a white
dwarf.  In fact the self-energy of such an object arising from
the exchange of massless neutrinos can be catastrophically large,
and this result eventually leads to the conclusion that neutrinos
cannot be massless.

	Since the large neutrino-exchange energy-density in
a neutron star arises from many-body interactions among neutrons,
it is appropriate to ask why many-body effects play so important
a role here, but are relatively unimportant in most other
circumstances.  The explanation can be found in 
Table~\ref{many-body table} which compares neutrino-exchange to
other known forces with respect to three conditions which
determine when many-body effects become significant.  In order
for there to be an enhancement effect arising from the binomial
coefficient ${N \choose k}$, it is necessary in the present
context to have a large number of particles interacting with
sufficient strength in a small volume.  

	The first condition is
that the force be long-range, which is what allows a given
particle to interact with many other particles in the first
place.  We see from Table~\ref{many-body table} that both the
strong interaction and the weak interaction mediated by
$Z^{0}$-exchange fail to meet this condition.  In the case of
strong interactions it is well known that the forces among
nucleons saturate, i.e., that each nucleon in a nucleus interacts
with only a limited number of other nucleons \cite{FH74,deB64}.
This is evidenced by the fact that the strong interaction binding
energy of a nucleus does not increase with baryon number $A$ as
${A \choose 2} = A(A - 1)/2$, as might be expected from a 2-body
interaction, but grows approximately as $A$ for $A > 4$.  In
nuclei this can be understood in terms of exchange forces and the
existence of a repulsive hard core, but for a macroscopic object
the finite range of nuclear forces is also significant.  

	The
second condition in Table~\ref{many-body table} is that the
force couple to a charge capable of having a large expectation value
in a macroscopic system such as a neutron star.  Evidently a
long-range force will not produce a significant many-body
interaction unless the constituents that are capable of
interacting with one another have a (large) net value of the
appropriate charge and a non-zero charge density.  
For the electromagnetic interaction a number
of independent arguments show that the net charge $Z$ of a
typical neutron star is less than $10^{36}e_{0}$, where $e_{0}$
is the charge of the electron, and may be as small as
$10^{17}e_{0}$ \cite{KKF94}.  The latter estimate is smaller than
$N$ by a factor of order $10^{-40}$, which by itself would be
sufficient to suppress many-body effects.  However, there are in
addition the suppression effects originally discussed by
Primakoff and Holstein \cite{PH39} and these, along with
the small net $Z$, suggest that many-body electromagnetic
interactions should be small in neutron stars.  

	The last condition in Table~\ref{many-body table}
requires that the interaction be sufficiently strong so that, when
many-body interactions are in fact possible, their net effect be
large.  This condition can be understood with reference to the
gravitational interaction as follows:  We note that the quantity
which determines when gravity is sufficiently strong that
many-body interactions would be important is the dimensionless
potential $\Phi$,
\begin{equation}
	\Phi = \frac{G_{N}M}{Rc^{2}},
\label{gravity Phi}
\end{equation}
where $G_{N}$ is the Newtonian gravitational constant, $M$ is the
mass of the neutron star, and $R$ is its radius.  Using the
numerical results in Sec.~V below we find that $\Phi \simeq 0.2$
for the binary pulsar PSR 1913+16, which is the ``typical''
neutron star we are considering.  The fact that $\Phi$ must be
small is
obvious since $\Phi \geq 1/2$ would correspond to a black
hole.  It follows that for ordinary matter many-body
gravitational effects will never dominate over the two-body
interaction, although they may lead to detectable effects in some
systems \cite{Will94}.  Another way of understanding this result
is to note that in natural units, where $G_{F}$ and $G_{N}$ have
the same dimensions,
\begin{equation}
	G_{N}/G_{F} \simeq 10^{-33}.
\label{GN/FG}
\end{equation}
If follows from Eq.~(\ref{GN/FG}) that if one were to reduce
$G_{F}$ by a factor of order $10^{33}$ so as to make the weak
neutrino-exchange interaction have a strength comparable to that
of gravity, then many-body neutrino-exchange effects would become
relatively unimportant, just as gravitational many-body effects
are.

	The preceding discussion can be summarized as follows:
We see from Table~\ref{many-body table} that each of the four known
fundamental forces fails to satisfy at least one of the conditions
that must be met for many-body effects to be important in a
macroscopic system such as a neutron star.  The only known interaction
which is of long-range, and where there is a strong coupling to a
charge for which a neutron star is non-neutral, is the force arising
from neutrino exchange.  It is for these reasons that many-body
exchange effects can be significant for this interaction,
while they are relatively unimportant for the others.

	To understand quantitatively how many body neutrino effects
can give rise to a catastrophically large energy-density,
we follow the discussion of Feinberg
and Sucher \cite{FS68} who note that the functional form of
$V_{ee}^{(2)}(r)$ can
be inferred on dimensional grounds.  Since the Standard  Model is
renormalizable, the only dimensional factors upon which the
static spin-independent potential can depend are $G_{F}$ and
$r$.  Evidently the exchange of a single $\nu\overline{\nu}$
pair must be proportional to $G_{F}^{2}$, from which it follows
that $V_{ee}^{(2)} \propto G_{F}^{2}/r^{5}$ in agreement with
Eq.~(\ref{V2}) below.  An analogous argument shows that for $k \geq 3$
the $k$-body neutrino-exchange contribution $W^{(k)}$ to the binding
energy of a neutron star must be proportional to
$(G_{F}^{k}/R^{2k +1}){N \choose k}$, where $R$ is the radius of
the neutron star, which is assumed for present purposes to
contain
only neutrons and to have
a uniform density.  Since ${N \choose k} \simeq N^{k}/k!$ for $k
\ll N$, it follows that
\begin{equation}
        W^{(k)} \sim \frac{1}{k!}\,\frac{1}{R}
                \left(\frac{G_{F}N}{R^{2}}\right)^{k}.
\label{approx Wk}
\end{equation}
For a typical neutron star $(G_{F}N/R^{2}) = \O(10^{13})$ [see
Sec.~V below], and hence it follows from Eq.~(\ref{approx Wk}) that for
$k \ll
N$ higher order many-body interactions make increasingly larger
contributions to $W^{(k)}$.  It can be shown that
$W^{(8)}$ would exceed the known mass-energy of a typical neutron
star, and
that $W = \sum_{k}W^{(k)}$ can exceed the total mass-energy of the
Universe, as we discuss below.

	The calculation of the self-energy of a neutron star or
white dwarf arising from neutrino exchange closely parallels the
calculation of the electrostatic energy of a spherical charge
distribution such as a nucleus arising from photon exchange.
For this reason, we begin in Sec.~II by reviewing the derivation
of the familiar electrostatic (Coulomb) result, which is given in
Eq.~(\ref{spherical Coulomb energy}).  For present purposes the electrostatic
derivation can be viewed as proceeding in three steps:  1) First
the 2-body potential is derived from covariant perturbation
theory.  2)  The 2-body potential is then integrated over a
spherical volume to obtain the contribution from a single pair of
charges. 3) The result of the preceding calculation is then
multiplied by the factor ${N \choose 2} = N(N-1)/2$ which
represents the number of pairs that can be formed from $N$
charges.  Particular attention is devoted to the question of
integrating the 2-body Coulomb potential over a spherical volume,
which is carried out in two different ways.  One method uses a
geometric probability technique, which is later generalized to
apply to the many-body case.

	Sec.~III presents the derivation of the $k$-body
potential $V^{(k)}$ arising from neutrino-exchange, using a
formalism developed by Schwinger \cite{SCH54} and originally
applied to this problem by Hartle \cite{HAR70}.  After
reproducing the 2-body result of FS and the 4-body potential
obtained by Hartle, the 6-body potential is derived [Eq.~(\ref{V6})]
and then used to generalize to the $k$-body case 
[Eqs.~(\ref{Vk})--(\ref{Pk})].  In Sec.~IV we discuss the
integration of the $k$-body results over a spherical volume of
radius $R$, which leads eventually to Eq.~(\ref{effective U0 k}).  
When this expression is multiplied by ${N \choose k}$ the result is
the net $k$-body contribution to the self-energy of the object in
question.  After the sum over $k$ is carried out, the self-energy
$W$ of a spherical collection of $N$ neutrons can be approximated
by the expression in Eq.~(\ref{WforN 5}).  We then proceed to
demonstrate that, barring accidental cancellations, $W/M \gg 1$
for compact objects such as neutron stars and white dwarfs, where
$M$ is the (known) mass of each object.  After considering in
Sec.~VI, and excluding, several alternative explanations for the
unphysically large value of $W/M$, we turn in Sec.~VII to
recalculate $W$ when the exchanged neutrinos have a nonzero mass
$m$.  As expected, the expression for $W$ acquires additional
factors proportional to $\exp(-mR)$ which suppress the
neutrino-exchange contribution.  One can then calculate the
minimum value of $m$ required to reduce $W$ to a physically
acceptable value, and the result is presented in Sec.~VIII.  From
Eq.~(\ref{final m limit}) we find for $m$,
\begin{equation}
	m \gtrsim 0.4 \,\,\mbox{eV/$c^{2}$},
\end{equation}
which is consistent with the upper limits quoted in Eq.~(\ref{m
nu limits}).

	Appendix~A contains a summary of our notation and metric
conventions, as well as a discussion of the neutrino couplings in
the Standard Model.  In Appendix~B we present Hartle's
(unpublished) derivation \cite{HARUNP} of the
Schwinger formula for $W$, along with some additional technical
details of its application to the present problem.  Appendix~C
contains a detailed discussion of the geometric probability
formalism used in integrating the many-body potentials over a
spherical volume.  In Appendix~D we discuss the formalism for
summing the many-body contributions, and in Appendix~E we
generalize the Schwinger-Hartle formalism to apply to massive
neutrinos.

\section{ELECTROSTATIC ENERGY OF A SPHERICAL CHARGE DISTRIBUTION}

	As noted in the Introduction, the present 
calculation of the energy-density
of a spherical neutron star arising from neutrino-exchange parallels
that of the electrostatic energy of a spherical charge distribution.
For this reason we review the derivation of the familiar result for
the Coulomb energy $W_{C}$ of a spherical nucleus \cite{BM69},
\begin{equation}
	W_{C} = \frac{3}{5} Z(Z-1) \frac{e_{0}^{2}}{R} ,
\label{spherical Coulomb energy}
\end{equation}
where $R$ is the effective Coulomb radius, $Ze_{0}$ is the nuclear charge,
and $e_{0}^{2}/\hbar c \simeq 1/137$.  (We use $e_{0}$ to denote
the charge of the electron to avoid confusion with the transcendental
number $e$: $\ln e = 1$.)  
As noted in the Introduction, we can
view the derivation of Eq.~(\ref{spherical Coulomb energy}) as 
proceeding in three steps: (1) Determine the 2-body interaction
$V_{C}(\vec{r}_{12})$ between two point charges; (2) calculate
the interaction energy $U_{C}$ between the two point charges 
which are assumed to be uniformly distributed inside a sphere of
radius $R$; (3) generalize the 2-body result $U_{C}$ to the
$Z$-body electrostatic energy $W_{C}$ by incorporating the 
appropriate combinatoric factors.

\subsection{Calculation of the Two-Body Potential ${\bif V}_{\bif C}$}  

The electrostatic potential energy is dominated by the two-body
contribution arising
from the one-photon-exchange amplitude. 
In this case the
evaluation of the expression for the potential energy $V_{C}(\vec{r}_{12})$
of two interacting particles is trivial, and the result is the familiar
Coulomb potential
\begin{equation}
  V_{C}(\vec{r}_{12}) =  \frac{e_{0}^{2}}{r_{12}} ,
\label{Coulomb PE}
\end{equation}
where $r_{12} = |\vec{r}_1 - \vec{r}_2|$ is the distance between the
charges.  Although there are also in principle contributions from many-body
electromagnetic interactions, these are small for reasons discussed
originally by Primakoff and Holstein \cite{PH39}.  By contrast the
weak energy arising from neutrino-exchange is dominated by many-body
interactions, and hence the weak potential 
is correspondingly more complicated.

\subsection{Integration Over a Sphere: ${\bif U}_{\bif C}$}  

This is the most difficult of the three steps in the
weak-interaction case.  For the Coulomb interaction
in Eq.~(\ref{Coulomb PE}) we are interested in 
evaluating the energy of a single pair of 
charges 1 and 2 ($e_{1} = e_{2} \equiv e_{0}$), 
having a uniform probability distribution in 
a volume $(4/3)\pi R^{3}$, 
so that the effective
number density $\rho$ produced by each charge is
\begin{equation}
 \rho_{1} = \rho_{2} \equiv \rho = \frac{1}{(4/3)\pi R^3} .
\label{charge density}
\end{equation}
From Eq.~(\ref{Coulomb PE}), the potential energy arising from 
the Coulomb interaction between the
charges $\rho_{1}d^{3}x_{1}$ and $\rho_{2}d^{3}x_{2}$ centered
at $\vec{r}_1$ and $\vec{r}_2$ respectively is 
\begin{equation}
  	d U_{C} =  (\rho_{1}d^{3}x_{1})
			(\rho_{2}d^{3}x_{2})V_{C}(\vec{r}_{12}).
\label{infinitesmal PE}
\end{equation}
The average electrostatic energy $U_{C}$ is then obtained by integrating
Eq.~(\ref{infinitesmal PE}) over the entire sphere:
\begin{eqnarray}
  	U_{C} & = & \int\rho_{1}d^{3}x_{1}
		 \int\rho_{2}d^{3}x_{2}\,
		V_{C}(\vec{r}_{12})
	\nonumber \\
	& = &
	e_{0}^{2}\,\rho^{2}
	\int^{R}_{0}dr_{2}\,r_{2}^{2} 
	\int^{R}_{0}dr_{1}\,r_{1}^{2}
	\int d\Omega_{1}\int d\Omega_{2}\,
	\left(\frac{1}{r_{12}}\right)
	\nonumber \\
	& = & \frac{6}{5}\frac{e_{0}^{2}}{R},
\label{first U_C}
\end{eqnarray}
where $r_{1} = |\vec{r}_{1}|$ and $r_{2} = |\vec{r}_{2}|$
 are measured from the center of the
sphere, and $d\Omega =
d\!\cos\theta d\varphi$.  
The result in Eq.~(\ref{first U_C}) represents 
the average interaction energy of a single
pair of charges having a uniform probability distribution in a spherical volume of radius $R$.

Although the 6-dimensional integral arising from the 2-body potential
energy in Eq.~(\ref{Coulomb PE}) is straightforward, 
its generalization to the
many-body potentials that arise from neutrino-exchange becomes
increasingly cumbersome.  We will return below to discuss an alternative
method for evaluating the integral over a sphere, which generalizes
more naturally to the many-body case.

\subsection{Combinatorics:  $\bif W_{\bif C}$}  

As noted above, the result in Eq.~(\ref{first U_C})
gives the energy $U_{C}$ for a single pair of charges.  
For a nucleus containing
$Z$ charges the number of pairs that can be formed is
$Z(Z-1)/2$, so that the final expression for the total
energy $W_{C}$ is
\begin{equation}
  	W_{C} = \frac{1}{2}Z(Z-1)U_{C}
	= \frac{3}{5}
     	Z(Z-1)\frac{e_{0}^{2}}{R},
\label{Z W_C}
\end{equation}
in agreement with Eq.~(\ref{spherical Coulomb energy}).  
For application to the $k$-body problem
$(k = 2,3,4,\ldots)$ the combinatoric factor $Z(Z-1)/2$ generalizes to
the binomial coefficient \cite{PH39} defined in Eq.~(\ref{bin
coeff})
which counts the number of combinations of $k$ objects that
can be formed from $N$ objects.  Evidently
\begin{equation}
	{Z \choose 2} = \frac{Z(Z-1)}{2},
\end{equation}
which reproduces the result in Eq.~(\ref{Z W_C}).

	As we noted in the Introduction,
the fact that the net $k$-body contribution is proportional to the binomial
coefficient ${N\choose k}$ was pointed out
by Primakoff and Holstein in their classic paper \cite{PH39} on many-body
electromagnetic forces.  Subsequently Chanmugam and Schweber \cite{CS70}
re-examined the question of many-body forces in electromagnetism.
They observed that since the binomial coefficient 
for small $k$ grows as
\begin{equation}
    {N \choose k} \simeq \frac{N^{k}}{k!},
\end{equation}
this factor may make many-body $(k > 2)$ electromagnetic effects
important in white dwarfs, where the electron density can be high.
For neutrino-exchange amplitudes the presence of the binomial coefficient
in the final expression for the interaction energy is what eventually
leads to the large neutrino-exchange energy-density  
referred to in the
Introduction.  Specifically, each neutron in a neutron star carries
a non-zero weak charge which couples the neutron to the $\nu\bar{\nu}$
current, so that for a neutron star $N$ is of order $10^{57}$.  By way
of comparison, the electric charge $Z$ of a neutron star is at most of
order $10^{36}$ \cite{KKF94}, as noted in the Introduction.  
It follows that the enhancement of the
many-body amplitudes arising from the binomial coefficient is a much
more significant effect for neutrino exchange than it would be for the
corresponding electrostatic case. 

	In the ensuing discussion it will be helpful to recall that the
binomial coefficient ${N \choose k}$
is a decreasing function of increasing $k$ for fixed $N$:  From 
Eq.~(\ref{bin coeff}),
\begin{equation}
	\frac{{N \choose k+1}}{{N \choose k}} 
             = \frac{N-k}{k+1}.
\label{ratio binomial coefficients}
\end{equation}
Hence the $k$-body interaction is proportional to a combinatoric factor
which, although large, is decreasing monotonically as $k$ increases.
From the discussion in Appendix~B we see that the weak interaction
energy arising from neutrino-exchange can be expressed as a sum over $k$-body
contributions $W^{(k)}$ as in Eq.~(\ref{W sum}), 
each of which will be proportional
to ${N \choose k}$.  The fact that
the ratio in Eq.~(\ref{ratio binomial coefficients}) is 
less than unity when $k > (N-1)/2$ helps to
explain why $\sum_{k} W^{(k)}$ is dominated by a few terms with $k
\simeq N$, as we
discuss in Sec.~V below.

\subsection{Alternate Method of Integration}

	Having outlined the steps that lead to Eq.~(\ref{spherical
Coulomb energy}), we return to discuss
an alternate method for evaluating the integral of the Coulomb potential
in Eq.~(\ref{Coulomb PE}) over a spherical volume.  
Let ${\cal P}_{3}(r)$ denote the normalized
probability density for finding two points randomly chosen in a
uniform 3-dimensional
sphere to be a distance $r \equiv r_{12}$ apart. 
As discussed in Appendix~C, the
average value $\langle g\rangle$ of any function $g(r)$ 
taken over a 3-dimensional spherical volume is then given by
\begin{equation}
	  \langle g \rangle = \int_{0}^{2R}dr {\cal P}_3(r) g(r),
\label{average g}
\end{equation}
where
\begin{equation}
	 \int_{0}^{2R}dr{\cal P}_3(r) = 1.
\end{equation}
The functional form of ${\cal P}_3(r)$, and its 
generalization ${\cal P}_n(r)$ for an
$n$-dimensional ball of radius R, are discussed in Appendix C. 
We henceforth drop the subscript 3 when working in 3-dimensions
(${\cal P} \equiv {\cal P}_{3}$).   From
Eq.~(\ref{P_3(s)}), with $r = 2Rs$, 
\begin{eqnarray}
   {\cal P}(r)
	& = & \frac{3r^2}{R^3} - \frac{9}{4} \frac{r^3}{R^4}
       + \frac{3}{16} \frac{r^5}{R^6}
	\nonumber \\
	& = &
     \frac{3r^2}{R^3} \left[ 1 - \frac{3}{2} \left(\frac{r}{2R}\right)
           +  \frac{1}{2} \left( \frac{r}{2R}\right)^3 \right]. 
\label{3D P}
\end{eqnarray}
Since $2R \geq r \geq 0$,
it is sometimes convenient to introduce the scaled dimensionless variable
$s$ defined above which satisfies $1 \geq s \geq 0$.  
Plots of ${\cal P}(s)$ and its derivative ${\cal P}'(s)$ are
given in Fig.~\ref{P plot figure}.
Eq.~(\ref{average g}) can be viewed as an 
{\em analytic} Monte Carlo calculation of $\langle g\rangle$.

	Returning to the Coulomb problem we wish to calculate 
$\langle e_{0}^{2}/r\rangle$ by
this method.  From Eq.~(\ref{average g}), 
\begin{equation}
	e_{0}^2 \langle 1/r \rangle = 
	e_{0}^2 \int_{0}^{2R} dr \left( \frac{3r^2}{R^3} -
    \frac{9r^3}{4R^4} + \frac{3r^5}{16R^6}\right) \frac{1}{r} =
     \frac{6}{5} \frac{e_{0}^2}{R},
\label{average 1/r}
\end{equation}
in agreement with Eq.~(\ref{first U_C}).  
The advantage of this approach 
is that, where applicable, it replaces the 6-dimensional 
integral in Eq.~(\ref{first U_C}) by the 1-dimensional
integral in Eq.~(\ref{average 1/r}).  
Other useful results are given in Appendix C.

\subsection{Application to Massive Electrodynamics}

	We conclude the previous discussion by using Eq.~(\ref{average g}) 
to calculate the
electrostatic energy of a spherical charge distribution in massive
electrodynamics, i.e., for the case of a photon with a non-zero mass $\mu$.
This calculation completes the analogy between the electrostatic energy
arising from photon exchange, and the weak energy arising from neutrino
exchange when $m \neq 0$.  When $\mu \neq 0$ the Coulomb potential
in Eq.~(\ref{Coulomb PE}) is replaced by the Yukawa potential \cite{KLOOR94}
\begin{equation}
	V_{Y}(r_{12}) = \frac{e_{0}^2}{r_{12}} e^{-\mu r_{12}},
\label{Yukawa PE}
\end{equation}
and the result in Eq.~(\ref{average 1/r}) generalizes to
\begin{equation}
	e_{0}^2 \langle e^{-\mu r}/r\rangle  = 
	\frac{6}{5}\frac{e_{0}^{2}}{R} F(\mu R),
\label{average Yukawa}
\end{equation}
where
\begin{equation}
   F(\mu R) = \frac{15}{4} \left[ \frac{2}{3(\mu R)^2} -
    \frac{1}{(\mu R)^3} + \frac{1}{(\mu R)^5} \right] -
    \frac{15}{4} e^{-2\mu R} \left[ \frac{1}{(\mu R)^3} +
     \frac{2}{(\mu R)^4} + \frac{1}{(\mu R)^5}\right].
\label{Fmu}
\end{equation}
The ``form factor'' $F(\mu R)$ incorporates all the modifications arising
from $\mu \not= 0$, and has the property that $F(0) = 1$.

	We note in passing that the result in Eqs.~(\ref{average
Yukawa}) and (\ref{Fmu}) is also of interest in
connection with recent work setting limits on new forces co-existing
with electromagnetism \cite{KLOOR94}.  Although limits on the photon mass
are quite stringent, the corresponding limits on a new vector force
co-existing with electromagnetism are less restrictive.  If $e_{0}$ in 
Eq.~(\ref{average Yukawa})
were replaced by the corresponding unit of charge $f$ for the new field,
then Eq.~(\ref{average Yukawa}) 
could be used to set limits on $f^{2}$ and $\mu$ in
appropriate systems, in a manner similar to that described below
for the neutrino mass $m$.	

\subsection{Quantum Mechanical Effects}

	The preceding calculation also serves to clarify another
issue which arises in neutrino-exchange, namely, why the charge
distribution can be treated classically even in an object such as
a nucleus or a neutron star where quantum effects are important.
In particular one may ask what role the Pauli exclusion principle
plays for the protons in a nucleus or the neutrons in a neutron
star in the presence of a long-range force.  
For present purposes we can invoke an argument due to Fermi
\cite{FERMI} to demonstrate that the self-energy of a nucleus or
neutron star arising from long-range forces can in fact be
approximated by the classical result, as we have done.
(We will return in
Sec.~V below to discuss the effects of the Pauli principle for
the neutrinos.)

\section{THE MANY-BODY POTENTIALS ARISING FROM NEUTRINO-EXCHANGE}

	Following the discussion in Sec. II, we present in this section a
derivation of the many-body  spin-independent
potentials arising from neutrino-exchange.
We will begin by using the Hartle formalism \cite{HAR70,HARUNP} to
re-derive the original 2-body result of Feinberg and Sucher (FS) arising
from the diagram in
Fig.~\ref{2-body neutrino exchange figure} \cite{FS68,FSA89,HS94}.  
It will then be shown that $k$-body potentials
where $k$ is odd make no contribution to the neutrino-exchange energy
of a spherical neutron star.  After reproducing the 4-body result of
Hartle \cite{HAR70}, we derive the 6-body potential which is then used to
infer the relevant combinatoric factors in the general $k$-body potential.
For the sake of definiteness we assume that the external fermions
are neutrons, as would be the case in a neutron star, and we
therefore suppress the corresponding subscripts on $V$ and $W$.

	Our starting point is the Schwinger formula in 
Eq.~(\ref{Schwinger W}), 
\begin{equation}
	W = \frac{i}{2\pi} \mbox{Tr} \left\{ \int_{-\infty}^\infty dE~\ln
     [1 + \frac{G_F a_n}{\sqrt{2}}  N_\mu \gamma_\mu (1+\gamma_5) S_F^{(0)}
      (E)]\right\},
\label{Schwinger W again}
\end{equation}
where Tr denotes the generalized trace defined by Eq.~(\ref{new trace}),
and $\ln[1+ \cdots]$ represents the infinite series
\begin{equation}
	\ln(1 + \Delta) = -\sum_{k = 1}^\infty (-1)^k \frac{\Delta^k}{k} .
\label{ln Delta}
\end{equation}
The factor $1/k$ can be understood as the product of $1/k!$ arising
from the perturbation expansion of $\exp (i{\cal L}_I(x))$
where ${\cal L}_I(x)$ is given in Eq.~(\ref{neutrino Lagrangian}), 
and $(k-1)!$ which represents
the number of ways that the $k$ external currents can be attached
to the
closed neutrino loop.  There is an additional factor of $k!$ which
counts the number of ways that the momenta carried by the external 
currents can be assigned to the $k$ vertices.  The net result is that
the expansion of $\ln[1 + \cdots]$ in the Schwinger formula gives rise
to 
$(k-1)!$ topologically distinct diagrams, but there is no overall $k$-dependent
numerical coefficient \cite{IZ80}.  
For $k=4$ there are 6 independent diagrams:  These are given
by the 3 shown in Fig.~\ref{First 4-body figure}
along with an additional 3 diagrams
obtained from those shown by reversing the direction of the internal
neutrino momentum, as in Fig.~\ref{Second 4-body figure}.  
For $k=2$, $(k-1)! = 1$ which agrees with our expectation
that there be only one diagram in order $G_F^2$, since both senses of the
neutrino momentum are topologically equivalent.

\subsection{The 2-Body Potential}

Combining Eqs.~(\ref{Schwinger W again}) and (\ref{ln Delta}) 
the $k = 2$ contribution to $W$ is given by
\begin{eqnarray}
	W^{(2)} & = &
	\frac{-i}{2\pi} \left(\frac{G_Fa_n}{\sqrt{2}}\right)^2
     \int d^3x_1d^3x_2 \int_{-\infty}^\infty dE 
	\nonumber \\
        &   & \times \mbox{tr} \left\{
     \gamma_\mu(1+\gamma_5)S_{F}^{(0)}(\vec{r}_{12},E)\gamma_\nu
      (1+\gamma_5)S_{F}^{(0)}(\vec{r}_{21},E)\right\} 
    N_\mu(x_1)N_{\nu}(x_2),
\label{2bodyW}
\end{eqnarray}
where $\vec{r}_{12} = (\vec{x}_1-\vec{x}_2)$, $N_\mu(x_1)$ and $N_\nu(x_2)$
denote the external neutron currents at 
$x_{1}$ and $x_{2}$, and $S_{F}^{(0)}(\vec{r}_{ij},E)$ is given
by Eq.~(\ref{final inverted SF}).  
The overall minus sign arises from the expansion in 
Eq.~(\ref{ln Delta}),
and tr denotes the trace over the Dirac matrices in $\left\{\cdots\right\}$.
The factors of $(1+\gamma_5)$ can be anticommutated past $S_F^{(0)}$ and,
using the relation
\begin{equation}
	(1+\gamma_5)^2 = 2(1+\gamma_5),
\label{1 + gamma5}
\end{equation}
we have 
\begin{equation}
	\mbox{tr} \left\{ \cdots \right\} 
	= 2 \mbox{tr} \left\{S_{F}^{(0)} 
     (\vec{r}_{21},E)\gamma_\mu S_{F}^{(0)} (\vec{r}_{12},E)
      \gamma_\nu (1+\gamma_5)\right\}.
\label{trace identity}
\end{equation}
Since we are interested in computing the self energy of a static collection
of neutrons (which are assumed to have no net polarization), the neutron
currents are given by
\begin{equation}
	N_\mu(x_1) = i\rho_1 \delta_{\mu 4};~~~~ 
	N_\nu(x_2) = i\rho_2\delta_{\nu 4},
\label{Nus}
\end{equation}
where $\rho_1 = \rho_2 = \rho$ is the number density of neutrons in the
neutron star, and the factors of $i$ arise from Eq.~(A4).  
Combining 
Eqs.~(\ref{trace identity}), (\ref{Nus}), 
and (A8) 
we see that the term containing 
$\gamma_5$,
whose trace is proportional to $\epsilon_{\mu\nu\lambda\rho}$, makes no
contribution.  The remaining terms give
\begin{eqnarray}
	W^{(2)} & = & 
	\frac{-(i)^3}{\pi} 
	\left( \frac{G_{F}a_{n}}{\sqrt{2}}\right)^2
       \int \rho_{1} d^{3}x_{1} 
	\int \rho_{2} d^{3}x_{2}
	\int_{-\infty}^{\infty} dE 
	\nonumber \\
	&  & \mbox{} \times 
    \mbox{tr}\left[\gamma_\alpha \gamma_4 \gamma_\beta \gamma_4\right] 
	\eta_\alpha(21)
    \Delta_F(\vec{r}_{21},E)
    \eta_\beta (12) \Delta_F (\vec{r}_{12},E).
\label{W2 a}
\end{eqnarray}
Using
\begin{equation}
  \mbox{tr}
	[\gamma_\alpha \gamma_4 \gamma_\beta \gamma_4] =
     4(2\delta_{\alpha 4} \delta_{\beta 4} - \delta_{\alpha \beta}),
\end{equation}
the expression in curly brackets in Eq.~(\ref{trace identity}) can be written as
\begin{eqnarray}
  \{ (\ref{trace identity}) \} 
&=& 4[2\eta_4(21)\eta_4(12) - \eta(21)\cdot \eta(12)]
	\nonumber \\
     & = & 4[E^2 - \vec{\partial}_{12} \cdot \vec{\partial}_{21}],
\label{curly brackets}
\end{eqnarray}
where $\vec{\partial}_{12} \equiv \partial/\partial \vec{r}_{12}$.  Combining
Eqs.~(\ref{curly brackets}) and (\ref{W2 a}) we have
\begin{eqnarray}
    W^{(2)} & = & 
	\frac{4i}{\pi} \left(\frac{G_Fa_{n}}{\sqrt{2}}\right)^2
     \int \rho_1 d^3 x_1 \int \rho_2 d^3x_2 \int_{-\infty}^\infty dE 
	\nonumber \\
    	    &  & \times
	\left\{ E^2 \Delta_F(\vec{r}_{21},E) \Delta_F (\vec{r}_{12},E)
      - \vec{\partial}_{12}\cdot \vec{\partial}_{21} \Delta_F (\vec{r}_{21},E)
     \Delta_F(\vec{r}_{12},E)\right\}
	\nonumber \\
    	 & = & 
	\frac{4i}{\pi} \left(\frac{G_Fa_{n}}{\sqrt{2}}\right)^2
      \int \rho_1 d^3x_1 \int \rho_2 d^3x_2 \int_{-\infty}^\infty dE
	\nonumber \\ 
     	 &    &  \times
	 \left( \frac{i}{4\pi}\right)^2 
	\left\{ E^2 \frac{e^{i|E|(r_{12}+r_{21}+i\epsilon)}}
            {r_{12}r_{21}}
    - \vec{\partial}_{12}\cdot \vec{\partial}_{21} 
    \left( \frac{e^{i|E|(r_{12}+r_{21}+i\epsilon)}}
	{r_{12}r_{21}}\right) \right\}.
\label{W2 b}
\end{eqnarray}
We note from Eq.~(\ref{final inverted SF}) 
that the operators $\vec{\partial}_{12}$ and
$\vec{\partial}_{21}$ act on the respective coordinates $\vec{r}_{12}$ and
$\vec{r}_{21}$ as if these were independent, notwithstanding the fact that
$\vec{r}_{12} + \vec{r}_{21} = 0$.  
This applies as well to all the
derivative terms that appear in the $k$-body amplitudes.

Following Hartle \cite{HARUNP} the integral over $E$ can be evaluated 
by considering the functions $\bar{I}_{n}(z)$ defined by
\begin{eqnarray}
	\bar{I}_{n}(z) & = &
		\int_{-\infty}^{\infty} dE~E^ne^{i|E|(z+ i\epsilon)} 
	\nonumber \\
 		& = & \left\{ \begin{array}{ll}
		2\int_{0}^{\infty} dE~E^n~e^{i|E|(z+i\epsilon)}
		\equiv I_{n}(z) & \,\,\, \mbox{even $n$}
		\\
	0 & \,\,\,\mbox{odd $n$},
                \end{array} \right.
\label{general In}
\end{eqnarray}
where $z = r _{12} + r_{21}$.  Since $|E|$ is an even 
function of $E$, $\bar{I}_{n}(z)$ is nonzero
only for even values of $n$.  An elementary integration gives
\begin{equation}
	I_{0}(z) = \frac{2i}{z+i\epsilon} ,
\label{I0}
\end{equation}
and differentiating Eqs.~(\ref{general In}) and 
(\ref{I0}) with respect to $z$ leads to
\begin{equation}
	-i \frac{dI_0(z)}{dz} = I_1(z) = \frac{-2}{(z+i\epsilon)^2}.
\end{equation}
Continuing in this way we find \cite{HARUNP}
\begin{equation}
	I_n(z) = \frac{2i^{n+1}n!}{(z+i\epsilon)^{n+1}}  .
\label{In}
\end{equation}
Combining Eqs.~(\ref{W2 b}) and (\ref{In}) allows $W^{(2)}$ to be
written as
\begin{eqnarray}
  	W^{(2)} & = & 
	\int \rho_1 d^3x_1 \int \rho_2 d^3x_2 \left\{ \left(\frac{-1}{\pi^3}\right)
      \left(\frac{G_Fa_{n}}{\sqrt{2}}\right)^2 
	\right.
	\nonumber \\
    	&   &  \times \left.
	\left[ \frac{1}{r_{12}r_{21}(r_{12}+r_{21})^3}
     + \frac{1}{2} \vec{\partial}_{12}\cdot \vec{\partial}_{21}
     \frac{1}{r_{12}r_{21}(r_{12}+r_{21})} \right] \right\}.
\label{W2}
\end{eqnarray}
We note from Eq.~(\ref{first U_C}) 
that the integrand in Eq.~(\ref{W2}) has the same form as in the
analogous electromagnetic case, with the quantity in curly brackets in
Eq.~(\ref{W2}) corresponding to the 2-body potential $V^{(2)}(r_{12})$.
Thus 
\begin{equation}
	V^{(2)}(r_{12}) = 
	\frac{-1}{\pi^3} \left(\frac{G_Fa_n}{\sqrt{2}}
     \right)^2 \left[ \frac{1}{r_{12}r_{21}(r_{12} + r_{21})^3}
    + \frac{1}{2} \vec{\partial}_{12}\cdot \vec{\partial}_{21}
     \frac{1}{r_{12}r_{21}(r_{12} + r_{21})} \right] .
\label{general V2}
\end{equation}
Since $r \equiv r_{12} = r_{21}$, first term in square brackets
in Eq.~(\ref{general V2}) reduces to $1/8r^5$.  In the second term we note that the
gradients act on a function which depends only on
$r_{12}$ and $r_{21}$, and hence we can write
\begin{equation}
	\vec{\partial}_{12} \cdot \vec{\partial}_{21} =
     \left( \hat{r}_{12} \frac{\partial}{\partial r_{12}}\right) \cdot
      \left(\hat{r}_{21} \frac{\partial}{\partial r_{21}}\right) =
      \hat{r}_{12} \cdot \hat{r}_{21} \frac{\partial}{\partial r_{12}}
      \frac{\partial}{\partial r_{21}} = -
      \frac{\partial}{\partial r_{12}} \frac{\partial}{\partial r_{21}} .
\label{12 gradients}
\end{equation}
Using Eq.~(\ref{12 gradients}) 
the expression in square brackets in Eq.~(\ref{general V2}) can
be written as
\begin{equation}
   [(\ref{general V2})] = \frac{1}{8r^5} - \frac{5}{8r^5} 
     		= \frac{-1}{2r^5} ,
\end{equation}
and hence,
\begin{equation}
	V^{(2)}(r) = + \frac{G_F^2 a_n^2}{4\pi^3} \frac{1}{r^5} .
\label{V2}
\end{equation}
Eq.~(\ref{V2}) gives the original FS result \cite{FS68,FSA89,HS94}
when we set $a_n = 1$, which is the value appropriate to the
charged-current model of the weak interaction assumed by FS.
 
For later purposes it is interesting to note that the functional 
form of $V^{(2)}(r)$ can be
inferred on dimensional grounds, as noted originally by Feinberg and
Sucher \cite{FS68}.  The only dimensional quantities upon which a static
neutrino-exchange potential can depend are $G_F$, $r$ and (possibly) the
masses of the external particles.  However, in the non-relativistic limit
appropriate to a static potential, bilinear covariants such as
$\bar{u}(p^\prime) \gamma_\lambda (1+\gamma_5)u(p)$ are independent of the
mass of the fermion characterized by the spinor $u(p)$.  Thus the only relevant
dimensional parameters are $G_F$ and $r$ and, since the 2-body operator is
proportional to $G_F^2$, it follows that $V^{(2)}(r) \propto G_F^2/r^5$.
Implicit in this argument is the assumption that no other dimensional
parameters are present and, since the standard model is renormalizable, this
will indeed be the case.  This argument holds even in the framework of the
(non-renormalizable) charged-current model originally assumed by FS, since
the regularization procedure employed by FS to extract the long-distance
behavior of $V^{(2)}(r)$ introduces no additional mass parameters.

	We conclude the discussion of the 2-body potential by demonstrating
quantitatively that its effects are too small to be detected at present
in any known system.
Let us consider the analog of Eq.~(\ref{V2}) for electrons, 
for which the ``background''
gravitational interaction would be smallest.  Reinstating $\hbar$ and $c$, and
substituting $a_n \rightarrow a_e = (2\sin^2\theta_W+ 1/2)$,
where \cite{PDG94}
\begin{equation}
	\sin^2\theta_W = 0.2319(5),
\end{equation}
we find
\begin{equation}
	V^{(2)}(r) = \frac{(2\sin^2\theta_W + 1/2)^2}{4\pi^3}
      \frac{G_F^2}{\hbar c} \frac{1}{r^5} .
\label{final V2}
\end{equation}
Using \cite{PDG94}
\begin{equation}
	\frac{G_F}{(\hbar c)^3} = 1.16639(2) \times 10^{-5} \mbox{GeV}^{-2}
\end{equation}
leads to
\begin{equation}
	V^{(2)}(r) = 3 \times 10^{-82} \frac{\mbox{eV}}{(r/1~\mbox{m})^5} .
\end{equation}
The magnitude of the corresponding force $\vec{F}_{12}(r) = -\vec{\nabla}V^{(2)}(r)$
is then given by
\begin{equation}
	|\vec{F}_{12}(r)| = \frac{2\times 10^{-100}}{(r/1~\mbox{m})^6}
	\,\,\,\mbox{Newtons} .
\end{equation}
To appreciate how weak this force is, we compare
$|\vec{F}_{12}|$ to the gravitational force $\vec{F}_{12}^{grav}$ at a
nominal separation $r = 1$~km,
\begin{equation}
	\frac{|\vec{F}_{12}(r =1~\mbox{km})|}
	{|\vec{F}_{12}^{grav} (r = 1~\mbox{km})|}
     = 4 \times 10^{-42}.
\end{equation}
As $r$ decreases $|\vec{F}_{12}(r)|$ increases more rapidly than does
$|\vec{F}_{12}^{grav}|$, and these forces become equal at $r = 5 \times 10^{-8}$m.
However, at this separation electromagnetic and weak forces arising
from $Z^{0}$-exchange would be much
larger than $|\vec{F}_{12}|$, and hence there appears to be no distance
scale over which the presence of the 2-body neutrino-exchange interaction
can be directly
detected \cite{FS68,HAR72}.  

\subsection{The $\bif k$-Body Contribution when $k$ is Odd}

In this subsection we show that there is no 3-body static potential arising
from neutrino-exchange.  This result is then generalized by demonstrating
that for odd values of $k \geq 5$, the $k$-body potential exists but
averages to zero when integrated over a spherical volume.  The net result
is that the neutrino-exchange energy of a spherical nucleus or neutron star
is given by a sum of $k$-body contributions where $k$ is even.

As noted previously, there are two independent diagrams for $k=3$ and these
are shown in Fig.~\ref{3-body figure}.  Expanding Eq.~(\ref{Schwinger W again}) 
to $\O(G_F^3)$ we find for the
contribution from diagram \ref{3-body figure}(a)
\begin{eqnarray}
   W_a^{(3)} & = &
	\frac{i}{2\pi} \left( \frac{G_Fa_n}{\sqrt{2}}\right)^3
     \int d^3x_1d^3x_2 d^3x_3 \int_{-\infty}^\infty dE 
	\nonumber \\
	&   & \times \left[
     \mbox{tr}
	\{ \gamma_\lambda (1+\gamma_5)S_F^{(0)} (12) \gamma_\mu (1+\gamma_5)
      S_F^{(0)} (23) \gamma_\nu (1+\gamma_5) S_F^{(0)} (31)\}
	\right. 
	\nonumber  \\
	&   &  \times \left.
      N_\lambda (x_1)N_\mu (x_2) N_\nu (x_3)\right],
\label{W3a}
\end{eqnarray}
where $S_F^{(0)}(12) \equiv S_F^{(0)} (\vec{r}_{12},E)$ etc.
Using Eq.~(\ref{1 + gamma5}) 
the expression in $\{\cdots\}$ for diagram \ref{3-body figure}(a) 
can be simplified
to
\begin{equation}
	\mbox{tr} 
	\{\mbox{diagram}~\ref{3-body figure}(a)\} = 
4~\mbox{tr}\{\gamma_\lambda S_F^{(0)} (12)
      \gamma_\mu S_F^{(0)} (23) \gamma_\nu S_F^{(0)} (31)
        (1-\gamma_5)\}.
\label{trace 3.4a}
\end{equation}
The contribution $W_b^{(3)}$ from diagram \ref{3-body figure}(b) 
is given by the
same expression as in Eq.~(\ref{W3a}) except that tr$\{\cdots\}$ is replaced by
\begin{equation}
	\mbox{tr}
	\{\mbox{diagram}~\ref{3-body figure}(b)\} 
= 4~\mbox{tr}\{ \gamma_\lambda S_F^{(0)} (13)
    \gamma_\nu S_F^{(0)} (32) \gamma_\mu S_F^{(0)} (21)
     (1 -\gamma_5)\} .
\label{trace 3.4b}
\end{equation}

	The familiar arguments used to derive Furry's theorem 
\cite{AB65} can now
be adapted to show that the 3-body contribution vanishes.  We introduce the
charge-conjugation matrix $C$ defined by
\begin{mathletters}
\begin{eqnarray}
  	C^{-1}\gamma_\mu C & = & - \gamma_\mu^T ,
\label{gamma C} \\
	 C^{-1}(1-\gamma_5) C & = & (1-\gamma_5)^T,
\label{1 - gamma5 C}
\end{eqnarray}
\end{mathletters}
where the superscript $T$ denotes the transpose matrix.  In the Dirac-Pauli
conventions $C$ is given (up to an overall phase) by $C = \gamma_2\gamma_4$.
By virtue of Eqs.~(\ref{gamma C}) and (\ref{1 - gamma5 C}), 
the neutrino propagator in Eq.~(\ref{final inverted SF}) satisfies
\begin{equation}
	C^{-1} S_F^{(0)} (\vec{r}_{ij},E) C = 
	-\gamma^T \cdot \eta (ij)
    \Delta_F(\vec{r}_{ij},E) = S_F^{(0)T} (-\vec{r}_{ij}, -E)
    \equiv S_F^{(0)T} (-ij).
\label{S C}
\end{equation}
Using Eqs.~(\ref{gamma C}), (\ref{1 - gamma5 C}), 
and (\ref{S C}), Eq.~(\ref{trace 3.4b}) 
can be rewritten in the form
\begin{eqnarray}
 	\lefteqn{ 4~\mbox{tr}
	\{ \gamma_\lambda S_F^{(0)} (13) \gamma_\nu S_F^{(0)} (32) \gamma_\mu
     S_F^{(0)} (21) (1-\gamma_5)\}} \hspace{.5in}
	\nonumber  \\
    & =  & 4 \mbox{tr}
	 \{(-\gamma_\lambda^T)S_F^{(0)T} (-13)(-\gamma_\nu^T) S_F^{(0)T}
     (-32) (-\gamma_\mu^T) S_F^{(0)T} (-21) (1-\gamma_5)^T\}
	\nonumber \\
    & = & -4~\mbox{tr}
	\{ \gamma_\lambda S_F^{(0)}(12) \gamma_\mu S_F^{(0)} (23)
       \gamma_\nu S_F^{(0)} (31) (1+\gamma_5)\}.
\label{SSS trace}
\end{eqnarray}
The last step of Eq.~(\ref{SSS trace}) follows by noting from Eq.~(\ref{S
C}) that we can replace
$-\vec{r}_{ij}$ by $\vec{r}_{ji}$ in $S_F^{(0)}(-ij)$, so that $S_F^{(0)}(-ij)$
is the same as $S_F^{(0)}(ji)$ except for the sign of $E$.  However, by
virtue of the symmetric limits of the integration over $E$, only even powers
of $E$ contribute when the products of the neutrino propagators are expanded.
Since these give the same contributions for $\pm E$, we can effectively
replace $S_F^{(0)}(-ij)$ by $S_F^{(0)}(ji)$ in Eq.~(\ref{SSS
trace}).  Combining 
Eqs.~(\ref{trace 3.4a})
and (\ref{SSS trace}) 
we see that the terms independent of $\gamma_5$ cancel exactly,
which is Furry's theorem, while the terms proportional to $\gamma_5$ add
yielding
\begin{eqnarray}
    W^{(3)} & = & W_a^{(3)} + W_b^{(3)} 
	\nonumber \\
	& = & \frac{-i}{2\pi}
     \left(\frac{G_Fa_n}{\sqrt{2}}\right)^3 \int d^3x_1d^3x_2d^3x_3
     \int_{-\infty}^\infty dE 
	\nonumber \\
	&   &  \times
	\left[
     8~\mbox{tr} \{ \gamma_\lambda S_F^{(0)} (12) \gamma_\mu S_F^{(0)}
      (23) \gamma_\nu S_F^{(0)} (31) \gamma_5 \} N_\lambda (x_1)
       N_\mu (x_2) N_\nu (x_3)\right].
\label{W3}
\end{eqnarray}
Using Eq.(A8) 
we see that the trace in Eq.~(\ref{W3}) is proportional to the
Levi-Civita tensor $\epsilon_{\mu\nu\lambda\rho}$, which reduces in the
non-relativistic limit to the permutation symbol $\epsilon_{ijl}$.  It
then follows that the 3-body potential $V^{(3)}(\vec{r}_{12},\vec{r}_{13},\vec{r}_{23})$
will contain terms of the form
\begin{equation}
	\epsilon_{ijl} (\vec{r}_{12})_i (\vec{r}_{31})_j
         (\vec{r}_{23})_l = \vec{r}_{12}\cdot (\vec{r}_{31}
          \times \vec{r}_{23}) = 0,
\label{dot/cross product}
\end{equation}
since $\vec{r}_{12} + \vec{r}_{31} + \vec{r}_{23} = 0$.  Thus there is
no 3-body potential arising from neutrino exchange.

	The preceding arguments can be generalized to show that one can 
ignore the contributions from 
$k$-body potentials when $k \geq 5$ is odd.  As we discuss in the following
subsection, the $k$-body potential arising from the expansion of
Eq.~(\ref{Schwinger W again}) 
receives contributions from $\frac{1}{2} (k-1)!$  topologically
distinct pairs of diagrams, 
each being sum of two diagrams, representing the two senses
of the neutrino-loop momentum.  For the $j$-th such diagram the sum of the
contributions from the two senses of the neutrino-loop momentum gives rise
to an analog of Eq.~(\ref{W3}) which, if $k$ is odd, has the form
\begin{eqnarray}
    W_j^{(k)} & = & 
	\frac{-i}{2\pi} \left(\frac{G_Fa_n}{\sqrt{2}}\right)^k
     \int d^3x_1d^3x_2 \cdots d^3x_k \int_{-\infty}^\infty dE 
	\nonumber \\
      &  &  \times 
	2^{k} \mbox{tr}\{ \gamma_\lambda S_F^{(0)} (12) \gamma_\mu
       S_F^{(0)} (23)\cdots S_F^{(0)} (k1)\gamma_5\} N_\lambda (x_1)
	\cdots N_\mu(x_k) .
\end{eqnarray}
Evaluation of tr$\{\cdots \}$ leads again to an expression which is proportional
to $\epsilon_{\mu\nu\lambda\rho}$, and hence to $\epsilon_{ijl}$ in the
nonrelativistic limit.  For $k \geq 5$ an expression such as 
Eq.~(\ref{dot/cross product}) need not
vanish, since $\epsilon_{ijl}$ can be contracted with 3 linearly
independent vectors, and hence the corresponding potential is in general
nonzero.  However, the integral of any such potential over a spherical
volume {\em is} zero:  In analogy to the electrostatic case discussed in
Sec. II, the integrated energy can depend only on $G_F$ and the radius $R$
of the sphere.  Since a nonzero scalar product cannot be formed utilizing
only $\epsilon_{ijl}$, $R$, and $\vec{R} = \hat{R}R$, the $k$-body contribution
to the energy of a spherical charge distribution must vanish when $k$ is odd.
This conclusion does not necessarily hold for an asymmetric charge distribution,
but in the present context any contributions from $k$-odd terms would be
proportional to the (presumably small) deviations of the matter distribution
in a white dwarf or neutron star from spherical symmetry.

\subsection{The 4-Body Potential}

	Following the discussion in Appendix B we note that the expansion of
Eq.~(\ref{Schwinger W}) in powers of $G_F$ 
leads in order $G_F^k$ to the one-loop
diagrams shown in Fig.~\ref{Second 4-body figure}, 
each having $k$ vertices.  Since there
are $(k-1)!/2$ distinct ways that the integers $1,2,3,\ldots,k$ can be
arranged on the perimeter of a circle, there are altogether $(k-1)!$
topologically distinct diagrams in order $k$, where the additional
factor of 2 takes account of the two senses of the neutrino-loop momentum.
In each order we will refer to the diagram in which the vertices are
labeled sequentially $1,2,3,\ldots,k$ as the``standard'' diagram.  When
computing the $k$-body potential, each of the $(k-1)!/2$ possible
sequences of integers leads to a distinct potential arising from the
sum of the two diagrams with opposite senses of the neutrino-loop
momentum.  Consider, for example, the three topologically distinct
diagrams corresponding to $k=4$ shown in Fig.~\ref{First 4-body
figure}.  Since the
dependence of any term in the potential on $\vec{r}_{ij}$ arises from
the neutrino propagator connecting $i$ and $j$, the standard diagram,
Fig.~\ref{First 4-body figure}(a), will be a function of only the variables
$\vec{r}_{12}, \vec{r}_{23}, \vec{r}_{34}$, and $\vec{r}_{41}$,
but not of $\vec{r}_{24}$ or $\vec{r}_{13}$, which are not connected
by neutrino propagators.  Thus the 3 pairs of diagrams represented
in Fig.~\ref{Second 4-body figure}
give rise to 3 distinct contributions to the total
potential in order $G_F^4$, as we show explicitly below.  However,
on symmetry grounds each of these 3 contributions leads to the same
result when integrated over a spherical volume.  Hence in practice it
suffices to evaluate the pair of diagrams containing the standard
sequence of vertices $1,2,3,\ldots,k$, in the anticipation that the integrated
result will eventually be multiplied by $(k-1)!/2$ to obtain the total
contribution in order $k$.  Since any specific set of $k$ particles can
be chosen in ${N \choose k}$
ways from among $N$ particles, there is an additional factor of
${N \choose k}$ present in the
final result, and it is this factor which is ultimately responsible
for the large neutrino-exchange energy-density.

We can summarize the preceding discussion and Appendix B as follows:
The expansion of the Schwinger formula in Eq.~(\ref{Schwinger W}) 
leads to a set
of irreducible 1-loop Feynman diagrams which describe the
$\O(G_F^k)$ contribution to the $k$-body potential.  The $k$-body
potential can also receive contributions from diagrams containing
higher powers of $G_F$, but these are suppressed both by $G_F$ and
by various mass factors.  Hence for practical
purposes we need consider only the one-loop diagrams arising from
the expansion of Eq.~(\ref{Schwinger W}).  
Each of these contains $k$ vertices, which
represent in configuration space the coordinates 
$\vec{r}_i(i=1,2,\ldots,k)$ of the $k$ particles.  The $k$ vertices
are connected by $k$ legs representing the variables 
$r_{ij} = |\vec{r}_{ij}| = |\vec{r}_i - \vec{r}_j|$.  There
are $k(k-1)/2$ such variables that arise, and $(k-1)!/2$
topologically distinct diagrams, although each diagram depends on
a subset containing only $k$ of the variables.  Notwithstanding
the fact that the variables $\vec{r}_{ij}$ satisfy a constraint
of the form
\begin{equation}
 \vec{r}_{12} + \vec{r}_{23} + \vec{r}_{34} + \cdots + \vec{r}_{k1} = 0,
\end{equation}
the evaluation of the $k$-body potential for $k > 2$ 
should be carried out treating
the $\vec{r}_{ij}$ as if they were in fact independent.

	We turn next to the detailed form of the 4-body
contribution.  It follows from the preceding discussion that we can confine
our attention to the standard diagrams, Figs.~\ref{Second 4-body
figure}(a) and \ref{Second 4-body figure}(a$'$),
which incorporate both senses of
the neutrino loop momentum.  Expanding Eq.~(\ref{Schwinger W}) to
$\O(G_F^4)$ and using Eq.~(\ref{1 + gamma5}) 
we find for the contribution from diagram \ref{Second 4-body
figure}(a),
\begin{eqnarray}
  	W_a^{(4)} & = & 
	\frac{-i}{2\pi} \left( \frac{G_F a_n}{\sqrt{2}}\right)^4
     \int d^3x_1d^3x_2d^3x_3d^3x_4 \int_{-\infty}^\infty dE 
	\nonumber \\
	&  &  \times
    2^3 \mbox{tr} \left\{ \gamma_\mu S_F^{(0)} (14) \gamma_\sigma S_F^{(0)}
      (43) \gamma_\lambda S_F^{(0)} (32) \gamma_\nu S_F^{(0)} (21)
      (1-\gamma_5)\right\} T_{\mu\nu\lambda\sigma} (x_1,x_2,x_3,x_4),
\label{W4a}
\end{eqnarray}
where
\begin{equation}
     T_{\mu\nu\lambda\sigma} (x_1,x_2,x_3,x_4) 
	\equiv  N_\mu (x_1) N_\nu(x_2)
       N_\lambda(x_3)N_\sigma(x_4).
\label{T4}
\end{equation}
By use of Eqs.~(\ref{gamma C}) and (\ref{1 - gamma5 C}), 
$\mbox{tr}\{\cdots\}$ in Eq.~(\ref{W4a}) can be written in the form
\begin{equation}
  \mbox{tr} \{\mbox{diagram}~\ref{Second 4-body figure}(\mbox{a})\} =
	\mbox{tr} \{S_F^{(0)}(12) \gamma_\nu
      S_F^{(0)} (23) \gamma_\lambda S_F^{(0)} (34) \gamma_\sigma
      S_F^{(0)} (41) \gamma_\mu (1-\gamma_5)\} .
\label{trace 3.3a}
\end{equation}
The contribution from diagram~\ref{Second 4-body figure}(a$'$), corresponding to the standard
diagram with the loop momentum reversed, has the same form as 
Eq.~(\ref{W4a})
except that $\mbox{tr}\{\cdots\}$ is replaced by
\begin{equation}
	\mbox{tr}\{\mbox{diagram}~\ref{Second 4-body figure}
(\mbox{a}')\} = 
	\mbox{tr} \{S_F^{(0)} (12) \gamma_\nu
    S_F^{(0)} (23) \gamma_\lambda S_F^{(0)} (34) \gamma_\sigma   
    S_F^{(0)} (41) \gamma_\mu (1+\gamma_5)\}.
\label{trace 3.3a'}
\end{equation}
We see that when the contributions in Eqs.~(\ref{trace 3.3a}) 
and (\ref{trace 3.3a'}) are added
the terms proportional to $\gamma_5$ now cancel, whereas they added in
the 3-body case.  
Similarly, the terms independent of $\gamma_5$, which cancelled previously
to yield Furry's theorem, now add to give
\begin{eqnarray}
	W^{(4)} & = & W_a^{(4)} + W_{a'}^{(4)} 
	\nonumber \\
	& = & \frac{-i}{2\pi}
   \left( \frac{G_Fa_n}{\sqrt{2}}\right)^4 2^4 \int d^3x_1...d^3x_4
    \int_{-\infty}^\infty dE 
	\nonumber \\
	&    &  \times
    \mbox{tr}\{S_F^{(0)} (12) \gamma_\nu S_F^{(0)} (23) \gamma_\lambda
     S_F^{(0)} (34) \gamma_\sigma S_F^{(0)} (41) \gamma_\mu\}
      T_{\mu\nu\lambda\sigma} (x_1,\ldots,x_4) .
\label{W4 1}
\end{eqnarray}
As in the 2-body case we are interested in the static potential which
arises from
\begin{equation}
	T_{\mu\nu\lambda\sigma} (x_1,\ldots,x_4) 
	= (i)^4 \rho_1\rho_2\rho_3\rho_4
     \delta_{\mu 4} \delta_{\nu 4} \delta_{\lambda 4} \delta_{\sigma 4},
\label{static T}
\end{equation}
and hence the trace in Eq.~(\ref{W4 1}) is proportional to
\begin{equation}
	\mbox{tr}
	\{ (\ref{W4 1})\} \sim 
	\mbox{tr}[\gamma \cdot \eta(12)\gamma_4 \gamma \cdot \eta
     (23) \gamma_4 \gamma \cdot \eta(34) \gamma_4 \gamma \cdot \eta (41)
    \gamma_4],
\label{trace proportion}
\end{equation}
where our notation is the same as in Eq.~(\ref{W2 a}).  It is convenient to
simplify the trace by writing
\begin{equation}
  \gamma_4 \gamma\cdot\eta (12) = 
	-(\vec{\gamma} \cdot \vec{\partial}_{12}
         + \gamma_4E)\gamma_{4} 
		\equiv -\gamma \cdot\bar{\eta}(12) \gamma_4,
\end{equation}
so that tr$[\cdots]$ in Eq.~(\ref{trace proportion}) assumes the form
\begin{eqnarray}
     \mbox{tr}[\cdots] & = & 
	\mbox{tr}[\gamma\cdot \bar{\eta}(12) \gamma\cdot\eta(23)
     \gamma\cdot \bar{\eta} (34) \gamma \cdot \eta (41)] 
	\nonumber \\
     & = & 
	4 \{ E^4 - E^2 (\vec{\partial}_{12}\cdot \vec{\partial}_{23}
     + \vec{\partial}_{23} \cdot \vec{\partial}_{34}
      + \vec{\partial}_{34} \cdot \vec{\partial}_{41}
      + \vec{\partial}_{12} \cdot \vec{\partial}_{34}
      + \vec{\partial}_{12} \cdot \vec{\partial}_{41}
      + \vec{\partial}_{23} \cdot \vec{\partial}_{41})
	\nonumber \\
     &  & \mbox{}
	 + [(\vec{\partial}_{12}\cdot \vec{\partial}_{23})
        (\vec{\partial}_{34}\cdot \vec{\partial}_{41})
       - (\vec{\partial}_{12} \cdot \vec{\partial}_{34})
        ( \vec{\partial}_{23}\cdot \vec{\partial}_{41})
       + (\vec{\partial}_{12} \cdot \vec{\partial}_{41})
        (\vec{\partial}_{23} \cdot \vec{\partial}_{34})] \}.
\label{long trace}
\end{eqnarray}
Combining Eqs.~(\ref{W4 1}), (\ref{static T}), and (\ref{long
trace}), we can write $W^{(4)}$ in the form
\begin{eqnarray}
   W^{(4)} & = & \frac{-i}{2\pi} \left( \frac{G_Fa_n}{\sqrt{2}}\right)^4
       2^4 \int \rho_1 d^3x_1 \int \rho_2 d^3x_2 \int
       \rho_3 d^3x_3 \int \rho_4d^3x_3 \int_{-\infty}^\infty
         dE
	\nonumber \\
	&   & \times
         \mbox{tr}[\cdots]
     \Delta_F (\vec{r}_{12},E) \Delta_F (\vec{r}_{23},E)
       \Delta_F (\vec{r}_{34},E) \Delta_F (\vec{r}_{41},E) ,
\label{W4 2}
\end{eqnarray}
where tr$[\cdots]$ denotes the expression on the right-hand-side of
Eq.~(\ref{long trace}).  Using Eq.~(\ref{final inverted SF}) 
the product of the functions
$\Delta_F(\vec{r}_{ij},E)$ can be written as
\begin{equation}
  \Delta_F (\vec{r}_{12},E) \Delta_F (\vec{r}_{23},E)
      \Delta_F(\vec{r}_{34},E) \Delta_F(\vec{r}_{41},E) =
     \left( \frac{i}{4\pi} \right)^4
      \frac{e^{i|E|(r_{12} + r_{23} + r_{34} + r_{41} + i\epsilon)}}
     {r_{12} r_{23} r_{34} r_{41}} . 
\end{equation}
If we denote the sum of the $r_{ij}$ and their
product by $S_4$ and $P_4$ respectively,
\begin{mathletters}
\begin{eqnarray}
    S_4 & = & r_{12} + r_{23} + r_{34} + r_{41}, 
\label{S4} \\
    P_4 & = & r_{12} r_{23} r_{34} r_{41}, 
\label{P4}
\end{eqnarray}
\end{mathletters}
then the integral over $E$ can then be expressed in terms
of $S_4$ by using Eqs.~(\ref{general In}) and (\ref{In}), and
then making the replacements
\begin{mathletters}
\begin{eqnarray}
     E^4 & \rightarrow & 2i \frac{4!}{S_4^5} ,\\
     E^2 & \rightarrow & -2i \frac{2!}{S_4^3}, \\
     E^0 & \rightarrow & 2i \frac{1}{S_4} .
\end{eqnarray}
\end{mathletters}
Combining Eqs.~(\ref{W4 2})--(\ref{P4}) we can extract the 
contribution to the 4-body potential
$V^{(4)}$ from diagrams 4(a) and 4(a$^{'}$) by writing
\begin{equation}
	W^{(4)}[\mbox{4(a) + 4(a$^{'}$)}]  =  
	\int \rho_1 d^3x_1 \int \rho_2 d^3x_2 \int \rho_3
      d^3x_3 \int \rho_4 d^3x_4 V^{(4)} 
      (\vec{r}_{12}, \vec{r}_{23}, \vec{r}_{34}, \vec{r}_{41}),
\label{W4}
\end{equation}
where
\begin{eqnarray}
    \lefteqn{ V^{(4)} (\vec{r}_{12}, \vec{r}_{23}, \vec{r}_{34}, \vec{r}_{41})
	= 
     \left( \frac{G_Fa_n}{\sqrt{2}} \right)^4 \frac{1}{4\pi^5}
    \biggr\{ \frac{4!}{P_4S_4^5}} \hspace{.8in} 
	\nonumber \\
	&   &  \mbox{} + 2!\left( 
    \vec{\partial}_{12}\cdot \vec{\partial}_{23}
   + \vec{\partial}_{23}\cdot \vec{\partial}_{34}
   + \vec{\partial}_{34}\cdot \vec{\partial}_{41}
   + \vec{\partial}_{12}\cdot \vec{\partial}_{34}
   + \vec{\partial}_{12}\cdot \vec{\partial}_{41}
   + \vec{\partial}_{23}\cdot \vec{\partial}_{41} 
	\right)
   \frac{1}{P_4S_4^3}  
	\nonumber \\
	&  & \mbox{}
    + \left[
	(\vec{\partial}_{12}\cdot \vec{\partial}_{23})
     (\vec{\partial}_{34} \cdot \vec{\partial}_{41})
     -(\vec{\partial}_{12} \cdot \vec{\partial}_{34})
      (\vec{\partial}_{23} \cdot \vec{\partial}_{41})
     + (\vec{\partial}_{12} \cdot \vec{\partial}_{41})
     (\vec{\partial}_{23} \cdot \vec{\partial}_{34})
	\right]
     \frac{1}{P_4S_4} \biggr\}.  
	\nonumber \\
\label{V4}
\end{eqnarray}
The expression in Eq.~(\ref{V4}) reproduces the 4-body result of the
Hartle \cite{HAR70} up to some minor misprints in that reference.
We note again that the gradient operators $\vec{\partial}_{ij}$
act on $P_{4}$ and $S_{4}$ as if all the coordinates $\vec{r}_{ij}$ are
independent, notwithstanding the fact that
$ \vec{r}_{12} + \vec{r}_{23} + \vec{r}_{34} + \vec{r}_{41} = 0$.

\subsection{The 6-Body Potential}

The general discussion that introduced the 4-body potential in the
preceding subsection can be taken over immediately for the 6-body
case.  Our purpose in deriving the 6-body potential $V^{(6)}$ is
to explicitly exhibit the various phases and combinatoric factors
in a way that will allow us in the next subsection to obtain the
general $k$-body result $V^{(k)}$.  From the preceding discussion
we note that for $k=6$ there are 60 pairs of topologically distinct
diagrams, with each pair representing the two possible senses of
the neutrino loop momentum.  The 6 legs of the standard diagram
correspond to the variables $\vec{r}_{12}, \vec{r}_{23},
\vec{r}_{34}, \vec{r}_{45}, \vec{r}_{56}$, and $\vec{r}_{61}$
which satisfy
\begin{equation}
	\vec{r}_{12} + \vec{r}_{23} + \vec{r}_{34} + \vec{r}_{45}
    + \vec{r}_{56} + \vec{r}_{61} = 0,
\end{equation}
and $S_4,P_4$ in Eqs.~(\ref{S4}) and (\ref{P4}) 
are now replaced by
\begin{mathletters}
\begin{eqnarray}
     S_6 & = & 
	r_{12} + r_{23} + r_{34} + r_{45} + r_{56} + r_{61},
\label{S6} \\
     P_6 & = & r_{12}  r_{23}  r_{34}  r_{45}  r_{56}  r_{61}.
\label{P6}
\end{eqnarray}
\end{mathletters}
Using the formalism of the previous subsection the expression
for $V^{(6)}(\vec{r}_{12},\ldots,\vec{r}_{61})$ can be written as follows:
\begin{equation}
	V^{(6)}(\vec{r}_{12},\ldots,\vec{r}_{61}) 
		= \frac{-i}{2\pi}
	 \left(\frac{i}{4\pi}\right)^6 (i)^7 2^7 \left(
 \frac{G_Fa_n}{\sqrt{2}}\right)^6 4 \Phi^{(6)}(S_6,P_6),
\label{V6}
\end{equation}
where
\begin{eqnarray}
	\Phi^{(6)}(S_6,P_6) 
	&  =  &  \frac{(i)^6 6!}{P_6S_6^7} - (i)^4 4!
\Biggl[ \vec{\partial}_{12} \cdot \vec{\partial}_{23} +
\vec{\partial}_{34}\cdot \vec{\partial}_{45} +
\vec{\partial}_{56} \cdot \vec{\partial}_{61} +
\vec{\partial}_{12} \cdot \vec{\partial}_{34} +
\vec{\partial}_{23} \cdot \vec{\partial}_{45} 
	\nonumber \\
	&   & \mbox{}
+ \vec{\partial}_{12} \cdot \vec{\partial}_{45}
+ \vec{\partial}_{23} \cdot \vec{\partial}_{34}
- \vec{\partial}_{12} \cdot \vec{\partial}_{56}
- \vec{\partial}_{23} \cdot \vec{\partial}_{61}
- \vec{\partial}_{12} \cdot \vec{\partial}_{61} 
	\nonumber \\
	&   & \mbox{}
- \vec{\partial}_{23} \cdot \vec{\partial}_{56}
+ \vec{\partial}_{34} \cdot \vec{\partial}_{61}
+ \vec{\partial}_{34} \cdot \vec{\partial}_{56}
+ \vec{\partial}_{45} \cdot \vec{\partial}_{61}
+ \vec{\partial}_{45} \cdot \vec{\partial}_{56} \Biggl]
\frac{1}{P_6S_6^5}  
	\nonumber \\
	&   & \mbox{}
+ (i)^2 2!\Biggl[ (\vec{\partial}_{12}\cdot \vec{\partial}_{23})
(\vec{\partial}_{45} \cdot \vec{\partial}_{56}) 
+ (\vec{\partial}_{12} \cdot \vec{\partial}_{23})
  (\vec{\partial}_{34} \cdot \vec{\partial}_{61})
+ (\vec{\partial}_{34} \cdot \vec{\partial}_{61})
  (\vec{\partial}_{45} \cdot \vec{\partial}_{56})
	\nonumber \\
	&   & \mbox{}
+ (\vec{\partial}_{12} \cdot \vec{\partial}_{23}) 
  (\vec{\partial}_{34} \cdot \vec{\partial}_{56})
+ (\vec{\partial}_{12} \cdot \vec{\partial}_{23})
  (\vec{\partial}_{45} \cdot \vec{\partial}_{61})
- (\vec{\partial}_{34} \cdot \vec{\partial}_{56})
  (\vec{\partial}_{45} \cdot \vec{\partial}_{61})
	\nonumber \\ 
	&   & \mbox{}
- (\vec{\partial}_{12} \cdot \vec{\partial}_{34})
  (\vec{\partial}_{23} \cdot \vec{\partial}_{61})
+ (\vec{\partial}_{12} \cdot \vec{\partial}_{34})
  (\vec{\partial}_{45} \cdot \vec{\partial}_{56})
+ (\vec{\partial}_{23} \cdot \vec{\partial}_{61})
  (\vec{\partial}_{45} \cdot \vec{\partial}_{56}) 
	\nonumber \\
	&   & \mbox{}
+ (\vec{\partial}_{12} \cdot \vec{\partial}_{34})
  (\vec{\partial}_{45} \cdot \vec{\partial}_{61})
- (\vec{\partial}_{12} \cdot \vec{\partial}_{34})
  (\vec{\partial}_{23} \cdot \vec{\partial}_{56})
- (\vec{\partial}_{23} \cdot \vec{\partial}_{56})
  (\vec{\partial}_{45} \cdot \vec{\partial}_{61})
	\nonumber \\
	&   & \mbox{}
+ (\vec{\partial}_{23} \cdot \vec{\partial}_{61})
  (\vec{\partial}_{34} \cdot \vec{\partial}_{56})
- (\vec{\partial}_{12} \cdot \vec{\partial}_{45})
  (\vec{\partial}_{23} \cdot \vec{\partial}_{61})
- (\vec{\partial}_{12} \cdot \vec{\partial}_{45})
  (\vec{\partial}_{34} \cdot \vec{\partial}_{56})
	\nonumber \\
	&   & \mbox{}
- (\vec{\partial}_{12} \cdot \vec{\partial}_{45})
  (\vec{\partial}_{23} \cdot \vec{\partial}_{56})
- (\vec{\partial}_{12} \cdot \vec{\partial}_{45})
  (\vec{\partial}_{34} \cdot \vec{\partial}_{61})
- (\vec{\partial}_{23} \cdot \vec{\partial}_{56})
  (\vec{\partial}_{34} \cdot \vec{\partial}_{61})
	\nonumber \\
	&   & \mbox{}
+ (\vec{\partial}_{23} \cdot \vec{\partial}_{34})
  (\vec{\partial}_{12} \cdot \vec{\partial}_{61})
+ (\vec{\partial}_{23} \cdot \vec{\partial}_{34})
  (\vec{\partial}_{45} \cdot \vec{\partial}_{56})
+ (\vec{\partial}_{12} \cdot \vec{\partial}_{61})
  (\vec{\partial}_{45} \cdot \vec{\partial}_{56})
	\nonumber \\
	&   & \mbox{}
+ (\vec{\partial}_{23} \cdot \vec{\partial}_{34})
  (\vec{\partial}_{12} \cdot \vec{\partial}_{56})
+ (\vec{\partial}_{23} \cdot \vec{\partial}_{34})
  (\vec{\partial}_{45} \cdot \vec{\partial}_{61})
- (\vec{\partial}_{12} \cdot \vec{\partial}_{56})
  (\vec{\partial}_{45} \cdot \vec{\partial}_{61})
	\nonumber \\
	&   & \mbox{}
+ (\vec{\partial}_{12} \cdot \vec{\partial}_{61})
  (\vec{\partial}_{34} \cdot \vec{\partial}_{56})
- (\vec{\partial}_{23} \cdot \vec{\partial}_{45})
  (\vec{\partial}_{34} \cdot \vec{\partial}_{56})
+ (\vec{\partial}_{23} \cdot \vec{\partial}_{45})
  (\vec{\partial}_{12} \cdot \vec{\partial}_{61})
	\nonumber \\
	&   & \mbox{}
+ (\vec{\partial}_{23} \cdot \vec{\partial}_{45})
  (\vec{\partial}_{12} \cdot \vec{\partial}_{56})
- (\vec{\partial}_{23} \cdot \vec{\partial}_{45})
  (\vec{\partial}_{34} \cdot \vec{\partial}_{61})
- (\vec{\partial}_{12} \cdot \vec{\partial}_{56})
  (\vec{\partial}_{34} \cdot \vec{\partial}_{61})
	\nonumber \\
	&   & \mbox{}
- (\vec{\partial}_{34} \cdot \vec{\partial}_{45})
  (\vec{\partial}_{23} \cdot \vec{\partial}_{56})
- (\vec{\partial}_{34} \cdot \vec{\partial}_{45})
  (\vec{\partial}_{12} \cdot \vec{\partial}_{61})
- (\vec{\partial}_{12} \cdot \vec{\partial}_{61})
  (\vec{\partial}_{23} \cdot \vec{\partial}_{56})
	\nonumber \\	
	&   & \mbox{}
+ (\vec{\partial}_{12} \cdot \vec{\partial}_{56})
  (\vec{\partial}_{23} \cdot \vec{\partial}_{61})
- (\vec{\partial}_{34} \cdot \vec{\partial}_{45})
  (\vec{\partial}_{12} \cdot \vec{\partial}_{56})
- (\vec{\partial}_{34} \cdot \vec{\partial}_{45})
  (\vec{\partial}_{23} \cdot \vec{\partial}_{61})
	\nonumber \\
	&   & \mbox{}
+ (\vec{\partial}_{12} \cdot \vec{\partial}_{45})
  (\vec{\partial}_{23} \cdot \vec{\partial}_{34})
+ (\vec{\partial}_{12} \cdot \vec{\partial}_{45})
  (\vec{\partial}_{56} \cdot \vec{\partial}_{61})
+ (\vec{\partial}_{23} \cdot \vec{\partial}_{34})
  (\vec{\partial}_{56} \cdot \vec{\partial}_{61})
	\nonumber \\
	&   & \mbox{}
+ (\vec{\partial}_{23} \cdot \vec{\partial}_{45})
  (\vec{\partial}_{56} \cdot \vec{\partial}_{61})
- (\vec{\partial}_{12} \cdot \vec{\partial}_{34})
  (\vec{\partial}_{23} \cdot \vec{\partial}_{45})
+ (\vec{\partial}_{12} \cdot \vec{\partial}_{34})
  (\vec{\partial}_{56} \cdot \vec{\partial}_{61})
	\nonumber \\
	&   & \mbox{}
+ (\vec{\partial}_{12} \cdot \vec{\partial}_{23})
  (\vec{\partial}_{34} \cdot \vec{\partial}_{45})
+ (\vec{\partial}_{12} \cdot \vec{\partial}_{23})
  (\vec{\partial}_{56} \cdot \vec{\partial}_{61})
+ (\vec{\partial}_{34} \cdot \vec{\partial}_{45})
  (\vec{\partial}_{56} \cdot \vec{\partial}_{61})\Biggr]
  \frac{1}{P_6S_6^3}
	\nonumber \\
	&   & \mbox{}
+ \Biggr[ 
-(\vec{\partial}_{12} \cdot \vec{\partial}_{23})
 (\vec{\partial}_{34} \cdot \vec{\partial}_{61})
 (\vec{\partial}_{45} \cdot \vec{\partial}_{56})
+(\vec{\partial}_{12} \cdot \vec{\partial}_{23})
 (\vec{\partial}_{34} \cdot \vec{\partial}_{56})
 (\vec{\partial}_{45} \cdot \vec{\partial}_{61})
	\nonumber \\
	&   & \mbox{}
+(\vec{\partial}_{12} \cdot \vec{\partial}_{34})
 (\vec{\partial}_{23} \cdot \vec{\partial}_{61})
 (\vec{\partial}_{45} \cdot \vec{\partial}_{56})
-(\vec{\partial}_{12} \cdot \vec{\partial}_{34})
 (\vec{\partial}_{23} \cdot \vec{\partial}_{56})
 (\vec{\partial}_{45} \cdot \vec{\partial}_{61})
	\nonumber \\
	&   & \mbox{}
-(\vec{\partial}_{12} \cdot \vec{\partial}_{45})
 (\vec{\partial}_{23} \cdot \vec{\partial}_{61})
 (\vec{\partial}_{34} \cdot \vec{\partial}_{56})
+(\vec{\partial}_{12} \cdot \vec{\partial}_{45})
 (\vec{\partial}_{23} \cdot \vec{\partial}_{56})
 (\vec{\partial}_{34} \cdot \vec{\partial}_{61})
	\nonumber \\
	&   & \mbox{}
-(\vec{\partial}_{23} \cdot \vec{\partial}_{34})
 (\vec{\partial}_{12} \cdot \vec{\partial}_{61})
 (\vec{\partial}_{45} \cdot \vec{\partial}_{56})
+(\vec{\partial}_{23} \cdot \vec{\partial}_{34})
 (\vec{\partial}_{12} \cdot \vec{\partial}_{56})
 (\vec{\partial}_{45} \cdot \vec{\partial}_{61})
	\nonumber \\
	&   & \mbox{}
+(\vec{\partial}_{23} \cdot \vec{\partial}_{45})
 (\vec{\partial}_{12} \cdot \vec{\partial}_{61})
 (\vec{\partial}_{34} \cdot \vec{\partial}_{56})
-(\vec{\partial}_{23} \cdot \vec{\partial}_{45})
 (\vec{\partial}_{12} \cdot \vec{\partial}_{56})
 (\vec{\partial}_{34} \cdot \vec{\partial}_{61})
	\nonumber \\
	&   & \mbox{}
+(\vec{\partial}_{34} \cdot \vec{\partial}_{45})
 (\vec{\partial}_{12} \cdot \vec{\partial}_{61})
 (\vec{\partial}_{23} \cdot \vec{\partial}_{56})
-(\vec{\partial}_{34} \cdot \vec{\partial}_{45})
 (\vec{\partial}_{12} \cdot \vec{\partial}_{56})
 (\vec{\partial}_{23} \cdot \vec{\partial}_{61})
	\nonumber \\
	&   & \mbox{}
-(\vec{\partial}_{12} \cdot \vec{\partial}_{45})
 (\vec{\partial}_{23} \cdot \vec{\partial}_{34})
 (\vec{\partial}_{56} \cdot \vec{\partial}_{61})
+(\vec{\partial}_{12} \cdot \vec{\partial}_{34})
 (\vec{\partial}_{23} \cdot \vec{\partial}_{45})
 (\vec{\partial}_{56} \cdot \vec{\partial}_{61})
	\nonumber \\
	&   & \mbox{}
-(\vec{\partial}_{12} \cdot \vec{\partial}_{23})
 (\vec{\partial}_{34} \cdot \vec{\partial}_{45})
 (\vec{\partial}_{56} \cdot \vec{\partial}_{61})\Biggr]
    \frac{1}{P_6S_6}
\end{eqnarray}
It is helpful to understand the sources of the various numerical factors in
$V^{(6)}$ as a prelude to discussing the $k$-body potential $V^{(k)}$.  The
overall coefficient $(-i/2\pi)$ arises from Eq.~(\ref{Schwinger
W}), with an additional
minus sign from the expansion of $\ln(1 + \Delta)$ in Eq.~(\ref{ln
Delta}).  Each
of the neutrino propagators in Eq.~(\ref{inverted SF 2}) 
contributes a factor
$(i/4\pi)$, and $I_n(z)$ in Eq.~(\ref{In}) is responsible for the factor
$(i)^7$.  The 6 factors of $(1+\gamma_5)$ eventually collapse to a
single factor by use of Eq.~(\ref{1 + gamma5}), 
leaving a numerical coefficient of $2^5$.
An additional factor of 2 arises from the diagram with the reversed loop
momentum and this, along with the factor of 2 from $I_n(z)$ in Eq.~(\ref{In}),
explains the factor $2^7$.  The Dirac trace leads to the coefficient
4 multiplying $\Phi^{(6)}$, and in 6th order the overall strength of
the 6-body interaction is given by $(G_Fa_n/\sqrt{2})^6$.  

\subsection{The $\bif k$-Body Potential}

The preceding discussion can be generalized in a straightforward manner to
the $k$-body potential $V^{(k)}$.  As we discuss in Sec.~V below, it is
sufficient for present purposes to consider only the single term in
$V^{(k)}$ which is
independent of derivatives, and hence we wish to exhibit this term in
complete detail.  We will also exhibit the coefficients of the derivative
terms, whose forms can be obtained by generalizing Eq.~(\ref{V6}) in an obvious
way.  Collecting together the various powers of $i$ and factors of 2,
we can write the contribution to $V^{(k)}$ from the standard diagram in
$\O(G_F^k)$ as follows:
\begin{equation}
   V^{(k)}(\vec{r}_{12},\ldots,\vec{r}_{k1}) = 
	\frac{1}{2\pi} (i)^{2k}
      2^{k+1} \left (\frac{1}{4\pi}\right)^k 
	\left(\frac{G_Fa_n}{\sqrt{2}}\right)^k
      4\Phi^{(k)} (S_k,P_k),
\label{Vk}
\end{equation}
where
\begin{eqnarray}
	\Phi^{(k)}(S_k,P_k) & = & 
	\frac{(i)^k k!}{P_kS_k^{k+1}}
	\nonumber \\
	&    &  \mbox{}
	 + (i)^{k-2}
     (k-2)! \left[ {\sum}^{(2)} 
	(\vec{\partial}_{ab} \cdot \vec{\partial}_{rs})\right]
     \frac{1}{P_kS_k^{k-1}} 
	\nonumber \\
	&  & \mbox{}
  + (i)^{k-4} (k-4)! \left[ {\sum}^{(4)} 
	(\vec{\partial}_{ab} \cdot \vec{\partial_{rs}})
     (\vec{\partial}_{cd} \cdot \vec{\partial}_{l m})\right]
     \frac{1}{P_kS_k^{k-3}} 
	\nonumber \\
	&   & \mbox{}
 	 +  \cdots
	+ \left[ {\sum}^{(k)}(\vec{\partial}_{ab} \cdot \vec{\partial}_{rs})
       (\vec{\partial}_{cd} \cdot \partial_{l m}) \cdots \right]
     \frac{1}{P_kS_k} ,
\label{Phi k}
\end{eqnarray}
\begin{eqnarray}
	S_k & = & r_{12} + r_{23} + r_{34} + \cdots + r_{k1} ,
\label{Sk} \\
	 P_k & = & r_{12} r_{23} r_{34} \cdots r_{k1}.
\label{Pk}
\end{eqnarray}
In Eq.~(\ref{Vk}) the notation 
${\sum}^{(2)}, {\sum}^{(4)},\ldots,{\sum}^{(k)}$ symbolically
represents the terms containing products of 
$2,4,\ldots,k$ derivatives, which act to
the right on the indicated functions of $P_k$ and $S_k$.  The numerical
coefficient of $\Phi^{(k)}$ can be understood in light of the preceding
discussion as follows:  The factor of $1/2\pi$ arises from the product of the
overall coefficient $(-i/2\pi)$ of $W$ in Eq.~(\ref{Schwinger W}),
Eq.~(\ref{ln Delta}), 
and a factor of $i$ extracted
from $I_n(z)$ in Eq.~(\ref{In}), where 
$n \rightarrow k$, $k-2$, $k-4,\ldots$ as appropriate
in $\Phi^{(k)}$.  The factor of $2^{k+1}$ is the product of $2^{k-1}$ arising
from $k$ factors of $(1+\gamma_5)$, and factors of 2 from $I_n(z)$ and from
the diagram with the reversed neutrino-loop momentum.  
The overall strength of the $k$-body contribution is determined by
$(G_Fa_n/\sqrt{2})^k$, and the neutrino propagators contribute the factor
$(1/4\pi)^k$ and the phase $(i)^k$.  This phase, when combined with the
$k$ powers of $i$ resulting from the generalization of
$T_{\mu\nu\lambda\sigma}(x_1,\ldots,x_4)$ in Eq.~(\ref{static T}), 
leads to the overall
phase $(i)^{2k} = + 1$ for even $k$.  As before, the Dirac trace contributes
the coefficient 4 of $\Phi^{(k)}$, and $I_n(z)$ gives the coefficients
$k!,(k-2)!,\ldots$ of the individual terms in $\Phi^{(k)}$.  Collecting
together the various factors in Eq.~(\ref{Vk}) we have (for $k$ even)
\begin{equation}
	V^{(k)} (\vec{r}_{12},\ldots,\vec{r}_{k1}) 
	= \frac{4}{\pi}
    \left( \frac{G_Fa_n}{2\pi\sqrt{2}}\right)^k \Phi^{(k)} (S_k, P_k) .
\label{final Vk}
\end{equation}
It is worth noting that the coefficient of $\Phi^{(k)}$ is real, as is each
of the terms in $\Phi^{(k)}$.  However, because the individual phase factors
$(i)^k, (i)^{k-2},\ldots$ can (depending on $k$) assume the values $(\pm 1)$,
various partial cancellations take place.  
Since the individual terms are unphysically large, these cancellations
do not eliminate the problem arising from the energy-density due to
neutrino exchange, but produce other novel unphysical effects, as we
discuss in Sec.~V below.  We will thus be led eventually
to the conclusion that
because all these effects are consequences of
the assumption that neutrinos are massless, there
must in fact be a lower bound on the mass of any
neutrino.

\section{INTEGRATION OF THE NEUTRINO-EXCHANGE POTENTIALS OVER A
SPHERICAL VOLUME}

Having derived the analytic forms of the $k$-body potentials in the previous
section, we turn to the problem of integrating these over a spherical volume,
in analogy to the electrostatic case.

\subsection{The 2-Body Potential}

	From Eq.~(\ref{final V2}) the 
2-body potential between neutrons arising from
neutrino exchange is
\begin{equation}
	V^{(2)}(r) = \frac{(G_Fa_n)^{2}}{4\pi^3 \hbar c} \frac{1}{r^5} \equiv
     \frac{\kappa}{r^5} ,
\label{V2 again}
\end{equation}
where $r$ is the distance between the neutrons.  In the absence of a
cutoff limiting how small $r$ can become, the integral of $V^{(2)}(r)$
over a spherical volume would be singular.  This contrasts with the
electrostatic case discussed in Sec. II, where the radial factors in
the volume element $ d^3x_1 d^3x_2$ in Eq.~(\ref{first U_C}) 
offset the contribution
from the potential $e_{0}^2/r_{12}$ and lead to a well-behaved result.
In practice a natural cutoff on $r$ exists in the systems of interest
to us, specifically in nuclei and in neutron stars.  In the former case
it is well known \cite{BM69} that the nucleon-nucleon interaction has
a strong repulsive component (the ``hard core'') which prevents the
nucleon-nucleon separation $r_c$ from becoming smaller than approximately
0.5~fm.  The dynamics of neutron stars, although less well known, would
also require a hard core.  Combining Eq.~(\ref{V2 again}) 
with Eqs.~(\ref{average g}) and (\ref{3D P})
we find for the 2-body contribution $U^{(2)}$ arising from neutrino
exchange
\begin{eqnarray}
	U^{(2)} & = & 
		\int_{r_c}^{2R} dr\, {\cal P}(r)
		V^{(2)}(r)
	\nonumber \\
	 & = & 
		\int_{r_c}^{2R} dr\, {\cal P}(r)
    \left( \frac{\kappa}{r^5}\right) 
	\nonumber \\
	 & = &  \kappa\left[
		\frac{3}{2R^{3}r_{c}^{2}}
		\left(1 - \frac{r_{c}^{2}}{4R^{2}}\right)
		- \frac{9}{4R^{4}r_{c}}
		\left(1 - \frac{r_{c}}{2R}\right)
		+ \frac{3}{16R^{6}}
		\left(2R - r_{c}\right)
		\right]
	\nonumber \\
	& \simeq & \frac{3\kappa}{2}
    \frac{1}{R^3 r_c^2} .
\label{approx U2}
\end{eqnarray}
Eq.~(\ref{approx U2}), which is the analog of 
Eq.~(\ref{first U_C}) in the electrostatic case,
gives the average interaction
energy for a single pair of neutrons having a uniform probability
distribution
in a spherical volume of radius $R$.

	It is worth commenting on the implication of the fact that the
integration of $V^{(2)}(r)$ over the spherical volume leads to
the introduction of an additional dimensional parameter, namely $r_{c}$.
We note to start with that since $r_{c} < R$ 
the presence of $r_c$ {\em strengthens}
the dimensional arguments given in the Introduction concerning the
approximate magnitude of the contributions from neutrino exchange.
Secondly, we anticipate the ensuing discussion of
the $k$-body contributions by noting that for $k \geq 4$ the integrations
of the potentials $V^{(k)}$ lead to expressions for $U^{(k)}$ which are
well-behaved as $r_c \rightarrow 0$.  Hence with the exception of
$U^{(2)}$, which makes a negligible contribution to the total
energy density, all $U^{(k)}$
for $k \geq 4$ depend only on $G_F$ and $R$, as noted in the
Introduction.

\subsection{The 4-Body Potential}

	The new feature which arises in the integration of
$V^{(4)}(r_{12},\ldots,r_{41})$, as given in Eqs.~(\ref{W4}) and
(\ref{V4}), is the presence of angular factors which result from
the scalar products of the derivative terms.  Although one might
be tempted to assume that any such factors average to zero when
integrated over a sphere, this is not the case as we discuss
below.  The evaluation of the terms involving derivatives can be
simplified by generalizing the result in Eq.~(\ref{12
gradients}).  To start with we recall from Eq.~(\ref{final
inverted SF}) that all the derivatives have their origin in the
neutrino propagator $S_{F}^{(0)}$, where $\vec{\partial}_{ij}$
acts on $r_{ij}$ appearing in $\Delta_{F}(\vec{r}_{ij},E)$.  In
principle the presence of all derivatives can be eliminated at
the outset by evaluating 
$\vec{\partial}_{ij}\Delta_{F}(\vec{r}_{ij},E)$ before carrying
out $\int dE$.  Although it is more convenient to evaluate $\int
dE$ first, one must recall that the $\vec{\partial}_{ij}$ which
then appear in $V^{(4)}$ are to be understood as acting on the
corresponding $\vec{r}_{ij}$ as if all the $\vec{r}_{ij}$ were
independent.  It is straightforward to show, for example, that
the same 2-body result is obtained if one evaluates
$\vec{\partial}_{ij}\Delta_{F}$ first, rather than leaving the
derivatives to the end as was done in arriving at Eq.~(\ref{general
V2}).

	Since a derivative such as $\vec{\partial}_{12}$ in
$V^{(4)}$ acts on a function $f(r_{12},\ldots)$ which depends
only on the magnitude of $\vec{r}_{12}$ (but not its direction)
we can write
\begin{equation}
	\frac{\partial}{\partial\vec{r}_{12}}
	f(r_{12},\ldots) =
	\hat{r}_{12}\frac{\partial}{\partial r_{12}}
	f(r_{12},\ldots).
\end{equation}
It follows that
\begin{equation}
	\vec{\partial}_{12}\cdot\vec{\partial}_{23}
	f(r_{12},r_{23},\ldots) 
	\equiv
	\frac{\partial}{\partial\vec{r}_{12}}
	\cdot
	\frac{\partial}{\partial\vec{r}_{23}}
	f(r_{12},r_{23},\ldots) 
	=	
	\hat{r}_{12}\cdot\hat{r}_{23}
	\frac{\partial}{\partial r_{12}}\frac{\partial}{\partial r_{23}}
	f(r_{12},r_{23},\ldots).
\label{1223 partials}
\end{equation}
Unlike the 2-body case, where the analog of Eq.~(\ref{1223
partials}) simplifies because $\hat{r}_{12}\cdot\hat{r}_{21} =
-1$, the angular factor $\hat{r}_{12}\cdot\hat{r}_{23}$ in
Eq.~(\ref{1223 partials}) is not a constant, and does not average
to zero when integrated over a sphere.  Moreover, the expression
in Eq.~(\ref{1223 partials}) cannot be calculated by separately
evaluating the angular factor and the function which it
multiplies, since these contributions are not independent. 

	To further explore the angular factors we consider
$\langle\hat{r}_{ij}\cdot\hat{r}_{lm}\rangle_{R}$, where the
notation $\langle\cdots\rangle_{R}$ denotes the average over a
spherical volume of radius $R$, and $i,j,l,m = 1,\ldots,4$.
$\langle\hat{r}_{12}\cdot\hat{r}_{34}\rangle_{R}$ and
$\langle\hat{r}_{12}\cdot\hat{r}_{23}\rangle_{R}$ have been
evaluated numerically \cite{Tu94} by randomly generating
$10^{8}$ sets of vectors $\vec{r}_{1}$, $\vec{r}_{2}$, 
$\vec{r}_{3}$, and $\vec{r}_{4}$ in a sphere of radius $R = 0.5$.
We find
\begin{mathletters}
\begin{eqnarray}
	\langle\hat{r}_{12}\cdot\hat{r}_{34}\rangle_{R}
		& = &
	\left\langle
	\frac{(\vec{r}_{1} - \vec{r}_{2})}
		{|\vec{r}_{1} - \vec{r}_{2}|}
		\cdot
	\frac{(\vec{r}_{3} - \vec{r}_{4})}
		{|\vec{r}_{3} - \vec{r}_{4}|}
	\right\rangle_{R}
	\simeq 0,
\label{1234 average} \\
	\langle\hat{r}_{12}\cdot\hat{r}_{23}\rangle_{R} 
		& = &
	\left\langle
	\frac{(\vec{r}_{1} - \vec{r}_{2})}
		{|\vec{r}_{1} - \vec{r}_{2}|}
		\cdot
	\frac{(\vec{r}_{2} - \vec{r}_{3})}
		{|\vec{r}_{2} - \vec{r}_{3}|}
	\right\rangle_{R}
		= -0.442.
\label{1223 average} 
\end{eqnarray}
\end{mathletters}
The result in Eq.~(\ref{1234 average}), which is zero to within
the expected statistical fluctuations, can be understood as
follows:  Given a sufficient number of trials (i.e., sets of
vectors) we would find that for each configuration of 4 particles
specified by $\vec{r}_{1}$, $\vec{r}_{2}$, 
$\vec{r}_{3}$, and $\vec{r}_{4}$ there will be another in which
the coordinates of particles 1 and 2 are same but those of
particles 3 and 4 are interchanged.  The contributions from these
two configurations to
$\langle\hat{r}_{12}\cdot\hat{r}_{34}\rangle_{R}$ evidently
cancel, which accounts for Eq.~(\ref{1234 average}).  By
contrast, the result in Eq.~(\ref{1223 average}) is less obvious,
but the fact that
$\langle\hat{r}_{12}\cdot\hat{r}_{23}\rangle_{R}$ is nonzero can
be understood as follows:  To start with the preceding argument
cannot be applied to 
$\langle\hat{r}_{12}\cdot\hat{r}_{23}\rangle_{R}$ since changing
the coordinates of particle 2 affects both $\hat{r}_{12}$ and
$\hat{r}_{23}$.  That such a scalar product is non-zero when
averaged over the sphere can be understood if we replace the unit
vectors by the corresponding dimensional vectors, $\hat{r}_{ij}
\rightarrow \vec{r}_{ij}$, etc.  Then
\begin{equation}
\langle\vec{r}_{12}\cdot\vec{r}_{23}\rangle_{R} =
\langle(\vec{r}_{1} - \vec{r}_{2}) \cdot
(\vec{r}_{2} - \vec{r}_{3})\rangle_{R} =
\langle
	\vec{r}_{1} \cdot \vec{r}_{2}
	- \vec{r}_{1} \cdot \vec{r}_{3}
	+ \vec{r}_{2} \cdot \vec{r}_{3}
	- r_{2}^{2}
\rangle_{R}.
\label{new 1223 average}
\end{equation}
The $\vec{r}_{i} \cdot \vec{r}_{j}$ terms in Eq.~(\ref{new 1223
average}) average to zero by an argument similar to that for
$\langle\hat{r}_{12}\cdot\hat{r}_{34}\rangle_{R}$:  Given a
sufficient number of trials, then for each random pair of vectors
$\vec{r}_{i}$ and $\vec{r}_{j}$, there will be another in which
$\vec{r}_{i}$ will be the same, but the coordinates of $r_{j}$
will be inverted ($\vec{r}_{j} \rightarrow -\vec{r}_{j}$).  The
sums of these two contributions evidently cancel, which leaves
\begin{equation}
\langle\vec{r}_{12}\cdot\vec{r}_{23}\rangle_{R} =
	\langle -r_{2}^{2}\rangle_{R} =
	-\frac{\int_{0}^{R}r_{2}^{2}\cdot4\pi r_{2}^{2}dr_{2}}
		{\int_{0}^{R}4\pi r_{2}^{2}dr_{2}}
	= -\frac{3}{5}R^{2}.
\label{simpler 1223 average}
\end{equation}
We note that since $\vec{r}_{2}$ is the coordinate of particle 2
measured from the center of the sphere, the upper limit on the
$r_{2}$ integration is $R$ and not $2R$, as in Eq.~(\ref{approx
U2}).
(In the 2-body case considered in the previous
subsection $r$ in Eq.~(\ref{approx U2}) denotes the distance between
two points whose maximum value is $2R$.)  The analytic result
in Eq.~(\ref{simpler 1223 average}) has been verified
numerically, and was used as a check on other numerical results.

	We conclude from the above that at least some of the
angular-dependent factors in $V^{(k)}$ for $k \geq 4$ are
non-zero.  One can show, moreover, that even an angular factor
such as $\hat{r}_{12}\cdot\hat{r}_{34}$, which averages to zero
over the sphere when considered by itself, can give a non-zero
contribution when multiplied by the functions which arise from
differentiating $S_{k}$ and $P_{k}$ in Eq.~(\ref{V4}).  To see
this consider the term in Eq.~(\ref{V4}) proportional to 
$\hat{r}_{12}\cdot\hat{r}_{34}$:
\begin{eqnarray}
	\lefteqn{
	\vec{\partial}_{12}\cdot\vec{\partial}_{34}
	\frac{1}{P_{4}S_{4}^{3}}  = 
	\hat{r}_{12}\cdot\hat{r}_{34}
	\frac{\partial}{\partial r_{12}}
	\frac{\partial}{\partial r_{34}}
	\left[
	\frac{1}
	{r_{12}r_{23}r_{34}r_{41}
	(r_{12} + r_{23} + r_{34} + r_{41})^{3}}
	\right]
	}
	\nonumber \\
	& = &
	\hat{r}_{12}\cdot\hat{r}_{34}
	\nonumber
	\\
	&  &
	\times \left[
	\frac{4(r_{12}^{2} + r_{34}^{2}) + r_{23}^{2} 
	+ r_{41}^{2} + 5(r_{12}r_{23} + r_{23}r_{34}
	+ r_{34}r_{41} + r_{41}r_{12})
	+ 20r_{12}r_{34} + 2r_{23}r_{41}}
	{r_{12}^{2}r_{23}r_{34}^{2}r_{41}
	(r_{12} + r_{23} + r_{34} + r_{41})^{5}}
	\right].
	\nonumber \\
\label{1234 derivatives}
\end{eqnarray}
Returning to the discussion following Eq.~(\ref{1223 average}) we
note that the contributions from two configurations which differ
by the interchange of the coordinates of particles 3 and 4 no
longer cancel in Eq.~(\ref{1234 derivatives}).  This is because
such an interchange effectively replaces the right-hand side of
Eq.~(\ref{1234 derivatives}) by 
\begin{equation}
	-\hat{r}_{12}\cdot\hat{r}_{34}
\left[
	\frac{4(r_{12}^{2} + r_{34}^{2}) + r_{24}^{2} 
	+ r_{31}^{2} + 5(r_{12}r_{24} + r_{24}r_{34}
	+ r_{34}r_{13} + r_{13}r_{12})
	+ 20r_{12}r_{34} + 2r_{24}r_{13}}
	{r_{12}^{2}r_{24}r_{34}^{2}r_{13}
	(r_{12} + r_{24} + r_{34} + r_{13})^{5}}
\right].
\label{interchanged 1234 term}
\end{equation}
With the coordinates $\vec{r}_{1}$ and $\vec{r}_{2}$ remaining
fixed, the expressions multiplying
$\hat{r}_{12}\cdot\hat{r}_{34}$ in Eqs.~(\ref{1234 derivatives})
and (\ref{interchanged 1234 term}) are not equal in general.
(The exception would be a 4-body configuration in which the
coordinates $\vec{r}_{1}$, $\vec{r}_{2}$, $\vec{r}_{3}$, and
$\vec{r}_{4}$ formed a regular tetrahedron.  However, in
3-dimensional space such a configuration is not possible for more
than 4-bodies.)  The above argument has been verified numerically
by a Monte Carlo simulation, as we discuss in more detail
elsewhere \cite{Tu94}.
	
	It follows from the preceding discussion that all of the
terms in Eq.~(\ref{V4}) are expected to contribute to $U^{(4)}$,
and this has been verified numerically by means of a Monte Carlo
simulation \cite{Tu94}.  Briefly, the symbolic program {\sc
Mathematica} 
\cite{Wolfram} was used at the outset to explicitly evaluate all
the derivatives appearing in Eq.~(\ref{V4}), as was done in
Eq.~(\ref{1234 derivatives}) and (\ref{interchanged 1234 term}).
The resulting expression for $V^{(4)}(\vec{r}_{12},\vec{r}_{23},
\vec{r}_{34}, \vec{r}_{41})$ was then evaluated for each of
$10^{8}$ configurations, where a configuration is obtained by
randomly generating a set of four 3-vectors
 $\vec{r}_{1}$, $\vec{r}_{2}$, $\vec{r}_{3}$, and
$\vec{r}_{4}$.  If 
$V^{(4)}_{i}(\vec{r}_{12}^{\,i},\vec{r}_{23}^{\,i},
\vec{r}_{34}^{\,i}, \vec{r}_{41}^{\,i})$ denotes the value of
$V^{(4)}$ obtained from the $i$th configuration, then
\begin{equation}
	U^{(4)} =
	\frac{1}{N_{r}}
	\sum_{i = 1}^{N_{r}}
	V^{(4)}_{i}(\vec{r}_{12}^{\,i},\vec{r}_{23}^{\,i},
	\vec{r}_{34}^{\,i}, \vec{r}_{41}^{\,i})
\end{equation}
where $N_{r} = 10^{8}$ in our calculations. We find numerically
\cite{Tu94},
\begin{equation}
	U^{(4)} = 
		\frac{4}{\pi R}
		\left(\frac{G_{F}a_{n}}{2\sqrt{2}\pi R^{2}}
		\right)^{4}
	(7.7).
\label{numerical U4}
\end{equation}

	There are several implications of the 4-body results
which will be useful in the ensuing discussion.  To start with
there is no theoretical reason why $U^{(4)}$ should vanish, and
there is no suggestion that it does either analytically or
numerically.  Secondly, the computational effort required to
evaluate $U^{(4)}$ is sufficiently great as to question the
feasibility of a calculation of even $U^{(6)}$ and $U^{(8)}$, let
alone $U^{(k)}$ for $k = \O(10^{57})$.  We note that a calculation
of $U^{(k)}$ starting from a generalized version of Eq.~(\ref{infinitesmal
PE}) would require a $3k$-dimensional integral, and hence this
approach is impractical for large $k$.  Fortunately it turns out
that for present purposes it suffices to determine the
$0$-derivative contributions $U_{0}^{(k)}$ to $U^{(k)}$, 
for which we can
obtain an approximate analytic result as we describe below.

	Having evaluated $U^{(4)}$ directly by a Monte-Carlo
simulation, we ask whether it is possible to generalize the
probability density function ${\cal P}(r)$ in Eq.~(\ref{3D
P}) to the 4-body case.  If this were possible it would allow the
Monte-Carlo evaluation to be carried out {\em analytically}, just
as in the 2-body case.  A major obstacle that must be confronted
in any attempt to generalize ${\cal P}(r)$ is that the
appropriate $k$-body probability density function must depend not
only on the separations $r_{ij} = |\vec{r}_{i} - \vec{r}_{j}|$ 
of $i$ and $j$, but also on angular factors such as $\hat{r}_{ij}
\cdot\hat{r}_{lm}$ which arise from the derivative terms.  Even
for the $0$-derivative term the generalization of ${\cal P}(r)$
to the $k$-body case is not known at present, as we discuss in
Appendix~C.  For this reason, we develop in Appendix C the ``mean
value approximation'' which allows the $k$-body integrals to be
done analytically, and which we use for $k \geq 6$.

\subsection{The $k$-Body Potential for $\bif k \geq 6$}

	As we have noted previously, $W = \sum_{k}W^{(k)} =
\sum_{k}U^{(k)}{N \choose k}$ in
Eq.~(\ref{W sum}) is dominated by terms with $k \simeq N =
\O(10^{57})$.  For values of $k$ this large, numerical evaluation
of $U^{(k)}$ is not possible, even for the 0-derivative terms.
However, there is a useful bound on the 0-derivative contribution
$U_{0}^{(k)}$ to $U^{(k)}$ which we derive in this subsection
which, along with the ``mean value approximation'' in Appendix~C,
forms
the basis for the ensuing discussion.  Before doing so we present
the numerical result for $U_{0}^{(6)}$  for
illustrative purposes,  
\begin{equation}
	U^{(6)}_{0} = 
		-\frac{4}{\pi R}
     \left( \frac{G_Fa_n}{2\sqrt{2}\pi R^{2}} \right)^6 C_{6},
\label{U6}
\end{equation}
\begin{equation}
	C_{6} = 0.268(34).
\label{C6}
\end{equation}

	We proceed to derive the previously-discussed bound on
$U^{(k)}_{0}$. Let
${\cal P}^{(k)}(r_{12},r_{23},\ldots,r_{k1})$ denote the
probability density for the separation of particles 1 and 2 to be
in the interval between $r_{12}$ and  $r_{12} + dr_{12}$, etc.
Thus, the differential probability $d^{k}\Pi^{(k)}$ of a
configuration specified by $\vec{r}_{1}, \vec{r}_{2}, \ldots,
\vec{r}_{k}$ is
\begin{equation}
	d^{k}\Pi^{(k)} =
		{\cal P}^{(k)}(r_{12},r_{23},\ldots,r_{k1})
	dr_{12}dr_{23}\cdots dr_{k1},
\label{differential k-prob-density}
\end{equation}
where
\begin{equation}
	\int^{2R}_{0}dr_{12}\cdots\int^{2R}_{0}dr_{k1}
		{\cal P}^{(k)}(r_{12},r_{23},\ldots,r_{k1})
	= 1.
\label{k-prob normalization}
\end{equation}
We know that the function
${\cal P}^{(k)}(r_{12},r_{23},\ldots,r_{k1})$ exists because it can
be determined numerically by randomly generating sets of vectors
$\vec{r}_{1}, \vec{r}_{2}, \ldots,\vec{r}_{k}$ in a sphere, and
then computing the probability $d^{k}\Pi^{(k)}$ of
any configuration, as in Eq.~(\ref{differential k-prob-density}).
In fact the first non-trivial generalization of ${\cal P}(r)$
in Eq.~(\ref{3D P}), namely 
${\cal P}^{(3)}(r_{12},r_{23},r_{31})$, has been inferred in
exactly this way \cite{Tu94}.  (As noted in
Sec.~III, the fact that there is no 3-body contribution to $W$
has to do with the functional form of $V^{(3)}$, not 
${\cal P}^{(3)}(r_{12},r_{23},r_{31})$.)  Combining Eqs.~(\ref{differential
k-prob-density}) and (\ref{Vk}) we find:
\begin{eqnarray}
	U_{0}^{(k)} & = & \frac{4}{\pi}
	\left(\frac{G_{F}a_{n}}{2\pi\sqrt{2}}\right)^{k}
	{i}^{k}k!
	\int^{2R}_{0}dr_{12}\cdots\int^{2R}_{0}dr_{k1}
		{\cal P}^{(k)}(r_{12},r_{23},\ldots,r_{k1})
	\nonumber \\
	&  &  ~~~~\times
	\left[
	\frac{1}{r_{12}r_{23}\cdots r_{k1}
	(r_{12} + r_{23} + \cdots + r_{k1})^{k + 1}}
	\right].
\label{U0 k}
\end{eqnarray}
We observe that the function multiplying ${\cal P}^{(k)}(\cdots)$ is
non-negative, and is a monotonically decreasing function of
$r_{ij}$ in the interval $[0,2R]$.  It follows that the minimum
value of $|U_{0}^{(k)}|$ is achieved when $r_{ij} = 2R$ for all
$i,j$.  Thus, 
\begin{equation}
	|U_{0}^{(k)}| \geq  \frac{4}{\pi}
	\left(\frac{G_{F}a_{n}}{2\pi\sqrt{2}}\right)^{k}
	k!
	\frac{1}{(2R)^{k}(2Rk)^{k + 1}}
	\int^{2R}_{0}dr_{12}\cdots\int^{2R}_{0}dr_{k1}
		{\cal P}^{(k)}(r_{12},r_{23},\ldots,r_{k1}).
\label{U0 k condition}
\end{equation}
Using the normalization condition, Eq.~(\ref{k-prob normalization}),
we have
\begin{equation}
	|U_{0}^{(k)}| \geq  \frac{2}{\pi R}
	\left(\frac{G_{F}a_{n}}{8\pi\sqrt{2}R^{2}}\right)^{k}
	\frac{k!}{k^{k + 1}}.
\label{U0 k constraint}
\end{equation}

	The result in Eq.~(\ref{U0 k constraint}) represents the
$\O(G_{F}^{k})$ contribution from the standard diagram (with both
senses of the loop momentum included) as can be seen from the
starting point in Eq.~(\ref{Vk}).  As noted in Section~ III and
in Fig.~\ref{Second 4-body figure},
there are in $\O(G_{F}^{k})$ a total of $(k - 1)!/2$ topologically
distinct pairs of diagrams, with each pair representing the sum
of the contributions from both senses of the loop momentum.  The
sum of all these $(k~-~1)!/2$ diagrams replaces the expression in
square brackets in Eq.~(\ref{U0 k}) by the appropriately
symmetrized [in the $k(k-1)/2$ variables $r_{ij}$] generalization
of the standard contribution exhibited there.
Since ${\cal P}^{(k)}(r_{12},r_{23},\ldots,r_{k1})$ is also a
symmetric function of $r_{ij}$, we conclude that each of the
$(k-1)!/2$ contributions to $U_{0}^{(k)}$ will be equal in
magnitude to that arising from Eq.~(\ref{U0 k}).  It follows that the
complete contribution to $U_{0}^{(k)}$ is simply $(k-1)!/2$ times
that given in Eq.~(\ref{U0 k constraint}),
\begin{equation}
	|U_{0}^{(k)}| \geq  
		\frac{2}{\pi R}
	\left(\frac{G_{F}a_{n}}{8\pi\sqrt{2}R^{2}}\right)^{k}
	\frac{k!(k - 1)!}{2k^{k + 1}}
		=
		\frac{1}{\pi R}
	\left(\frac{G_{F}a_{n}}{8\pi\sqrt{2}R^{2}}\right)^{k}
	\frac{(k!)^{2}}{k^{k + 2}}.
\label{better U0 k constraint}
\end{equation}

	It is instructive to consider the expression for
$U_{0}^{(4)}$ in greater detail to understand the origin of the
factor $(k-1)!/2$.  For brevity define
\begin{equation}
	u = r_{12} 
	\phantom{space}
	v = r_{23}
	\phantom{space}
	w = r_{34}
	\phantom{space}
	x = r_{41}
	\phantom{space}
	y = r_{24}
	\phantom{space}
	z = r_{13}.
\label{uvxyz}
\end{equation}
From Eq.~(\ref{U0 k}) the complete expression for $U_{0}^{(4)}$
is then given by
\begin{eqnarray}
	U_{0}^{(4)} & = & \frac{4}{\pi}
	\left(\frac{G_{F}a_{n}}{2\pi\sqrt{2}}\right)^{4}
	4!
	\int^{2R}_{0}du\,dv\,dw\,dx\,dy\,dz\,
		{\cal P}^{(4)}(u,v,w,x,y,z)
	\nonumber \\
	&  &  \times
	\left[
	\frac{1}{uvwx(u + v + w + x)^{5}}
	+
	\frac{1}{vxyz(v + x + y + z)^{5}}
	+
	\frac{1}{uwyz(u + w + y + z)^{5}}
	\right],
\label{more complete U0 4}
\end{eqnarray}
where the three terms in square brackets arise from diagrams (a),
(b), and (c) respectively in Fig.~\ref{First 4-body figure}.
We note that ${\cal P}^{(4)}(u,v,w,x,y,z)$ is necessarily a symmetric
function of its arguments, and that the expression in square
brackets is also a symmetric function of the same arguments.  It
then follows that whatever the explicit functional form of ${\cal
P}^{(4)}(u,v,w,x,y,z)$, each of the 3 terms in square brackets
contributes equally to the integral.  For the $k$-body
contribution the expression in square brackets would contain
$(k-1)!/2$ terms, which is the origin of the factor $k!/2k$ in
Eq.~(\ref{better U0 k constraint}).

	As we have noted previously the many body contributions
are dominated by terms with $k \simeq N$, and for these terms the
bound in Eq.~(\ref{better U0 k constraint}) is useful.
However, if we wish to obtain a closed-form expression for the
sum, then the bound in Eq.~(\ref{better U0 k constraint}) cannot
be taken over directly, since the individual terms in the series
alternate in sign, as can be seen from Eq.~(\ref{U0 k}).  We thus
wish to replace the bound by a more precise estimate of the
integral, and this can be done using the ``mean value
approximation,'' as discussed in Appendix~C.   From the preceding
discussion, $U_{0}^{(k)}$ can be obtained by multiplying the
expression in Eq.~(\ref{approximate U0 k}) by $k!/2k$ which
gives, for even $k$,
\begin{equation}
	U_{0}^{(k)} \simeq 
	\frac{2i^{k}}{\pi R}
	\left(\frac{G_{F}a_{n}}{2\pi\sqrt{2}R^{2}}\right)^{k}
	\frac{(k!)^{2}}{k^{k + 2}}.
\label{effective U0 k}
\end{equation}
Eq.~(\ref{effective U0 k}) is the starting point of our
discussion in the next section of the combinatoric factors
arising in a system of $N$ particles.

\section{Combinatorics for Many-Body Systems and Numerical
Results}

\subsection{Combinatorics}

	The expression for $U_{0}^{(k)}$ in Eq.~(\ref{effective U0
k}) represents the $k$-body contribution to the neutrino-exchange
energy in the approximation of retaining only the 0-derivative
terms.  As we noted in Sec.~II, for a system of $N$ particles
there are ${N \choose k}$ such (identical) terms, and hence the
total neutrino-exchange energy for a spherical distribution of
$N$ particles is given by
\begin{eqnarray}
	W & \simeq & U^{(2)}{N \choose 2} +
		\sum_{k = 4 \atop even}^{N}
		U_{0}^{(k)}{N \choose k}
	\nonumber \\
	& = & W^{(2)} +
	\sum_{k = 4 \atop even}^{N}
	\frac{2i^{k}}{\pi R}
	\left(
	\frac{G_{F}a_{n}}{2\pi\sqrt{2}R^{2}}
	\right)^{k}
	\frac{(k!)^{2}}{k^{k + 2}}
	{N \choose k}.
\label{WforN 1}
\end{eqnarray}
$U^{(2)}$ is the 2-body contribution and, since it can be
evaluated exactly, the full expression in Eq.~(\ref{approx U2})
will be used.  When calculating the neutrino-exchange energy in a
white dwarf or a neutron star, $W^{(2)} = U^{(2)}{N \choose 2}$
is completely negligible and hence it can be dropped from the sum
over $k$ when convenient.  

	The sum over $k$ in Eq.~(\ref{WforN 1}) can be evaluated
in closed form by making use of the Stirling approximation for
$k!$, which is valid for large $k$:
\begin{equation}
	k! \simeq \sqrt{2\pi}k^{k + 1/2}e^{-k}.
\label{sterling k}
\end{equation}
Combining Eqs.~(\ref{WforN 1}) and (\ref{sterling k}) we have
\begin{equation}
	W \simeq   W^{(2)} +
	\sqrt{\frac{2}{\pi}}
	\,
	\frac{2}{R}
	\,
	\sum_{k = 4 \atop even}^{N}
	\frac{i^{k}k!}{k^{3/2}}
	\left(
	\frac{G_{F}a_{n}}{2\pi\sqrt{2}eR^{2}}
	\right)^{k}
	{N \choose k},
\label{WforN 2}
\end{equation}
where $\ln e = 1$.
The remaining $k!$ will be replaced shortly by an integral
representation which, when evaluated by the saddle-point method,
amounts to a second application of the Stirling approximation.
The sum over $k$ can be further simplified by noting that were we
to apply the Stirling approximation again at this stage we could
write
\begin{equation}
	\frac{k!}{k^{3/2}} \simeq 
		\sqrt{2\pi}
		\frac{k^{k + 1/2}}{k^{3/2}}e^{-k}
	= \sqrt{2\pi}
	  \exp[(k + 1/2)\ln k - k - (3/2)\ln k].
\label{approx k! k3/2}
\end{equation}
Since we are interested in evaluating the sum for $k \leq N =
\O(10^{57}$), the term $(3/2)\ln k$ could be dropped relative to
$k\ln k$ in Eq.~(\ref{approx k! k3/2}), and hence the factor
$k^{3/2}$ in Eq.~(\ref{WforN 2}) could also be dropped. 
However, a better approximation is to simply replace $k^{3/2}$ by
$N^{3/2}$, noting that $W$ is dominated by terms with $k \simeq
N$.  (This approximation also gives the smallest estimate for
$W$.)
Although
the factor $k^{3/2}$ is negligible from a quantitative point of
view, it can (with some effort) be reinstated if necessary, as we
demonstrate in Appendix~D.  
In the approximation of replacing $k^{3/2}$ by $N^{3/2}$,
$W$ becomes
\begin{equation}
	W \simeq W^{(2)} +
		\sqrt{\frac{2}{\pi}}
		\,
		\frac{2}{R}
		\,
		\frac{1}{N^{3/2}}
		\,
		\sum_{k = 4 \atop even}^{N}
		\,k!
		(i\Gamma_{R})^{k}
		{N \choose k},
\label{WforN 3}
\end{equation}
where
\begin{equation}
	\Gamma_{R} =
	\frac{G_{F}a_{n}}{2\pi\sqrt{2}eR^{2}}.
\label{Gamma R}
\end{equation}
Although the lower limit on the sum in Eq.~(\ref{WforN 3}) is $k
= 4$, the summation can be extended down to $k = 0$ by subtracting
the $k = 0$ and $k = 2$ contributions at the end.  We thus
consider the sums,
\begin{equation}
	{\sum}^{(e)} =
		\sum_{k = 0 \atop even}^{N}
		\,k!
		(i\Gamma_{R})^{k}
		{N \choose k},
\label{extended sum}
\end{equation}
\begin{equation}
	{\sum}^{(\pm)} =
		\sum_{k = 0}^{N}
		\,k!
		(\pm i\Gamma_{R})^{k}
		{N \choose k},
\label{pm sum}
\end{equation}
where in ${\sum}^{(\pm)}$  the summation extends over all $k$.
The quantity we want is ${\sum}^{(e)}$, and from
Eqs.~(\ref{extended sum}) and (\ref{pm sum}) it is given by
\begin{equation}
	{\sum}^{(e)} = \frac{1}{2}\left[{\sum}^{(+)} +
		{\sum}^{(-)}\right].
\label{e sum}
\end{equation}
To evaluate ${\sum}^{(\pm)}$ we introduce into Eq.~(\ref{pm sum})
the familiar integral representation for $k!$,
\begin{equation}
	k! = \int^{\infty}_{0} du\,e^{-u}u^{k},
\end{equation}
which gives
\begin{equation}
	{\sum}^{(\pm)} = \int^{\infty}_{0}du\,e^{-u}\, 
		\sum_{k = 0}^{N}
		(\pm iu\Gamma_{R})^{k}
		{N \choose k}.
\label{new pm sum}
\end{equation}
The sum of the binomial series in Eq.~(\ref{new pm sum}) can be
expressed in closed form \cite{GR80},
\begin{equation}
	\sum_{k = 0}^{N} x^{k}a^{N - k}{N \choose k}
		= (a + x)^{N},
\label{binomial sum}
\end{equation}
where $a = 1$ in the present case.  Combining Eqs.~(\ref{new pm
sum}) and (\ref{binomial sum}) gives
\begin{equation}
	{\sum}^{(\pm)} = \int^{\infty}_{0}du\,e^{-u}\,(1 \pm
		iu\Gamma_{R})^{N}.
\label{newer pm sum}
\end{equation}
The integral in Eq.~(\ref{newer pm sum}) can be readily evaluated
by the saddle-point method \cite{ARF85 p428}.  We introduce the
variable $t$ defined by
\begin{equation}
	1 \pm iu\Gamma_{R} = \pm iNt,
\label{defining t}
\end{equation}
where the upper (lower) sign applies to ${\sum}^{(+)}$ (${\sum}^{(-)}$).
${\sum}^{(+)}$ can then be written in the form
\begin{equation}
	{\sum}^{(+)} = g(N)\int^{t_{max}}_{t_{min}}dt\,
		e^{Nf(t)},
\label{+ sum}
\end{equation}
where
\begin{equation}
	g(N) = \frac{N}{\Gamma_{R}}
		\exp\left[-\frac{i}{\Gamma_{R}}
			+ N\ln(iN)\right],
\label{gN}
\end{equation}
\begin{equation}
	f(t) = \left(-\frac{t}{\Gamma_{R}} + \ln t\right),
\label{ft}
\end{equation}
and $t_{min}$ ($t_{max}$) corresponds to $u = 0$ ($u = \infty$) in
Eq.~(\ref{defining t}).  Using Ref.~\cite{ARF85 p428}, ${\sum}^{(+)}$
is then given by 
\begin{equation}
	{\sum}^{(+)} \simeq 
		\frac{\sqrt{2\pi}g(N)e^{Nf(t_{0})}}
		{|Nf''(t_{0})|^{1/2}},
\label{+ sum 1}
\end{equation}
where $t_{0} = \Gamma_{R}$ is the saddle point, and the primes
indicate differentiation.  Combining Eqs.~(\ref{gN})--(\ref{+ sum
1}), we find
\begin{eqnarray}
	{\sum}^{(+)} & \simeq &
		\sqrt{2\pi N}
		\exp\left\{
		N\left[\ln\left(\Gamma_{R}N\right) - 1\right]
		+ i(N\pi/2 - 1/\Gamma_{R})\right\}
	\nonumber \\
	& = & 
		\sqrt{2\pi N}
		(\Gamma_{R}N/e)^{N}
		e^{i(N\pi/2 - 1/\Gamma_{R})}.
\label{+ sum 2}
\end{eqnarray}
Since ${\sum}^{(-)} = ({\sum}^{(+)})^{*}$, as can be seen from
Eq.~(\ref{new pm sum}), ${\sum}^{(e)}$ in Eqs.~(\ref{extended
sum}) and (\ref{e sum}) is given by
\begin{equation}
	{\sum}^{(e)} \simeq 
		(-1)^{N/2}
		\sqrt{2\pi N}
		(\Gamma_{R}N/e)^{N}
		\cos(1/\Gamma_{R}).
\label{new e sum}
\end{equation}
Noting that only even values of $N$ contribute to $\sum_{k}$, it
follows that the factor $(-1)^{N/2}$ is always real and
alternates in sign.  We also note that because the expression on
the right hand side of Eq.~(\ref{+ sum 1}) is the origin of both
the Stirling approximation for $k!$ and the result for ${\sum}^{(e)}$
above, it follows that both factors of $k!$ in Eq.~(\ref{WforN
1}) have been treated in a consistent way.  Combining Eqs.~(\ref{WforN
3}), (\ref{extended sum}), and (\ref{new e sum}) the neutrino-exchange
energy $W$ can be written in the form
\begin{equation}
	W \simeq W^{(2)} +
		\sqrt{\frac{2}{\pi}}
		\,
		\frac{2}{R}
		\,
		\frac{1}{N^{3/2}}
		\left({\sum}^{(e)} - {\sum}^{(0)}
			- {\sum}^{(2)}\right),
\label{WforN 4}
\end{equation}
where ${\sum}^{(0)}$ and ${\sum}^{(2)}$ are the $k = 0$ and $k =
2$ contributions to the sum over $k$ in Eq.~(\ref{extended sum})
which must be subtracted out.  From Eq.~(\ref{extended sum}),
\begin{equation}
	{\sum}^{(0)} + {\sum}^{(2)} =
		1 - N(N - 1)\Gamma_{R}^{2},
\end{equation}
and hence
\begin{eqnarray}
	W & \simeq &
		(-1)^{N/2}
		\frac{4}{RN}
		\left(\frac{\Gamma_{R}N}{e}\right)^{N}
		\cos(1/\Gamma_{R})
	\nonumber \\
	&   & \mbox{}
		+
		\left[ W^{(2)} 
		-
		\sqrt{\frac{2}{\pi N^{3}}}
		\left(\frac{2}{R}\right)
		+
		\frac{N(N - 1)}{2e^{2}}
		\sqrt{\frac{1}{2\pi^{5}N^{3}}}
		\left(\frac{G_{F}^{2}}{R^{5}}\right)
		\right].
\label{WforN 5}
\end{eqnarray}
For purposes of setting a bound on the neutrino mass the
expression in square brackets is negligible and can be dropped.
Using Eq.~(\ref{Gamma
R}) we can then write Eq.~(\ref{WforN 5}) as 
\begin{equation}
	W \simeq (-1)^{N/2}\frac{4}{RN}
		\left(
		\frac{G_{F}a_{n}N}
		{2\pi\sqrt{2}e^{2}R^{2}}
		\right)^{N}
		\cos(1/\Gamma_{R}).
\label{WforN 6}
\end{equation}

	Although the expression in Eq.~(\ref{WforN 6}) is the
result of summing over all $k \leq N$ in Eq.~(\ref{WforN 3}), it
can be shown that this expression is effectively the contribution
of a few terms which dominate the sum over $k$.  One way of
seeing this is to consider the ratio of successive terms in
Eq.~(\ref{WforN 3}).  Dropping the factor $k^{-3/2}$ as
discussed above we
find
\begin{equation}
	\left|\frac{W^{(k + 2)}}{W^{(k)}}\right|
		\simeq \Gamma_{R}^{2}
			(N - k)(N - k - 1).
\label{W ratio}
\end{equation}
As we demonstrate in Eq.~(\ref{Lambda_R}) below, the product $\Gamma_{R}N$
is of order $10^{11}$, and hence for small $k$ the ratio in
Eq.~(\ref{W ratio}) is much larger than unity.  It follows that
successive terms in $\sum_{k}$ initially make increasingly larger
contributions.  As noted in Sec.~II, however, the rate at which
this ratio increases is itself a decreasing function for
increasing $k$, and eventually a value $k_{max}$ is reached for
which the ratio is unity.  From Eq.~(\ref{W ratio}) $k_{max}$ is
given by
\begin{equation}
	\Gamma_{R}^{2}(N - k_{max})^{2} \simeq 1,
\end{equation}
\begin{equation}
	k_{max} \simeq N - 1/\Gamma_{R}.
\label{k_max}
\end{equation}
As noted in Sec.~II, the fact that the individual terms in
$\sum_{k}$ reach a maximum is a consequence of the behavior of ${N
\choose k}$ for large $k$, and of Eq.~(\ref{ratio binomial
coefficients}) in particular.  To calculate $W^{(max)}$, the
value of $W^{(k)}$ corresponding to $k_{max}$, we can set
$k_{max} \simeq N$ and $(1/\Gamma_{R}) \simeq 0$ since $N \gg
1/\Gamma_{R}$ (see below).  From Eq.~(\ref{WforN 3}) we then have
\begin{equation}
	W^{(max)} \simeq 
		(-1)^{N/2}
		\frac{4}{RN}
		\left(
		\frac{\Gamma_{R}N}{e}\right)^{N},
\label{W max}
\end{equation}
where 
the Stirling approximation for $N!$ has
been used.  Eq.~(\ref{W max}) agrees with the result obtained
previously in Eq.~(\ref{WforN 6}), up to the factor
$\cos(1/\Gamma_{R})$ which is the ``memory'' of the sum over $k$
as we discuss below.  It follows from Eq.~(\ref{W ratio}) that for
practical purposes one can approximate $W$ reasonably well by the
single term $W^{(max)}$.

\subsection{Numerical Results}

	We proceed to evaluate $W$ numerically for a typical
neutron star which we take to be the observed pulsar in the
Hulse-Taylor binary system PSR 1913+16 \cite{HT75,TW89,W92}.  The mass
$M_{1}$ of this pulsar is accurately known \cite{TW89,W92},
\begin{equation}
	M_{1} = 1.4411(7)M_{\odot},
\end{equation}
and hence the mass of a typical neutron star will be taken to be
\begin{equation}
	M = 1.4M_{\odot} = 2.8 \times 10^{33}\,\,\mbox{g}.
\label{neutron star mass}
\end{equation}
To calculate the number of neutrons $N$ we ignore the
contribution to $M$ from gravitational binding energy [see
Eq.~(\ref{V_GRAV}) below], and assume
that the neutron star is composed exclusively of neutrons.  Using
Eq.~(\ref{neutron star mass}) then leads to 
\begin{equation}
	N = 1.7 \times 10^{57}.
\label{neutron star N}
\end{equation}
The radius $R$ of the neutron star, although not directly
observable, can be inferred in various models.  We assume the
nominal value $R = 10$ km $\equiv R_{10}$ which corresponds to a
mass density $\rho_{m}$ and a number density $\rho$
given by
\begin{equation}
	\rho_{m} =  6.7 \times 10^{14}\,\,\mbox{g\,cm}^{-3},
\label{neutron star mass density}
\end{equation}
\begin{equation}
	\rho =  4.0 \times 10^{38}\,\,\mbox{cm}^{-3}.
\label{neutron star number density}
\end{equation}
In what follows we will assume these
values of $R$ and $\rho$, which are typical of the results
that arise in existing models of neutrons stars \cite{ST83,AA89}.
Combining Eqs.~(\ref{neutron star mass})--(\ref{neutron star mass
density}) with $|a_{n}| = 1/2$ and reinstating $\hbar$ and $c$, we find
\begin{equation}
	\frac{(G_{F}/\hbar c)N}{R_{10}^{2}}
		= 7.6 \times 10^{12},
\end{equation}
\begin{equation}
	\Gamma_{R} \equiv \frac{(G_{F}/\hbar c)|a_{n}|}
		{2\pi\sqrt{2}eR_{10}^{2}}
		= 9.4 \times 10^{-47}
		= \frac{1}{1.1 \times 10^{46}},
\label{Gamma_R}
\end{equation}
\begin{equation}
	\Lambda_{R} 
		\equiv 
	\frac{(G_{F}/\hbar c)N|a_{n}|}
	{2\pi\sqrt{2}e^{2}R_{10}^{2}}
		= 1.2 \times 10^{11}|a_{n}|
		= 5.8 \times 10^{10},
\label{Lambda_R}
\end{equation}
\begin{eqnarray}
	W & = & (-1)^{N/2}
		\frac{4\hbar c}{R_{10}}
	      \left(6.0\times 10^{-58}\right)
	    \left(5.8 \times 10^{10}\right)^{1.7 \times 10^{57}}
	   \cos(1/\Gamma_{R})
	\nonumber \\
	& & 
\label{W estimate}
	\\
	  & = &  
		10^{(2\times 10^{58} - 57 - 10)}
		(-1)^{N/2}
	   \cos(1/\Gamma_{R})
		\,\,\,\mbox{eV}.
	\nonumber
\end{eqnarray}
In Eq.~(\ref{W estimate})
the three terms in the exponent of 10 arise, respectively,
from $(\Gamma_{R}N/e)^{N}$, $1/N$, and $4\hbar c/R_{10}$ (in eV).  
For the ratio $W/Mc^{2}$ we find
\begin{equation}
	\frac{W}{Mc^{2}} = 	
		(-1)^{N/2}
		10^{(2\times 10^{58} - 57 - 76)}
	   \cos(1/\Gamma_{R}).
\label{W/mc2}
\end{equation} 
Leaving aside the cosine factor, to which we will return
shortly, we see that the neutrino-exchange energy is significantly larger
than the known mass-energy of the 1913+16 pulsar.  
For later purposes we note that in a neutron star $W^{(k)}/Mc^{2}$
exceeds unity for $k \geq 8$, which is a relatively low order
perturbation.  Using Eq.~(\ref{WforN 1})
we find that for the
0-derivative contribution,
\begin{equation}
	W^{(8)} = \frac{2i^{8}\hbar c}{\pi R_{10}}
		\frac{(8!)^{2}}{8^{10}}
		\left[
		\frac{(G_{F}/\hbar c)|a_{n}|}
			{2\pi\sqrt{2}R_{10}^{2}}
		\right]^{8}
		{N \choose 8}
		= 
		5.0 \times 10^{77} \,\,\mbox{eV},
\end{equation}
\begin{equation}
	\frac{W^{(8)}}{Mc^{2}} = 3.2 \times 10^{11}.
\label{W8}
\end{equation}
Another quantity of interest is
the ratio of $W$ to the total mass of the Universe $M_{U}$.
We find:
\begin{equation}
	M_{U} \simeq 4 \times 10^{55}\,\,\mbox{g},
\label{mass of universe}
\end{equation}
\begin{equation}
	\frac{W}{M_{U}c^{2}} \simeq 10^{(2\times 10^{59} - 57 -
99)}.
\label{W/MU}
\end{equation}
It follows from Eqs.~(\ref{W/mc2}) and (\ref{W/MU}) that the
neutrino-exchange energy in a single neutron star, when calculated
in the Standard Model, exceeds the total mass-energy of the
Universe, and this may be termed the
``neutrino-exchange energy-density catastrophe.'' 
In estimating $M_{U}$, we assume the average
density $\rho_{U}$ of the Universe to be $\rho_{U} \simeq 1
\times 10^{-29}$ g cm$^{-3}$, which is the critical density
corresponding to a Hubble constant of 
($80 \pm 17$) km s$^{-1}$ Mpc$^{-1}$
\cite{Weinberg,Freedman}, and we have taken the radius of the
Universe to be
$\simeq 1 \times
10^{10}$~ly \cite{Freedman}.

	We return to  discuss the sign of $W$ in Eqs.~(\ref{WforN 6})
and (\ref{W/mc2}).  
As noted previously, the factor $(-1)^{N/2}$ 
alternates in sign, and hence the sign of $W$ depends on $N$, as
well as on $\Gamma_{R}$ through $\cos(1/\Gamma_{R})$.  
Either sign of $W$ leads to a
catastrophic outcome:  $W > 0$ would correspond to a large
repulsive force against which the neutron star would be unstable,
whereas $W < 0$ would cause the neutron star to collapse to a
black hole.  In either case neutron stars as we know them would not
exist.
In this context it is helpful to recall that the
classical gravitational potential energy $V_{GRAV}$ is
approximately given by 
\begin{equation}
	V_{GRAV} \simeq -\frac{3}{5}\frac{G_{N}M^{2}}{R},
\label{V_GRAV}
\end{equation}
where $G_{N} = 6.67259(85) \times 10^{-11}$
m$^{3}$kg$^{-1}$s$^{-2}$ is the Newtonian gravitational constant.
For the assumed values of $M$ and $N$ in Eqs.~(\ref{neutron star
mass}) and (\ref{neutron star N}) we find
\begin{eqnarray}
	\frac{|V_{GRAV}|}{Mc^{2}} & \simeq & 0.124,
\label{V_Grav/M}
\\
	\frac{|V_{GRAV}|}{N} & \simeq & 117  \,\,\,\mbox{MeV/nucleon}.
\label{V_Grav/N}
\end{eqnarray}
It follows from Eqs.~(\ref{W/mc2}), (\ref{V_Grav/M}), and
(\ref{V_Grav/N}) that the neutrino-exchange energy, whatever its
sign, would dominate over $V_{GRAV}$, which is the largest 
contribution to the binding energy of a neutron star.  We note in
passing that since $V_{GRAV}$ is negative, the effect of
including gravitational binding energy in Eq.~(\ref{neutron star
mass}) would be to {\em increase} $N$ in Eq.~(\ref{neutron star
N}), which would make the neutrino-exchange energy density even
larger than the value quoted in Eqs.~(\ref{W estimate}) and
(\ref{W/mc2}).

	We turn next to the factor $\cos(1/\Gamma_{R}) =
\cos(2\pi\sqrt{2}eR^{2}/G_{F}a_{n})$.  This is similar to the
oscillatory factors that arise in other many-fermion systems, an
example being the Ruderman-Kittel interaction \cite{RK}. To better
understand the significance of this factor it is instructive to
ask what the expression for $W$ would be if all terms in
$\sum_{k}$ had the same sign (which we assume to be positive for
illustrative purposes).  The analog of Eq.~(\ref{extended sum}) would
then be 
\begin{equation}
	\widetilde{\sum}^{(e)} =
		\sum_{k = 0 \atop even}^{N}
		\,k!
		\Gamma_{R}^{k}
		{N \choose k},
\label{same extended sum}
\end{equation}
and hence from Eqs.~(\ref{extended sum}) and (\ref{same extended
sum}),
\begin{equation}
	\widetilde{\sum}^{(e)} =
	\widetilde{\sum}^{(e)}(\Gamma_{R}) =
	{\sum}^{(e)}(\Gamma_{R} \rightarrow -i\Gamma_{R}).
\label{same e sum 2}
\end{equation}
It follows from Eq.~(\ref{same e sum 2}) that $W$ in Eq.~(\ref{WforN
6}) would be replaced by $\widetilde{W}$ where
\begin{equation}
	\widetilde{W} \simeq \frac{4}{RN}
		\left(
		\frac{G_{F}a_{n}N}
		{2\pi\sqrt{2}e^{2}R^{2}}
		\right)^{N}
		\cosh(1/\Gamma_{R}),
\label{tilde WforN 6}
\end{equation}
and hence the difference
between $W$ and $\widetilde{W}$ is the replacement
\begin{equation}
	(-1)^{N/2}\cos(1/\Gamma_{R}) \rightarrow
	\cosh(1/\Gamma_{R}).
\end{equation}
Since $\cosh(1/\Gamma_{R})$ never vanishes, $\widetilde{W}$ is always
nonzero for the values of the parameters we are assuming.  By
contrast $\cos(1/\Gamma_{R})$ {\em can} be zero for some set of
parameters.  Hence $W$ can in principle vanish, but only when
$2/(\pi\Gamma_{R})$ coincides with an odd integer to $\sim 10^{59}$
decimal places.  Not only would this be unphysical for even a
single neutron star, it could hardly be supposed that
$\cos(1/\Gamma_{R})$ would vanish in this way for each of the more
than 500 known pulsars \cite{T93}.  From the recent catalog of
558 pulsars by Taylor {\em et al.} \cite{T93}, we note that there is
a significant variation in the period $P$ and spin-down rate
$\dot{P}$ both of which depend on the internal structure of the
pulsars.  Absent an overarching (and presently unknown) symmetry
principle which would force $W$ to vanish, it is difficult to
imagine that this could happen coincidentally in every case.  This
argument is further strengthened by noting that the catalog of
Taylor {\em et al.} specifically excludes accretion-powered
systems which are detected by their X-ray emissions.  For these,
the in-falling matter would continuously change the parameters of
the pulsar at a level that would preclude the possibility that
$\cos(1/\Gamma_{R})$ would always be zero.

	In the preceding discussion we have considered the
effects of the neutrino-exchange energy in an idealized
non-rotating neutron star.  Including the effects of rotation
would further bolster the preceding arguments, since the
variation in $P$ implies a variation in the internal structure of
the pulsars, as noted above.  Proceeding further one could also
ask whether a neutron-star could even come into existence in
the presence of an energy density as large as would arise from
neutrino-exchange.  Although the answer to this question could
lead to a more stringent limit on neutrino masses than is
obtained from our ``static'' picture, such dynamical
considerations are beyond the scope of the present paper.
However, even in the absence of a detailed calculation one can
argue that normal stars could not evolve into neutron stars in
the presence of neutrino-exchange.  This follows by noting that
as soon as a star evolves to the stage where $(\Gamma_{R}N/e)$ in
Eq.~(\ref{WforN 5}) exceeds unity, its mass-energy not only
becomes unphysically large, but also alternates in sign as each
pair of neutrons is accreted.  Hence if at some stage
$(-1)^{N/2}$ were $-1$, the addition of 2 neutrons would
require overcoming an unphysically large repulsive energy
barrier, which thus acts to prevent further accretion.

	Having demonstrated that the oscillatory factor
$\cos(1/\Gamma_{R})$ cannot prevent the neutrino-exchange 
energy-density catastrophe, we study some of the implications which
follow from the presence of this factor in the expression for
$W$.  To start with, the sign of the 0-derivative contribution to
$W$ in Eqs.~(\ref{WforN 6}) and (\ref{W estimate}) depends on the
product $(-1)^{N/2}\cos(1/\Gamma_{R})$.  To determine the sign
and magnitude of $\cos(1/\Gamma_{R})$, $\Gamma_{R}$ itself would
have to be known to an unphysically large number of decimal
places.  (For illustrative purposes, if $\Gamma_{R}$ in
Eq.~(\ref{Gamma_R}) were {\em exact}, then
$\cos(1/\Gamma_{R}) = +0.613$ \cite{Maple}.)  Hence as a practical
matter, the overall sign of $W$ for a given neutron star is
difficult to determine, but is also relatively unimportant given
that either sign leads to a difficulty.  There is, however, an
interesting implication of the variation in sign of $W$ as
a function of $N$ and $\Gamma_{R}$:  Rather than limit our
attention to the parameters of the $\sim 500$--600 known neutron
stars that are pulsars, we can view any spherical volume of
radius $r$ inside a neutron star as a collection of $N(r)$
neutrons to which the preceding arguments apply.  In order to have
a stable static neutron star all such subvolumes must be in
equilibrium, notwithstanding the variation of the sign and
magnitude of $W(r)$ with $r$.  By pursuing this argument, we are
led in Section VIII to the bound on $m$ quoted in
Eq.~(\ref{final m limit}) below.

	We conclude the present discussion by presenting a
physical interpretation of the factor
$(-1)^{N/2}\cos(1/\Gamma_{R})$.  Combining Eqs.~(\ref{W max})
and (\ref{tilde WforN 6}) we can write 
\begin{equation}
	W \simeq W^{(max)}\cos(1/\Gamma_{R}),
\label{approx W max}
\end{equation}
\begin{equation}
	\widetilde{W} \simeq 
		(-1)^{N/2}
	W^{(max)}\cosh(1/\Gamma_{R}).
\label{approx tilde W}
\end{equation}
We see from Eqs.~(\ref{approx W max}) and (\ref{approx tilde W})
that the overall scale of both $W$ and $\widetilde{W}$ is primarily
determined by the single term $W^{(max)}$ in Eq.~(\ref{W max}).
The factor $\cos(1/\Gamma_{R})$ in Eq.~(\ref{approx W max}),
which acts to suppress $W$, can then be understood as the
``memory'' of the cancellations arising from the remaining terms
in ${\sum}^{(e)}$.  By way of contrast the factor
$\cosh(1/\Gamma_{R})$ in Eq.~(\ref{approx tilde W}) enhances the
contribution from $W^{(max)}$, and thus reflects the effect of
coherently adding all the same-sign contributions in
$\widetilde{\sum}^{(e)}$.  For practical purposes the magnitude of
the neutrino-exchange energy-density is thus determined by
$W^{(max)}$, for which the bound in Eq.~(\ref{better U0 k
constraint}) can be used.

	While on the subject of cancellations it is worthwhile to
consider another possible source of cancellations, namely the
effects of the Pauli exclusion principle for neutrinos.  The
Pauli effect for neutrinos can enter in two ways:  Firstly, the
exchanged $\nu\bar{\nu}$ pairs can interfere with one another and,
secondly, the exchanged neutrinos could interfere with a possible
neutrino sea produced by the neutron star or white dwarf.  With
respect to the first possibility, the exchanged neutrinos do in
fact interfere with one another, but these effects are rigorously
taken into account from the outset in the Schwinger-Hartle
formalism, and at all stages in the present derivation.  Since
the Schwinger formula in Eq.~(\ref{Schwinger W}) is simply a
convenient way of summing all the one-loop Feynman amplitudes
(which clearly incorporate the Pauli exclusion principle), all
such effects are automatically included.  For example, the factor
of $(-1)$ from the closed neutrino loop, which is a
consequence of the anticommutativity of the neutrino field
operators, is already built in.  Moreover, the expression for $W$
in Eq.~(\ref{WforN 6}) is the result of coherently summing the
$k$-body contributions, with all the relevant phases determined
by the Pauli principle.  Hence all coherent effects of the Pauli 
exclusion principle for the exchanged neutrinos are accounted for
in the present results.

	Consider next the possibility that neutrino-exchange is
somehow suppressed by the presence of a sea of physical neutrinos
in a neutron star.  It is easy to demonstrate that such a
suppression cannot take place, because the density of physical
neutrinos in a neutron star is essentially zero  \cite{ST 11}.
From Ref.~\cite{ST 11} we note that the mean free path
$\lambda_{n}$ for the scattering of $\nu_{e}$ from neutrons is
given by
\begin{equation}
	\lambda_{n} \simeq 300\,\mbox{km}\,
		\frac{\rho_{nuc}}{\rho_{m}}
		\left(\frac{100\,\mbox{keV}}{E_{\nu}}
		\right)^{2}.
\label{lambdan}
\end{equation}
Here $\rho_{nuc}$ is the nuclear density, which is taken to be
$\rho_{nuc} = 2.8 \times 10^{14}$ g cm$^{-3}$, and $\rho_{m}$ is
the neutron star density.  For $E_{\nu} = 10$ eV, the energy scale set by
$G_{F}\rho_{n}$, we find $\lambda_{n} \simeq 1 \times 10^{10}$
km, which means that a neutron star is transparent to low energy
neutrinos \cite{ST 11}.  Since low-energy neutrinos cannot be
trapped inside a neutron star, there is no possibility of a
neutrino sea in a neutron star.  The few neutrinos that are
present (in transit) at any instant are obviously incapable of
reducing the effective Fermi constant by a factor of order
$10^{11}$, which is what would be needed to resolve the
energy-density problem.  Finally we recall from Table~\ref{white
dwarf table} that neutrino-exchange leads to a large energy-density
in white dwarfs as well as in neutron stars.  Hence
this difficulty cannot be resolved by invoking a physical
neutrino sea, since the mean free path of a neutrino in a white
dwarf would be even larger than in a neutron star, so even fewer
neutrinos could be trapped to form a neutrino sea.


	Returning to Eqs.~(\ref{WforN 6}) and (\ref{W max}) we
note that these results
arise from the 0-derivative contribution $U_{0}^{(k)}$ in
Eq.~(\ref{WforN 1}).  We consider next the
contributions to $W$ from the derivative terms in 
Eqs.~(\ref{Vk})--(\ref{Pk}).  Since these terms introduce no new
dimensional factors, their contribution to $W$ must have the same
dependence on $(G_{F}a_{n}/R^{2})^{k}$ as in Eq.~(\ref{WforN 1}),
although the $k$-dependent coefficients will in general be
different.  We consider three possibilities:  a) The net
contribution from the derivative terms is {\em smaller} than that
from the $0$-derivative terms.  In this case Eq.~(\ref{WforN 1})
is an adequate approximation to $W$, and the previous conclusions
apply as before.  b) The net contribution from the derivative
terms is {\em larger} than that from the 0-derivative terms, a
possibility suggested by the 4-body results. 
In this case the
neutrino-exchange energy-density catastrophe poses an even
greater difficulty than before, and resolving the problems raised
by the 0-derivative contribution is a necessary first step.  It
is evident that whatever $k$-dependent coefficient replaces
$(k!)^{2}/(k^{k + 2})$ in Eq.~(\ref{WforN 1}), the $k$-body
contribution arising from the derivative terms will necessarily
be proportional to the same factor
\begin{equation}
	\left(\frac{(G_{F}/\hbar c)
		a_{n}}{2\pi\sqrt{2}R^{2}}\right)^{k}
	{N \choose k},
\end{equation}
that appears in the 0-derivative contribution.  Hence whatever
mechanism resolves the neutrino-exchange energy-density
catastrophe for the 0-derivative terms, will also work for the
derivative terms, as we discuss in greater detail in Sec.~VIII
below. c) The last possibility is that the derivative terms
cancel the 0-derivative contribution so as to reduce 
$|W/Mc^{2}|$ in Eq.~(\ref{W/mc2})
to a number of order unity (or more realistically to $\sim 0.1$).
Such a cancellation would have to occur in each of the 500--600
known pulsars, and (as we discuss below) in each of $\sim$2000
known white dwarfs.  Since the derivative terms have a different
dependence on $\vec{r}_{ij}$ than the 0-derivative terms, they
would also depend differently on the matter distribution in the
neutron star, were we to allow for a varying density in a more
realistic calculation.  Thus, to avoid the neutrino-exchange
energy-density problem would require at a minimum an almost exact
cancellation between the derivative and 0-derivative terms in
each pulsar, notwithstanding the fact that the former depend
differently from the latter on the matter distribution, which 
itself varies from one pulsar to another.  As noted
previously, such cancellations would be unphysical absent some
(presently unknown) symmetry principle.  We are thus led to the
conclusion that the derivative terms can exacerbate the
energy-density problem arising from the 0-derivative terms, but
they cannot resolve it in the framework of the present
(Standard Model) calculation.

	As noted in the preceding paragraph, neutrino-exchange
leads to an energy-density catastrophe in white dwarfs as well as
in neutron stars.  Although neutron stars provide the most
stringent lower bound on $m$, the weaker bound from white
dwarfs is nevertheless important because its existence makes it
less likely that this problem
can be resolved without introducing massive neutrinos.  For
example, the fact that white dwarfs contain roughly comparable
numbers of electrons, protons, and neutrons, rules out any
attempt to address the problem
in neutron stars by a mechanism which somehow
suppresses the neutrino-neutron coupling.  A 1987 catalog by
McCook and Scion \cite{McCook} lists 1279 white dwarfs which have
been identified spectroscopically, and the current catalog
contains more than 2000 entries.  To demonstrate that the
energy-density arising from neutrino-exchange in white dwarfs is
also unphysically large, we consider two representative white
dwarfs whose masses and radii are reasonably well-known
\cite{ST83b}.  The relevant parameters for these two white
dwarfs, Sirius B and 40~Eri~B, are summarized in Table~\ref{white
dwarf table}.
Following Ref.~\cite{ST83b} we assume that the interiors of these
white dwarfs are predominantly carbon and oxygen, in which case
these stars contain approximately equal numbers of neutrons,
protons, and electrons as noted previously.  
The combinatorics of the many-body
diagrams involving three types of particles are somewhat
complicated, but for present purposes it is sufficient to evaluate
the purely electronic contribution, noting from Appendix~A 
that electrons have
the strongest coupling to neutrinos.  The previous formalism for
neutron stars can then be taken over immediately and leads to
the results given in the last two lines of Table~\ref{white
dwarf table} where,
\begin{equation}
	\Lambda_{R} =
	\frac{(G_{F}/\hbar c)
		N_{e}a_{e}}{2\pi\sqrt{2}e^{2}R^{2}}.
\label{Lambda R}
\end{equation}
Here $N_{e}$ is the total number of electrons, $a_{e} =
(2\sin^{2}\theta_{W} + 1/2) \simeq 0.964$ is the $e$-$\nu$ coupling
constant [see Eq.~(\ref{ae})], and we assumed for illustrative
purposes that 
the electron density is spatially constant.  It
follows from Table~\ref{white dwarf table}
that for both Sirius B and 40 Eri B the energy
arising from neutrino-exchange exceeds the mass-energy of the
Universe, just as in the case of neutron stars.
	
	The results in Table~\ref{white dwarf table}
raise the possibility that the existence of a large neutrino-exchange
energy-density may be not be limited to neutron stars
and white dwarfs.  From Eq.~(\ref{WforN 6}) we see that for a
single particle species $j = n, e,$ or $p$, 
this happens when 
\begin{equation}
	\Lambda_{R} = 
	\frac{(G_{F}/\hbar c)
		N_{j}a_{j}}{2\pi\sqrt{2}e^{2}R^{2}}
		> 1,
\label{critical Lambda R}
\end{equation}
where $a_{j}$ is the coupling to neutrinos.  For the Sun the
contribution to $\Lambda_{R}$ from electrons alone slightly
exceeds unity, which suggests that the very
existence of the Sun, and by extension life on Earth, is evidence
for the necessity of massive neutrinos.  Whether or not the Sun
itself proves to be unstable against neutrino-exchange
can only be determined by more detailed calculations. However,
the possibility that other stars exist for which this is true
indicates that the problems arising from neutrino-exchange
can only be resolved by a universal
mechanism that would apply to large numbers of rather different
celestial objects.

\section{Alternatives to Massive Neutrinos}

	We consider in this section several alternatives to the
introduction of massive neutrinos as a means for resolving the
neutrino-exchange energy-density problem.  It will be argued
that no presently known mechanism except for massive neutrinos is
compatible with existing experimental and theoretical
constraints.

\subsection{Deviations of the Neutrino Couplings from the
Standard Model Predictions}

	As noted in the Introduction, the magnitude of $W$ in
Eq.~(\ref{W/mc2}) is determined by dimensional arguments and
combinatoric considerations, and hence is insensitive to the
detailed form of the neutrino coupling.  To date there are no
known discrepancies between the predictions of the Standard Model
and experiment (see Appendix A), and the agreement is at the
level of a few percent or better \cite{PDG94}.  It would thus be
difficult to understand how the effective Fermi constant
$G_{F}|a_{n}|$ could differ from the Standard Model value by a
factor of order $10^{11}$, which is what would be required to
avert a large energy-density.  Similar considerations
apply to the possibility that the neutral current interaction
contains small admixtures of $S$, $P$, and $T$ in addition to the
usual $V$, $A$ currents.

	In this connection we recall that the mass density
$\rho_{m}$ in Eq.~(\ref{neutron star number density}), 
$\rho_{m} = 6.7 \times 10^{14}$
g~cm$^{-3}$, is comparable to nuclear density, $\rho_{nuc} = 2.8
\times 10^{14}$ g~cm$^{-3}$.  It follows that the success of
conventional theory in explaining weak interactions in nuclei
\cite{BS73} strongly suggests that the weak neutrino-hadron and
neutrino-electron couplings in neutron stars do not differ
significantly from the predicted Standard Model values.  This is
supported by the general agreement between theory and observation
for such processes as supernova formation \cite{supernova}.  We
further observe that even smaller renormalization effects are
expected in white dwarfs, where a large neutrino-exchange
energy-density nonetheless exists.  Finally, we note
that various astrophysical arguments support both the spatial and
temporal constancy of $G_{F}$ \cite{SS93}.  Taken together these
arguments make it unlikely that the present difficulties can be
resolved by modifying the neutrino coupling constants.

\subsection{Cancellations Among $\nu_{e}$, $\nu_{\mu}$, and
$\nu_{\tau}$}

	The expression for $W$ in Eq.~(\ref{WforN 6}) represents the
contribution from a single neutrino species,
$\nu_{e}\bar{\nu}_{e}$ for example.  It might be argued perhaps
the contributions from $\nu_{e}\bar{\nu}_{e}$,
$\nu_{\mu}\bar{\nu}_{\mu}$, and $\nu_{\tau}\bar{\nu}_{\tau}$
could cancel amongst themselves in such a way as to reduce $W$ to
a physically reasonable value.  However, such a scenario is in
conflict with $e$-$\mu$-$\tau$ universality, which implies that
the couplings of $\nu_{e}$, $\nu_{\mu}$, and $\nu_{\tau}$ to $n$,
$p$, or $e$ have the same signs as well as the same magnitudes,
and hence cannot produce the necessary cancellation.
Experimental support for $e$-$\mu$-$\tau$ universality comes from
a number of sources including the equality of the partial decay
rates $\Gamma(Z^{0}\rightarrow \ell^{+}\ell^{-})$ where $\ell =
e,\mu,\tau$ \cite{PDG94}.

\subsection{Breakdown of Perturbation Theory}

	It might be suggested that the fact that $W$ is unphysically
large indicates that our use of the Schwinger formula in
Eq.~(\ref{Schwinger W}) is not valid for some reason.  This is a
possibility that cannot be completely excluded at present,
particularly in light of recent interest in the breakdown of
perturbation theory in high orders \cite{PT}.  We can argue,
however, that this is not likely to lead to a resolution of the
energy-density difficulties for reasons we now discuss.  To start
with, the expression for $W$ in Eq.~(\ref{WforN 6}) is the sum of a
finite number of contributions, each of which is well-behaved.
Moreover, for $k \geq 6$ the one-loop $k$-body diagrams are
finite, even in the absence of any regularization scheme.  It
follows that the magnitude of $W$ cannot be explained as an
artifact of the manner in which the potentials are extracted from
one-loop amplitudes.  Finally, the actual $k$-body
neutrino-exchange potentials which arise from the
Schwinger-Hartle formalism are themselves well-behaved; it is
only when particular values of the parameters $G_{F}$, $N$, $R$,
$\rho$, and $m$ are used (e.g., $m = 0$) that unphysical
results emerge.  This suggests that the resolution of the present
difficulties lies in establishing what the correct value of $m$
is rather than in finding a possible breakdown of perturbation
theory.  The previously discussed calculation of the Coulomb
energy of a nucleus provides a useful analogy.  Suppose that it
was believed, for whatever reason, that for a nucleus with $A$
nucleons the nuclear radius $R$ in Eq.~(\ref{Z W_C}) was given by
$R = A^{1/3}r_{0}$, with $r_{0} = 0.01$ fm.  The resulting
Coulomb repulsion would be large enough to destabilize all
nuclei, so matter as we know it could not exist.  Although one
could attempt to resolve this ``Coulomb catastrophe'' by
examining higher-order effects in perturbation theory, a more
reasonable approach would be to first ask whether another value
of the parameter $r_{0}$, say $r_{0} \sim 1$ fm, was compatible
with experiment.  Analogously, we demonstrate in Sec.~VIII below
that if $m \gtrsim 0.4$ eV/$c^{2}$, then neutrino-exchange no
longer leads to an energy-density catastrophe.  This path has
clear observational implications, and should these be shown to be
incompatible with experiment, then a re-examination of
perturbation theory would be appropriate along the lines we now
consider.

	As we have already noted, the final expression for $W$ in
Eq.~(\ref{WforN 6}) is the result of summing the contributions from
a finite number of many-body potentials, each of which is
well-behaved.  It follows that if there is indeed a breakdown
of perturbation theory, it cannot be due simply to a failure of
the perturbation series to converge.  One possibility which has
been discussed recently \cite{PT} is the rapid growth of the
cross sections for producing large numbers of scalar particles.
It has been conjectured \cite{PT} that the combinatoric factor $n!$
which appears in the amplitude for producing $n$ particles in the
final state could overcome the smallness of the coupling constant
in weakly coupled theories, and thus produce observably large
effects at high energies.  This has led to an examination of the
need for unitarizing the tree-level amplitudes from which these
effects arise,
and more generally, to a discussion of whether
such effects can in fact be seen.

	Although the origin of the combinatoric factor $n!$ which
is responsible for the growth of the tree-level contributions is
similar to that for
the factor $(k - 1)!/2$ in Eq.~(\ref{better U0 k constraint}), the
neutrino-exchange problems depend critically on the presence of
the factor $N!$ in ${N \choose k}$, which has no analog in the
work described in Ref.~\cite{PT}.  Moreover, we have seen in
Eq.~(\ref{W8}) that the large neutrino-exchange energy-density
arises in $\O(G_{F}^{8})$, which is a relatively low
order of perturbation theory compared to the effects considered
in Ref.~\cite{PT}.  Finally, the self-energy of a neutron star
arising from neutrino-exchange is a low-energy phenomenon, in
contrast to the production processes in Ref.~\cite{PT}.  Hence
the unitarization mechanism considered in that context would not
be relevant here.  For all these reasons it appears unlikely that a
breakdown of perturbation theory along the lines contemplated in
Ref.~\cite{PT} could resolve the neutrino-exchange energy-density
catastrophe in a neutron star or white dwarf.  Nonetheless this
remains an interesting avenue for future exploration.  Since at
present the only {\em known} viable mechanism is the assumption
that neutrinos are massive, we explore the implications of $m
\neq 0$ in the following sections.

\section{Many-Body Interactions From the Exchange of Massive
Neutrinos}

	In this section we apply the formalism for massive
neutrinos developed in Appendix E to the problem of calculating
the many-body neutrino-exchange potentials.  Our objective is to
demonstrate that when $m \neq 0$, the neutrino-exchange energy
can be approximated by replacing
\begin{equation}
	1/R^{2} \rightarrow e^{-mR}/R^{2},
\end{equation}
in Eq.~(\ref{WforN 6}).  As noted in Appendix E this leads to
``saturation'' of the neutrino-exchange interaction, and
ultimately to a bound on $m$ as we describe below.

	When $m \neq 0$ the couplings of neutrinos to other
matter may be different from the predictions of the Standard
Model.  Although the details of whatever Hamiltonian would emerge
cannot be fully discerned at the present time, it is sufficient
for present purposes to consider a simple phenomenological model
in which $\nu_{e}$, $\nu_{\mu}$, and $\nu_{\tau}$ remain as the
eigenstates of the Hamiltonian, with each having a nonzero mass.
The modification of the neutrino couplings to matter can then be
parameterized as in Eq.~(\ref{b parameter}) below.

\subsection{The 2-Body Potential when $m \neq 0$}

	The expression for $W^{(2)}$ when $m \neq 0$ can be
obtained from the $m = 0$ expression in
Eq.~(\ref{2bodyW}) by replacing $S_{F}^{(0)}(\vec{r}_{12},E)$ with the
expression given in Eq.~(\ref{SF xE 2}).  In addition we allow for
the possibility that the neutrino coupling may not be pure
$V$-$A$ when $m \neq 0$ by substituting 
\begin{equation}
	(1 + \gamma_{5}) \rightarrow (1 + b\gamma_{5}),
\label{b parameter}
\end{equation}
where $b$ is a constant (which may be different for $\nu_{e}$,
$\nu_{\mu}$, and $\nu_{\tau}$).  If there is also a
modification of the overall strength of neutrino coupling, this
can be accommodated by appropriately redefining $G_{F}$.  After
calculating the trace of the Dirac $\gamma$-matrices, the analog
of Eq.~(\ref{W2 b}) can be written in the form
\begin{eqnarray}
	W^{(2)} & = &
		\frac{2i}{\pi}
		\left(\frac{G_{F}a_{n}}{\sqrt{2}}\right)^{2}
		\int\rho_{1}\,d^{3}x_{1}
		\int\rho_{2}\,d^{3}x_{2}
		\int^{\infty}_{-\infty}dE
		\left\{(1 + b^{2})E^{2}
		- [(1 + b^{2})
		\vec{\partial}_{12}\cdot
		\vec{\partial}_{21}
		- (1 - b^{2})m^{2}]\right\}
		\nonumber \\
	&   & \mbox{} \times
	\left(\frac{i}{4\pi}\right)^{2}
		\frac{1}{r_{12}r_{21}}
		\exp[i(r_{12} + r_{21})
		\sqrt{E^{2} - m^{2}}].
\label{massive W2 a}
\end{eqnarray}
Following the discussion in Sec.~III, and using Eq.~(\ref{general
Fn}),
the 2-body potential $V^{(2)}$ is given by
\begin{equation}
	V^{(2)} = -i\frac{G_{F}^{2}a_{n}^{2}}{16\pi^{3}}
	\left\{(1 + b^{2})F_{2}(z) -
	[(1 + b^{2})
		\vec{\partial}_{12}\cdot
		\vec{\partial}_{21}
	- (1 - b^{2})m^{2}]F_{0}(z)
	\right\}\frac{1}{r_{12}r_{21}},
\label{massive V2 a}
\end{equation}
where $z \equiv r_{12} + r_{21}$.  The expression in curly
brackets in Eq.~(\ref{massive V2 a}) can be simplified by using
Eq.~(\ref{12 gradients}) and the recurrence relation in Eq.~(\ref{F2})
which leads to 
\begin{equation}
	V^{(2)} = i\frac{G_{F}^{2}a_{n}^{2}}{16\pi^{3}}
	\left\{(1 + b^{2})
	\frac{F_{0}^{(2)}(z)}{r_{12}r_{21}}  -
	(1 + b^{2})
		\frac{\partial}{\partial r_{12}}	
		\frac{\partial}{\partial r_{21}}	
		\left[
		\frac{F_{0}(z)}{r_{12}r_{21}}
		\right]
		-  2m^{2}
		\frac{F_{0}(z)}{r_{12}r_{21}}
	\right\}.
\label{massive V2 b}
\end{equation}
After the derivatives in Eq.~(\ref{massive V2 b}) are explicitly
evaluated, and use is made of Eq.~(\ref{F0n}), we find
\begin{equation}
	V^{(2)} = -\frac{G_{F}^{2}a_{n}^{2}}{4\pi^{3}}
	\left\{(1 + b^{2})
	\left[\frac{m^{2}}{r^{3}}K_{1}^{(1)}(2mr) -
	\frac{m}{2r^{4}}K_{1}(2mr)\right]
	- \frac{m^{3}}{r^{2}}K_{1}(2mr)\right\},
\label{massive V2 c}
\end{equation}
where $r = r_{12} = r_{21}$.  Note that the contributions
proportional to $K_{1}^{(2)}$, which arise from the derivatives
and from $F_{0}^{(2)}$, cancel against each other.  Using
Eq.~(\ref{K1'b}) $K_{1}^{(1)}(2mr)$ can be eliminated in favor of
$K_{0}(2mr)$ and $K_{1}(2mr)$, and this leads to the final
expression for $V^{(2)}$ when $m \neq 0$:
\begin{equation}
	V^{(2)} = \frac{G_{F}^{2}a_{n}^{2}}{4\pi^{3}}
	\left\{(1 + b^{2})
	\left[\frac{m}{r^{4}}K_{1}(2mr) +
	\frac{m^{2}}{r^{3}}K_{0}(2mr)\right]
	+ \frac{m^{3}}{r^{2}}K_{1}(2mr)\right\},
\label{massive V2}
\end{equation}
The result in Eq.~(\ref{massive V2}), which is exact, can be
compared to the corresponding result in Eq.~(\ref{V2}) for the
massless case by expanding $K_{1}(x)$ and $K_{0}(x)$ for $x
\simeq 0$:
\begin{eqnarray}
	K_{0}(x) & \simeq & -\ln x, 
\label{K0 limit}
	\\
	K_{1}(x) & \simeq & 1/x.
\label{K1 limit}
\end{eqnarray}
The only contribution which survives as $m \rightarrow 0$ is from
the term proportional to $1/r^{4}$, and this gives
\begin{equation}
	V^{(2)}(r) \simeq \frac{G_{F}^{2}a_{n}^{2}}{8\pi^{3}}\,
		\frac{(1 + b^{2})}{r^{5}}.
\label{massless V2}
\end{equation}
When $b = 1$, which is the value appropriate to the massless
case, Eq.~(\ref{massless V2}) reproduces Eq.~(\ref{V2}) as
expected.  As noted in Appendix E, however, the regime of
interest here is when $2mr \gg 1$, in which case the asymptotic
expression in Eq.~(\ref{asymptotic Kn}) for $K_{\nu}(2mr)$ leads to the
anticipated exponential fall-off of $V^{(2)}(r)$.

\subsection{Absence of Odd-$k$ Contributions when $m \neq 0$}

	In this subsection we demonstrate that for $m \neq 0$ it
remains the case that there are no contributions to $W$ for odd
$k$.  Using Eq.~(\ref{gamma C}) we have
\begin{equation}
	C^{-1}(\gamma \cdot \eta - m)C =
		(-\gamma \cdot \eta -m)^{T},
\end{equation}
from which it follows that
\begin{equation}
	C^{-1}S_{Fm}^{(0)}(\vec{r}_{ij},E)C =
		S_{Fm}^{(0)T}(-\vec{r}_{ij},-E)
		=
		S_{Fm}^{(0)T}(\vec{r}_{ji},-E).
\end{equation}
This is the same relation that holds in the massless case [see
Eq.~(\ref{S C})], and because the sign of $E$ is immaterial for
the reasons discussed previously, we can write as before,
\begin{equation}
	C^{-1}S_{Fm}^{(0)}(\eta_{ij})C =
		S_{Fm}^{(0)T}(\eta_{ji}).
\label{CetaC}
\end{equation}
Under $C$ the modified neutrino coupling transforms as 
\begin{equation}
	C^{-1}(1 + b\gamma_{5})C = (1 + b\gamma_{5})^{T},
\end{equation}
and since both the coupling and the neutrino propagator behave
under $C$ just as they do in the massless case, the previous
conclusion that there are no odd-$k$ contributions follows when
$m \neq 0$ as well.

	It is instructive to illustrate the preceding conclusion
by considering the 5-body contribution
as an example.  The contribution from the standard diagram 
is given in the static limit by 
\begin{eqnarray}
	W^{(5)}_{a} & = & 
		\frac{-i}{2\pi}
		\left(\frac{G_{F}a_{n}}{\sqrt{2}}\right)^{5}
		\int_{-\infty}^{\infty}dE \,
		\mbox{Tr}
		\left\{
		\gamma_{\mu}(1 + b\gamma_{5})S_{Fm}^{(0)}(51)
		\gamma_{\sigma}(1 + b\gamma_{5})S_{Fm}^{(0)}(45)
		\gamma_{\rho}(1 + b\gamma_{5})S_{Fm}^{(0)}(34)
	\right. \nonumber \\
		&    & \left. \mbox{} \times
		\gamma_{\lambda}(1 + b\gamma_{5})S_{Fm}^{(0)}(23)
		\gamma_{\nu}(1 + b\gamma_{5})S_{Fm}^{(0)}(12)
		\right\}
		T_{\mu\sigma\rho\lambda\nu}
		(x_{1},x_{2},x_{3},x_{4},x_{5}).
\label{W5a}
\end{eqnarray}
In the static limit the expression in curly brackets is given by	
\begin{eqnarray}
\mbox{tr}\left[\mbox{Eq.~(\ref{W5a})}\right]
	& = &
	\mbox{tr}\left[
	(1 - b\gamma_{5})\bar{S}_{Fm}^{(0)}(51)
	(1 + b\gamma_{5})S_{Fm}^{(0)}(45)
	(1 - b\gamma_{5})\bar{S}_{Fm}^{(0)}(34)
	\right.
	\nonumber \\
	 &  & \left. \mbox{} \times
	(1 + b\gamma_{5})S_{Fm}^{(0)}(23)
	(1 - b\gamma_{5})\gamma_{4}S_{Fm}^{(0)}(12)
	\right],
\label{W5a trace a}
\end{eqnarray}
where $\bar{S}^{(0)}_{Fm}(51) \equiv
\gamma_{4}S_{Fm}^{(0)}(51)\gamma_{4}$.  Combining 
Eqs~(\ref{CetaC})--(\ref{W5a trace a}), the trace in Eq.~(\ref{W5a
trace a}) can be re-expressed in the form
\begin{eqnarray}
\mbox{tr}\left[\mbox{Eq.~(\ref{W5a})}\right]
	 & = &
	-\mbox{tr}\left[
	(1 - b\gamma_{5})S_{Fm}^{(0)}(-12)\gamma_{4}
	(1 - b\gamma_{5})S_{Fm}^{(0)}(-23)
	(1 + b\gamma_{5})\bar{S}_{Fm}^{(0)}(-34)
	\right.
	\nonumber \\
	 &  & \left. \mbox{} \times
	(1 - b\gamma_{5})S_{Fm}^{(0)}(-45)
	(1 + b\gamma_{5})\bar{S}_{Fm}^{(0)}(-51)
	\right].
\label{W5a trace b}
\end{eqnarray}
The contribution from the diagram with the reversed loop
momentum
can be written down in a similar manner, and the analog of
the right-hand side
Eq.~(\ref{W5a trace b}) is
\begin{equation}
	\mbox{tr}\left[
	(1 + b\gamma_{5})S_{Fm}^{(0)}(21)\gamma_{4}
	(1 + b\gamma_{5})S_{Fm}^{(0)}(32)
	(1 - b\gamma_{5})\bar{S}_{Fm}^{(0)}(43)
	(1 + b\gamma_{5})S_{Fm}^{(0)}(54)
	(1 - b\gamma_{5})\bar{S}_{Fm}^{(0)}(15)
	\right].
\label{W5a' trace a}
\end{equation}
Since $S_{Fm}^{(0)}(-ij) = S_{Fm}^{(0)}(ji)$ it follows that when
the contributions in Eqs.~(\ref{W5a trace b}) and (\ref{W5a'
trace a})
are added only those terms containing odd powers
of $b$ survive.  All such terms can be eventually reduced to the
trace of a product of $\gamma$-matrices containing a single
$\gamma_{5}$, and from Eq.~(\ref{trace identities}) it is seen that any such
trace is always proportional to $\epsilon_{\mu\nu\lambda\rho}$.
By the arguments given previously in the massless case, all terms
proportional to $\epsilon_{\mu\nu\lambda\rho}$ average to
zero when integrated over a spherically symmetric matter
distribution.  This leads to the conclusion that there is no
5-body contribution to $W$, and by extension no contribution for
any odd $k$, even when $m \neq 0$.

\subsection{The 4-Body Potential for $m \neq 0$}
	
	In this subsection we obtain the explicit form of the
4-body potential, which is then used to infer the dependence of
the general $k$-body potential on the neutrino mass $m$.  This
result forms the basis for the bound on $m$ that we derive in the
following section.

	As in the 2-body case we begin with Eq.~(\ref{Schwinger
W}) and
substitute $(1 + b\gamma_{5})$ for $(1 + \gamma_{5})$, and the
massive neutrino propagator in Eq.~(\ref{SF xE 2}) for the massless
one.  The formalism of Sec.~III can then be taken over directly
and, after the Dirac traces are evaluated, the expression for the
0-derivative contribution from Fig.~\ref{First 4-body figure}(a) to the 4-body
potential is given by
\begin{eqnarray}
	V_{0}^{(4)}(\vec{r}_{12},\vec{r}_{23},\vec{r}_{34},\vec{r}_{41})
		& = &
	-\frac{i}{2\pi}
	\left(\frac{G_{F}a_{n}}{\sqrt{2}}\right)^{4}
	\,8\,
	\int^{\infty}_{-\infty}dE\,
	\left[E^{4}c_{4}(b) + m^{2}E^{2}c_{2}(b)
		+ m^{4}c_{0}(b)\right]
	\nonumber \\
	&   & \mbox{} \times
	\Delta_{Fm}(\vec{r}_{12},E) 
	\Delta_{Fm}(\vec{r}_{23},E) 
	\Delta_{Fm}(\vec{r}_{34},E) 
	\Delta_{Fm}(\vec{r}_{41},E),
\label{massive V04}
\end{eqnarray} 
where
\begin{equation}
	\Delta_{Fm}(\vec{r}_{12},E) 
		= 
	\frac{i}{4\pi r_{12}}
	\exp\left(ir_{12}\sqrt{E^{2} - m^{2}}\right),
\label{DeltaFm}
\end{equation}
\begin{equation}
	c_{4}(b) = (1 + b^{2})^{2} + 4b^{2} 
	\,\,\,\,;\,\,\,\,
	c_{2}(b) = 2(1 - b^{2})(3 + b^{2})
	\,\,\,\,;\,\,\,\,
	c_{0}(b) = (1 - b^{2})^{2}.
\label{cb}
\end{equation}
The expression for $V_{0}^{(4)}$ given in Eqs.~(\ref{massive
V04})---(\ref{cb}) is the sum of the contributions from the
standard diagram with both senses of the neutrino loop momentum.
This introduces a factor of 2 in Eq.~(\ref{massive V04}), and
combined with a factor of 4 from the Dirac trace, accounts for
the factor of 8.  As will be clear from the ensuing discussion,
it is sufficient for present purposes to evaluate the 
0-derivative contribution to $V^{(4)}$, from which the dependence
of $V^{(k)}$ on $m$ can be inferred.  From Eqs.~(\ref{massive
V04}) and (\ref{DeltaFm}),
\begin{eqnarray}
	\Delta_{Fm}(\vec{r}_{12},E) 
	\Delta_{Fm}(\vec{r}_{23},E) 
	\Delta_{Fm}(\vec{r}_{34},E) 
	\Delta_{Fm}(\vec{r}_{41},E)
	 & = &
	\left(\frac{i}{4\pi}\right)^{4}
	\frac{1}{r_{12}r_{23}r_{34}r_{41}}
	\nonumber \\
	&  &  \mbox{} \times
	\exp\left[i\sqrt{E^{2} - m^{2}}
	(r_{12} + r_{23} + r_{34} + r_{41})\right],
	\nonumber \\
\label{DeltaFms}
\end{eqnarray}
which allows $V_{0}^{(4)}$ to be expressed in the form
\begin{equation}
	V_{0}^{(4)} = -i
		\frac{4}{\pi}
	\left(\frac{G_{F}a_{n}}{\sqrt{2}}\right)^{4}
	\left(\frac{i}{4\pi}\right)^{4}
	\frac{1}{r_{12}r_{23}r_{34}r_{41}}
	\left[c_{4}(b)F_{4}(z) + m^{2}c_{2}(b)F_{2}(z)
		+ m^{4}c_{0}(b)F_{0}(z)\right].
\label{massive V04 b}
\end{equation}
Here $z = (r_{12} + r_{23} + r_{34} + r_{41})$, and the functions
$F_{n}(z)$ are defined in Eq.~(\ref{general Fn}).  Combining
Eq.~(\ref{massive V04 b}) with Eqs.~(\ref{F2}) and (\ref{F4 2})
then leads to
\begin{eqnarray}
	V_{0}^{(4)} & =  & -i
		\frac{4}{\pi}
	\left(\frac{G_{F}a_{n}}{\sqrt{2}}\right)^{4}
	\left(\frac{i}{4\pi}\right)^{4}
	\frac{1}{r_{12}r_{23}r_{34}r_{41}}
	\nonumber \\
	&  &  \mbox{} \times
	\left\{c_{4}(b)F_{0}^{(4)}(z) - 
	[2c_{4}(b) + c_{2}(b)] m^{2}F_{0}^{(2)}(z)
		+ [c_{4}(b) + c_{2}(b) + c_{0}(b)]
			m^{4}F_{0}(z)\right\}.
\label{massive V04 c}
\end{eqnarray}
It is worth noting that there are mass-dependent
contributions even in the limit of a pure $V$-$A$ coupling ($b =
1$).  This follows from the observation that although
$c_{2}(1) = c_{0}(1) = 0$, each of the mass terms receives a
contribution from $c_{4}(1) = 8$.  Combining Eqs.~(\ref{massive
V04 c}) and (\ref{F2}) the complete expression for $V_{0}^{(4)}$
can be written as
\begin{eqnarray}
	V_{0}^{(4)} & =  & -i
		\frac{4}{\pi}
	\left(\frac{G_{F}a_{n}}{\sqrt{2}}\right)^{4}
	\left(\frac{i}{4\pi}\right)^{4}
	\frac{1}{r_{12}r_{23}r_{34}r_{41}}
	\nonumber \\
	&  &  \mbox{} \times
	\left\{
	c_{4}(b)\,2i\,
	\left[
                \left(\frac{2m^{4}}{z} + \frac{12m^{2}}{z^{3}}
                \right)K_{0}(mz)
                +
                \left(m^{5} + \frac{7m^{3}}{z^{2}} +
                        m\frac{4!}{z^{4}}\right)
                K_{1}(mz)\right]
	\right.
	\nonumber \\
	&  &  \mbox{} 
	- [2c_{4}(b) + c_{2}(b)]\,
		 2i\left[\frac{m^{4}}{z}K_{0}(mz) +
                \left(m^{5} + m^{3}\frac{2!}{z^{2}}\right)
                K_{1}(mz)\right]
	\nonumber \\
	&  &  \mbox{} 
	\left.+  
	 [c_{4}(b) + c_{2}(b) + c_{0}(b)]\,
	2im^{5}K_{1}(mz)\right\}.
\label{massive V4}
\end{eqnarray}
As in the
2-body case we wish to check $V_{0}^{(4)}$ in the limit $m
\rightarrow 0$.  Using Eqs.~(\ref{K0 limit}) and (\ref{K1 limit}) we note
that since $x^{n}\ln x \rightarrow 0$ as $x \rightarrow 0$ all
the terms containing $K_{0}(mz)$ vanish in the $m = 0$
limit.  Among the terms containing $K_{1}(mz)$ only the
term proportional to $1/z^{4}$ survives, and when $b = 1$ this
gives
\begin{equation}
	V_{0}^{(4)}\,\, \stackrel{m = 0}{\longrightarrow} \,\,
		\frac{4}{\pi}
		\left(\frac{G_{F}a_{n}}{2\pi\sqrt{2}}\right)^{4}
	\frac{4!}{r_{12}r_{23}r_{34}r_{41}
	(r_{12} + r_{23} + r_{34} + r_{41})^{5}},
\end{equation}
which agrees with the 0-derivative contribution in
the massless case as given in Eq.~(\ref{V4}).

	When $m \neq 0$ the spatial dependence of $V_{0}^{(4)}$
receives contributions from terms proportional to $m^{5}$,
$m^{4}/z$, $m^{3}/z^{2}$, and $m/z^{4}$, each multiplying either
$K_{0}(mz)$ or $K_{1}(mz)$.  We note from Eq.~(\ref{asymptotic
Kn}) that
when $mz \gg 1$, $K_{\nu}(mz)$ can be approximated by
\begin{equation}
	K_{\nu}(mz) \simeq \sqrt{\frac{\pi}{2mz}}e^{-mz},
\end{equation}
so that in the asymptotic regime $V_{0}^{(4)}$ contains terms of
the form
\begin{equation}
	\sqrt{\frac{\pi mz}{2}}e^{-mz}
	\left(
	\frac{m^{4}}{z},
	\frac{m^{3}}{z^{2}},
	\frac{m^{2}}{z^{3}},
	\frac{m}{z^{4}},
	\frac{1}{z^{5}}
	\right).
\end{equation}
For values of $mz$ where the exponential makes a significant
contribution to $W$ one can approximate $\sqrt{\pi mz/2}$ by
unity.  It follows that for the term proportional to $1/z^{5}$,
which is the origin of the $m = 0$ result in Eq.~(\ref{massive V2
c}),
the most important consequence of a nonzero neutrino mass is that
\begin{equation}
	\frac{1}{z^{5}} \rightarrow \frac{e^{-mz}}{z^{5}},
\end{equation}
as noted in Appendix E.  Since $z = (r_{12} + r_{23} + r_{34} +
r_{41})$ the effect of the mean value approximation is to replace
$z$ by $4R$ so that
\begin{equation}
	\frac{1}{r_{12}r_{23}r_{34}r_{41}}
	\left(
	\frac{e^{mz}}{z^{5}}
	\right)
	\rightarrow
	\frac{1}{4^{5}}
	\frac{1}{R}
	\left(\frac{e^{-mR}}{R^{2}}\right)^{4}.
\label{mv arrow}
\end{equation}
We conclude from Eq.~(\ref{mv arrow}) that for the contribution
to $W^{(4)}$ arising from $1/z^{5}$, the primary effect of a
nonzero neutrino mass is that $1/R^{2}$ is replaced by
$\exp(-mR)/R^{2}$ as expected.

	It is straightforward to show that this result can be
generalized to the $k$-body case.  In order $G^{k}_{F}$ the
contribution which reproduces the $m = 0$ result arises from the
terms proportional to $E^{k}$, which leads to the function
$F_{k}(z)$ in Eq.~(\ref{general Fn}).  The generalizations of the
recurrence relations in Eqs.~(\ref{F2}) and (\ref{F4 2})
eventually express $F_{k}(z)$ in terms of the $k$-th derivative
$K_{1}^{(k)}(mz)$ of $K_{1}(mz)$ by using Eq.~(\ref{F0n}).  These
derivatives can be evaluated from Eq.~(\ref{K1'b}),
\begin{equation}
	K_{1}^{(1)}(z) = -K_{0}(z) - \frac{1}{z}K_{1}(z)
\end{equation}
by repeatedly using Eq.~(\ref{K1'a}).  Among the terms that
contribute to $K_{1}^{(k)}$ will be one which arises from
successive differentiations of $1/z$, and this produces a term
proportional to $(-1)^{k}k!/z^{k}$.  When all the appropriate
factors are included we find
\begin{eqnarray}
	E^{k} \rightarrow 
		F_{k}(z) 
		\rightarrow 
		(-i)^{k}
		F_{0}^{(k)}(z)
		& = &
		(-i)^{k}2im^{k + 1}K_{1}^{(k)}(mz)
	\nonumber \\
		& = &
		\frac{(i)^{k}k!}{z^{k + 1}}
		2imzK_{1}(mz) + \cdots
	\nonumber \\
		& \simeq &
		2i^{k + 1}\frac{k!}{z^{k + 1}}
		\left(
		\sqrt{\frac{\pi mz}{2}}e^{-mz}\right)
		+ \cdots,
\label{Ek arrow}
\end{eqnarray}
where the dots indicate the remaining contributions to
$K_{1}^{(k)}(mz)$.  The coefficient of the expression in
parentheses is the massless result in Eq.~(\ref{In}), while the
remaining factor is the leading $m \neq 0$ modification of the
massless result in the asymptotic regime.  It follows that in the
mean value approximation, where $z \simeq kR$, the modification
of the massless result can be approximated by the factor
\begin{equation}
	\sqrt{\frac{\pi mz}{2}}e^{-mz}
		\simeq
	\sqrt{\frac{\pi mkR}{2}}e^{-mkR}.
\label{mv exp}
\end{equation}
When calculating $W$ the contribution from the exponential is
dominated by values of the argument near unity, in which case
$(\pi mkR/2)^{1/2}$ is also of order unity and can be dropped.
We conclude that when $m \neq 0$ the contribution to
$U_{0}^{(k)}$ from the term
we are considering in Eq.~(\ref{effective U0 k}) is approximately
given by
\begin{equation}
	U_{0}^{(k)} \simeq 
		\frac{2i^{k}}{\pi R}\,
	\left(
	\frac{G_{F}a_{n}e^{-mR}}{2\pi\sqrt{2}R^{2}}\right)^{2}
	\,
	\frac{(k!)^{2}}{k^{k + 2}}.
\label{U0(k)}
\end{equation}

	The contributions from the other $z$-dependent terms in
Eq.~(\ref{massive V4}) can be treated in a similar manner.  Since each
term contains either $K_{0}(mz)$ or $K_{1}(mz)$, both of which
are proportional to $e^{-mz}$, it follows that each contribution
to $U_{0}^{(k)}$ will contain the damping factor $\exp(-mkR)$
which leads to saturation of the neutrino-exchange forces.  The
contributions to $U_{0}^{(k)}$ from this factor are of the form
\begin{equation}
	\frac{C_{k}}{R}
	\left(\frac{G_{F}a_{n}e^{-mR}}{R^{2}}
	\right)^{k - \alpha - \beta}
	\left(\frac{G_{F}a_{n}me^{-mR}}{R}
	\right)^{\alpha}
	\left(G_{F}a_{n}m^{2}e^{-mR}
	\right)^{\beta},
\label{Ck}
\end{equation}
where $\alpha$ and $\beta$ are integers, and where $C_{k}$ is a
$k$-dependent coefficient which can in principle be determined
for each such term.
In practice this would be not only tedious but also
unnecessary, since any combination of such terms leads to
roughly the same limit on $m$.  For purposes
of deriving the limit on $m$ it is helpful to note that for each
$k$ there must be at least one term corresponding to $\alpha =
\beta = 0$, since this is the only term which reproduces the
known contribution to $U_{0}^{(k)}$ in the $m = 0$ limit.  In the
next section we use the preceding results to derive a lower bound
on the mass of neutrinos.
	
\section{Bound on the Neutrino Mass}

	In this section we derive the actual bound on $m$ which
follows from the assumption that the mechanism for resolving the
neutrino-exchange energy-density catastrophe is a nonzero value
of $m$.  A simple bound on the mass $m$ of any neutrino can 
be inferred from the observation that if the Compton wavelength
of the neutrino were larger than the radius $R_{10}$ of the
neutron star, then neutrino-exchange forces would behave as if
neutrinos were massless.  This leads to the estimate,
\begin{equation}
	\frac{\hbar}{mc} \lesssim R_{10}
	\,\,\, \Rightarrow \,\,\, mc^{2}
	\gtrsim \frac{\hbar c}{R_{10}} = 2 \times 10^{-11}\,\mbox{eV}.
\label{first m}
\end{equation}
However, a mass this small would be insufficient to prevent a
smaller subvolume of the neutron star from having an unphysically
large energy density.  Specifically, if the neutrino-exchange
energy in a given subvolume were to exceed the known mass (and by
extension all other interaction energies) contained in that
subvolume, then the subvolume would become unstable against small
perturbations, in analogy to Earnshaw's theorem for
electrostatics \cite{HARN49,FLS63}.  Following Ref.~\cite{HARN49}, let
$\Phi(\vec{x}_{0})$ denote the potential energy of a neutron located
in the neutron star at a point $\vec{x}_{0}$, at which no other
sources are present.  If the neutron is displaced infinitesimally
from $\vec{x}_{0}$ to $\vec{x}_{0} + \delta\vec{x}$, where
$\delta\vec{x} = (\delta x_{1},\delta x_{2}, \delta x_{3})$, then the change
in its potential energy is given by
\begin{equation}
	\Delta\Phi \equiv \Phi(\vec{x}_{0} + \delta\vec{x})
		- \Phi(\vec{x}_{0})
		\simeq
	\delta\vec{x}\cdot\vec{\nabla}\Phi(\vec{x}_{0})
		+ \frac{1}{2}\sum_{i,j = 1}^{3}
		\delta x_{i} \delta x_{j}
		\frac{\partial^{2}\Phi(\vec{x}_{0})}
		{\partial x_{i} \partial x_{j}}.
\label{DeltaP}
\end{equation}
For the neutron to be in equilibrium to start with, it must be
the case that the net field at $\vec{x}_{0}$ must vanish, and
since this is proportional to $\vec{\nabla}\Phi(\vec{x}_{0})$,
the term proportional to $\delta\vec{x}$ is zero.  By an
appropriate choice of coordinate system the remaining term in
Eq.~(\ref{DeltaP}) can be diagonalized so that
\begin{equation}
	\Delta\Phi \simeq \frac{1}{2}
		\left[
	(\delta x_{1})^{2}
	  \frac{\partial^{2}\Phi(\vec{x}_{0})}{\partial x_{1}^{2}}
	+
	(\delta x_{2})^{2}
	  \frac{\partial^{2}\Phi(\vec{x}_{0})}{\partial x_{2}^{2}}
	+
	(\delta x_{3})^{2}
	  \frac{\partial^{2}\Phi(\vec{x}_{0})}{\partial x_{3}^{2}}
	\right].
\label{DeltaPb}
\end{equation}
For a displacement $\delta\vec{x} =
(\delta l,\delta l,\delta l)$, $\Delta\Phi$ would then be
given by
\begin{equation}
	\Delta\Phi \simeq \frac{1}{2}(\delta l)^{2}
		\nabla^{2}\Phi(\vec{x}_{0}).
\label{DeltaPc}
\end{equation}
If $\Phi(\vec{x}_{0})$ were the electromagnetic potential then
$\nabla^{2}\Phi(\vec{x}_{0}) = 0$ if there are no sources at
$\vec{x}_{0}$.  Since the condition for a stable equilibrium is
that $\Delta\Phi > 0$ for any displacement $\delta\vec{x}$, it
follows that a collection of charges cannot be in stable
equilibrium under purely electrostatic forces, which is
Earnshaw's theorem.  For many-body neutrino-exchange forces
$\nabla^{2}\Phi(\vec{x}_{0})$ is non-zero in general, but it can
be either positive or negative depending on the product
$(-1)^{N/2}\cos(1/\Gamma_{R})$ in
Eq.~(\ref{WforN 6}).  Since $\Delta\Phi > 0$ cannot be ensured
for an arbitrary distribution of neutrons, it follows from
Eq.~(\ref{DeltaPc}) that in general such a distribution is
unstable if neutrino-exchange forces are the dominant or
exclusive forces present.  This is the same situation that 
obtains for electromagnetism, and hence a similar conclusion
follows:  In the electrostatic case a stable equilibrium requires
the presence of other (non-electrostatic) forces.  In the present
circumstance neutrino-exchange forces must be similarly balanced by
other forces (e.g., gravitational, strong), but for this to be
the case, the magnitude of the neutrino-exchange force must be
comparable to that  of the other forces.  Since the energy
density arising from the gravitational or strong forces in any
subvolume is smaller than the mass contained in that subvolume,
we can then assume that the same must hold true for the contribution
from neutrino exchange.

	It follows from the preceding discussion that the
neutrino-exchange energy $W(r)$ inside a volume of radius $r \leq
R_{10}$ must be smaller than the mass $M(r)$ inside that volume.
From Eqs.~(\ref{WforN 6}) and (\ref{U0(k)}) we note that for the 
$\alpha = \beta = 0$ term in Eq.~(\ref{Ck}),
the above condition leads to
\begin{equation}
	\frac{1}{M(r)}
	\left[\frac{4}{r}\,
	       \frac{1}{N(r)}\,
	\left(\frac{G_{F}|a_{n}|N(r)e^{-mr}}
		{2\pi\sqrt{2}e^{2}r^{2}}
	\right)^{N(r)}\right]
	< 1,
\label{m constraint}
\end{equation}
where $\cos(1/\Gamma_{R})$ has been approximated by unity.  It is
convenient to rewrite Eq.~(\ref{m constraint}) by introducing the
length scale $L$ defined by
\begin{equation}
	L \equiv \left(\frac{\sqrt{2}G_{F}|a_{n}|\rho}
		{3e^{2}}\right)^{-1}
	= 1.7 \times 10^{-5}\,\mbox{cm}
	= \frac{1}{1.1\,\mbox{eV}}.
\label{L}
\end{equation}
In the approximation of neglecting the binding energy of the
neutron star, so that $M(r) \simeq N(r)m_{n}$, Eq.~(\ref{m
constraint}) can be rewritten in the form
\begin{equation}
	\left(\frac{r}{L}\,e^{-mr}\right)^{N(r)}
		<
	\frac{rN^{2}(r)}{4\ell_{n}},
\label{new m constraint}
\end{equation}
where $\ell_{n} \equiv \hbar/m_{n}c = 2.1 \times 10^{-14}$ cm is
the Compton wavelength of the neutron.  Taking the logarithm of
both sides of Eq.~(\ref{new m constraint}) then leads to the
condition
\begin{equation}
	m > \frac{1}{r}
		\left[\ln\left(\frac{r}{L}\right)
		- \frac{1}{N(r)}
		\ln\left(\frac{rN^{2}(r)}{4\ell_{n}}
		\right)\right].
\label{m constraint 1}
\end{equation}
Since $N(r) = 4\pi r^{3}\rho/3$, the right hand side can be
expressed directly in terms of $r/L \equiv x$, so that Eq.~(\ref{m
constraint 1}) becomes
\begin{equation}
	Lm > \frac{1}{x}
		\left[ \ln x
		- \frac{1}{x^{3}}
		\left(
		2 \times 10^{-23} +
		5 \times 10^{-25}\ln x\right)\right].
\label{Lm constraint}
\end{equation}
This equation must hold for all values of $r \leq R_{10}$, 
which is the subvolume radius, and hence for all $x \leq
R_{10}/L$.  From Eq.~(\ref{new m constraint}) it follows that when $x =
r/L < 1$ the inequality holds even when $m = 0$, and hence $x <
1$ is uninteresting.  For $x > 1$ the coefficient of $1/x^{3}$ in
Eq.~(\ref{Lm constraint}) is small compared to $\ln x$ and hence
the inequality becomes
\begin{equation}
	Lm \gtrsim \frac{1}{x}\ln x.
\end{equation}
Since this inequality must hold for all $x$, $Lm$ must exceed the
largest value that $(1/x)\ln x$ can assume, which is $1/e$.  This
gives
\begin{equation}
	Lm \gtrsim 1/e,
\end{equation}
and,
\begin{equation}
	mc^{2} \gtrsim 
		\frac{\sqrt{2}G_{F}|a_{n}|\rho}
		{3e^{3}}
		= 0.4 \,\mbox{eV}.
\label{final m limit}
\end{equation}
We note that $m$ is proportional to the product $G_{F}\rho$ which
is the only relevant quantity having the dimensions of mass that
can be formed from the available dynamical variables.  Since the
product $G_{F}\rho$ also arises in the
Mikheyev-Smirnov-Wolfenstein (MSW) mechanism \cite{MS,W} for
neutrino oscillations in matter, a few comments are in order
relating the present work and the MSW effect.  The effective
energy $E_{eff}$ for a real neutrino of mass $m$ and momentum $p$
propagating through a neutron star is given by \cite{BV}
\begin{equation}
	E_{eff} \simeq p + m^{2}/2p + \sqrt{2}G_{F}\rho,
\label{MSW E}
\end{equation}
and hence real neutrinos can be viewed for some purposes as if
they had a small mass $\sqrt{2}G_{F}\rho$.  This heuristic
picture helps to explain why the index of refraction for
neutrinos differs from unity, in analogy to the index of
refraction for light propagating in a medium.  However, if we
pursue the electromagnetic analogy we note that even in a
dielectric medium electrostatic forces arising from the exchange
of {\em virtual} photons still obey Coulomb's law, albeit with an
attenuated strength.  In the present case, the fact that
neutrino exchange would retain its long-range character is
significant since this implies that the combinatoric factor
entering in Eq.~(\ref{WforN 1}) would remain as ${N \choose k}$.
This in turn implies that unless the effective Fermi constant in
the medium differed from the vacuum value by a factor
$\O(10^{11})$, the neutrino-exchange energy density problem would
still exist.  In fact one would not expect $G_{F}$ to be
significantly modified by the presence of a medium because there
is no analog for neutrino exchange of a polarization charge in
electromagnetism.  Stated another way, there is no mechanism for
shielding the neutrino-exchange force \cite{HS94}.

	Returning to Eq.~(\ref{final m limit}) we note that the
lower bound applies
separately to $\nu_{e}$, $\nu_{\mu}$, and $\nu_{\tau}$, and is
compatible with the upper bounds quoted in Eq.~(\ref{m nu
limits}) for the three neutrino species, as shown in
Fig.~\ref{neutrino bound figure}.  For $\nu_{e}$ the upper and
lower bounds are sufficiently close to suggest that direct
evidence for $m_{\nu_{e}} \neq 0$ could be forthcoming in the
foreseeable future.  Indeed it may well be the case that the
anomalies in the flux of solar neutrinos discussed in Sec.~I
could be a signal for a non-zero neutrino mass compatible with
the bound in Eq.~(\ref{final m limit}).  In addition, the
implication that both $\nu_{\mu}$ and $\nu_{\tau}$ must also be
massive may help to solve the ``missing mass'' problem discussed
earlier.

\acknowledgements

	I wish to thank Jim Hartle for making available to me his
unpublished notes on neutrino exchange, and for his permission to
reproduce parts of these notes in Sec.~III and  Appendix~B.  
I am very deeply indebted
to Dennis Krause for critically reading this manuscript, and for
numerous suggestions which have helped to elucidate both the
underlying physics and its presentation. Many of my colleagues
have given generously of their time to clarify various issues
that have arisen in the course of this work.  I am particularly
indebted to Sid Bludman, Yasunori Fujii, 
Vagos Hadjimichael, Boris Kayser, Serguei
Khlebnikov, Al Overhauser, Peter Rosen, Herman Rubin, Daniel
Sudarsky, Joe Sucher, and Carrick Talmadge.  I also wish to thank
Shu-Ju Tu and Sho Kuwamoto for their help in carrying out the
numerical work, and David Schleef, Michelle Parry, and Alex Pozamantir for
discussions on geometric probability.  This work is supported by
the U.S. Department of Energy under Contract No.
DE-AC02-76ER01428.

\pagebreak

\appendix

\section{NOTATION, METRIC CONVENTIONS, and Standard Model
Couplings}

	In this Appendix we summarize our metric conventions and those
for the Dirac equation.  We have employed the Pauli metric
conventions of Akhiezer and Berestetskii \cite{AB65}, 
deWit and
Smith \cite{dS86}, 
Luri\'e \cite{LUR68}, 
and Sakurai \cite{SAK67}.
Reference \cite{dS86} contains an excellent summary of these
conventions, along with tables relating the Pauli metric
conventions to those of Bjorken and Drell \cite{BD64} who use
real 4-vector notation.
In the Pauli conventions, the Dirac equation in configuration space for
a particle of mass $m$  is given by ($c = \hbar = 1$),
\begin{equation}
	(\gamma \cdot \partial + m) \psi(x) = 0,
\label{dirac equation}
\end{equation}
where the  Dirac matrices $\gamma_\mu = \gamma_\mu^\dagger$ satisfy
\begin{equation}
 \everymath={\displaystyle}
   \begin{array}{l}
     \{ \gamma_\mu, \gamma_\nu \} = 2\delta_{\mu\nu},~~~~~~~~~~~~~~
       \{ \gamma_5,\gamma_\mu\} = 0, \\ 
     \gamma_5 = \gamma_5^\dagger = \gamma_1 \gamma_2 \gamma_3 \gamma_4 = 
       (1/4!) {\epsilon}_{\mu\nu\lambda\rho} 
	\gamma_\mu \gamma_\nu \gamma_\lambda
      \gamma_\rho.
   \end{array}
\end{equation}
Here ${\epsilon}_{\mu\nu\lambda\rho}$ is the completely antisymmetric
permutation symbol, and the dagger denotes the Hermitian adjoint.
In discussing the many-body contributions from systems containing
an odd number of particles, the following trace identities involving
${\epsilon}_{\mu\nu\lambda\rho}$ are useful (tr denotes the trace
over Dirac indices):
\begin{eqnarray}
    \mbox{tr} (\gamma_\mu\gamma_\nu \gamma_\lambda \gamma_\rho \gamma_5) 
		& = &
        4 {\epsilon}_{\mu\nu\lambda\rho},
	\nonumber \\
    \mbox{tr} (\gamma_\rho\gamma_\sigma \gamma_\tau \gamma_\mu \gamma_\nu
         \gamma_\lambda \gamma_5) 
		& = &
		4 (\delta_{\rho\sigma} 
        {\epsilon}_{\tau\mu\nu\lambda} - \delta_{\rho\tau}
              {\epsilon}_{\sigma\mu\nu\lambda} + \delta_{\sigma\tau}
            {\epsilon}_{\rho\mu\nu\lambda}
\label{trace identities}
	\\
	&   & \mbox{}
		+
		\delta_{\mu\nu}\epsilon_{\rho\sigma\tau\lambda}
		- 
		\delta_{\mu\lambda}\epsilon_{\rho\sigma\tau\nu}
		+	
		\delta_{\nu\lambda}\epsilon_{\rho\sigma\tau\mu}),
	\nonumber
\end{eqnarray}
where $\rho,\sigma,\tau,\mu,\nu,\lambda = 1,2,3,4$.
Other useful trace identities are given in Ref.~\cite{dS86}.

	For purposes of deriving the Schwinger formula \cite{HAR70,SCH54,HARUNP}
for the
weak energy $W$ in Appendix B, the effective low-energy
Lagrangian describing the coupling of neutrinos to quarks and
leptons is required.  Using Ref. \cite{PDG94} the neutrino-quark
interaction is given by
\begin{equation}
	{\cal L}_{I}^{\nu q} = \frac{G_{F}}{\sqrt{2}}
		\ell_{\mu}(x)\sum_{j}
		\left[\epsilon_{L}(j)i\bar{q}_{j}(x)
		\gamma_{\mu}(1 + \gamma_{5})q_{j}(x)
		+
		\epsilon_{R}(j)i\bar{q}_{j}(x)
		\gamma_{\mu}(1 - \gamma_{5})q_{j}(x)
		\right],
\label{nu q L}
\end{equation}
\begin{equation}
	\ell_{\mu}(x) = i\bar{\psi}(x)\gamma_{\mu}
		(1 + \gamma_{5})\psi(x),
\end{equation}
where
$\psi(x)$ and $q_{j}(x)$ are the field operators for the neutrino
and for quark species $j$ respectively.  In the absence of
radiative corrections the parameters $\epsilon_{L}(j)$ and
$\epsilon_{R}(j)$ are given in terms of the weak mixing angle
$\theta_{W}$ by \cite{PDG94}
\begin{eqnarray}
	\epsilon_{L}(u) = \frac{1}{2} -
		\frac{2}{3}\sin^{2}\theta_{W} 
		& \phantom{space} &
	\epsilon_{R}(u) = 
		-\frac{2}{3}\sin^{2}\theta_{W}, 
\nonumber \\
		& \phantom{space} &
\\
	\epsilon_{L}(d) = -\frac{1}{2} +
		\frac{1}{3}\sin^{2}\theta_{W} 
		& \phantom{space} &
	\epsilon_{R}(d) = 
		\frac{1}{3}\sin^{2}\theta_{W}. 
\label{epsilon identities}
\nonumber 
\end{eqnarray}	
The neutrino-electron coupling can be similarly expressed as
\begin{equation}
	{\cal L}_{I}^{\nu e}(x) = \frac{G_{F}}{\sqrt{2}}
		\ell_{\mu}(x)
	\left[i\bar{e}(x)\gamma_{\mu}
	\left(g_{V}^{\nu e} + g_{A}^{\nu e}\gamma_{5}
	\right)e(x)
	\right],
\label{nu e L}
\end{equation}
where $e(x)$ is the electron field operator, and the constants
$g_{V}^{\nu e}$ and $g_{A}^{\nu e}$ are given by
\begin{equation}
	g_{V}^{\nu e} = \frac{1}{2} + 2\sin^{2}\theta_{W}
	\phantom{somespace}
	g_{A}^{\nu e} = -\frac{1}{2}.
\label{gVA}
\end{equation}
Note that in (\ref{gVA}) the charged-current contribution has
been included along with that from the neutral current
\cite{LEW80}.

	For present purposes, we are interested in the coupling
of neutrinos to a static source of unpolarized neutrons, protons,
or electrons.  Since the only relevant contributions in this
circumstance are proportional to $\bar{q}_{j}\gamma_{4}q_{j}$ and
$\bar{e}\gamma_{4}e$, the effective vector charges $a_{i}$
describing the couplings to neutrons, protons, and electrons are
\begin{mathletters}
\label{a's}
\begin{eqnarray}
	a_{n} & = &
		2[\epsilon_{L}(d) + \epsilon_{R}(d)]
		+
		[\epsilon_{L}(u) + \epsilon_{R}(u)]
		=
		-\frac{1}{2},
\\
	a_{p} & = &
		2[\epsilon_{L}(u) + \epsilon_{R}(u)]
		+
		[\epsilon_{L}(d) + \epsilon_{R}(d)]
		=
		\frac{1}{2} - 2\sin^{2}\theta_{W}
		= 0.036,
\\
	a_{e} & = & g_{V}^{\nu e} =
		\frac{1}{2} + 2\sin^{2}\theta_{W}
		= 0.964
\label{ae}
\end{eqnarray}
\end{mathletters}
From Table~26.2 of Ref. \cite{PDG94} we note that the agreement
between theory and experiment for the Standard Model couplings in
Eqs.~(\ref{epsilon identities})--(\ref{a's}) is typically at the
level of a few percent.  It follows from the preceding discussion
that the effective low-energy neutrino-neutron coupling can be
written in the form given in Eq.~(\ref{effective Lagrangian}) below.

\pagebreak

\section{THE SCHWINGER FORMULA FOR W}

As discussed in Sec. II, our derivation of the $k$-body
$(k=1,\ldots ,N)$ neutrino-exchange potential utilizes
the formalism developed by Hartle for the 4-body case
\cite{HAR70,HARUNP}, which follows in turn from the Schwinger
formula \cite{SCH54}, Eq.~(\ref{Schwinger W}) below.  
To help clarify the assumptions, notation,
and metric conventions that underlie our results, we present
here a derivation of the Schwinger formula due to Hartle
\cite{HARUNP}. Other useful results can be found in
Ref.~\cite{DR85} which deals with the related question of
effective Lagrangians in quantum electrodynamics.

We are interested in computing the weak-interaction energy
$W$ of a collection of $N$ particles (e.g. a nucleus or a neutron
star) due to neutrino exchange.  Following the discussion in
the Introduction, $W$ can be viewed as the weak-energy analog
of the static Coulomb energy $W_{C}$ of a collection of electric
charges, which for a nucleus can be approximated by the 2-body
contribution in Eq.~(\ref{Z W_C}).  As we have noted previously, one
of the novel features of neutrino-exchange is that $W$ is dominated
by many-body contributions.  We have shown in Sec.~V that if
$W$ is expressed in the form
\begin{equation}
	W = \sum_{k=2}^N W^{(k)},
\label{W sum}
\end{equation}
where $W^{(k)}$ is the $k$-body contribution, then the dominant contributions
to $W$ arise from terms with $k \simeq N$.

Consider for the sake of concreteness an idealized non-rotating
spherical neutron
star containing $N$ neutrons.  To derive the Schwinger formula for $W$,
the weak interaction energy can be viewed (to lowest order in the
Fermi constant $G_{F}$) as the energy difference between a neutrino
propagating in the ``vacuum'' $|\hat{0}\rangle$ inside the neutron star,
and one propagating in the usual matter-free vacuum $|0\rangle$.  Thus,
\begin{equation}
	W = \langle \hat{0}|H|\hat{0}\rangle 
	- \langle 0|H_{0}|0\rangle \equiv {\cal E}- {\cal E}_{0},
\label{U difference}
\end{equation}
where $H_{0}$ is the free Hamiltonian for the propagating neutrino, and
$H$ includes the interactions with the neutrons.  If $\psi(x)$ denotes the
field operator for the interacting neutrino then
\begin{equation}
	H = \int d^3x {\cal H}(x) = \int d^{3}x~\pi(x) \dot{\psi}(x) =
    i\int d^{3}x~\psi^{\dagger}(x) \partial_{t}\psi(x),
\label{H}
\end{equation}
where $\partial_t \equiv \partial/\partial t$.  From Eqs.~(\ref{U
difference}) and (\ref{H}),
we then have ($\bar{\psi} = \psi^{\dagger}\gamma_{4}$)
\begin{equation}
	{\cal E}= i \int d^3x \,
	\langle\hat{0}| \bar{\psi}(x) \gamma_4
         \partial_t \psi(x) |\hat{0}\rangle ,
\label{E}
\end{equation}
and an analogous formula for ${\cal E}_{0}$.  We can express ${\cal E}$
in terms of the neutrino propagator by writing Eq.~(\ref{E}) in the form
\begin{eqnarray}
    {\cal E} & = & 
	i \int d^3x \left\{ \partial_t 
	\langle\hat{0}|\bar{\psi}(x^\prime)
     	\gamma_4 \psi (x) |\hat{0}\rangle 
	\right\}_{x^\prime \rightarrow x} 
	\nonumber \\
      	     & = & 
	i\int d^3 x \left\{ (\gamma_4)_{\alpha\beta} \partial_t
        \langle\hat{0}|\bar{\psi}_\alpha (x^\prime) \psi_\beta (x)
          |\hat{0}\rangle  \right\}_{x^\prime \rightarrow x},
\end{eqnarray}
where $\alpha$ and $\beta$ are spinor indices and $\partial_t$ acts only
on unprimed variables.  We can assume without loss of generality that
$t' > t$ in which case,
\begin{eqnarray}
    \bar{\psi}_\alpha (x') \psi_\beta (x) 
	& = & \theta (t' - t)
     \bar{\psi}_\alpha(x') \psi_\beta(x) - \theta (t-t')
     \psi_\beta(x) \bar{\psi}_\alpha(x^\prime)
	\nonumber \\
     &\equiv & 
	\mbox{T}[\bar{\psi}_\alpha(x') \psi_\beta(x)]
     = - \mbox{T}[\psi_{\beta}(x) \bar{\psi}_{\alpha}(x')].
\end{eqnarray}
If we define the neutrino propagator $S_F(x,x')$ by
\begin{equation}
	[S_F(x,x')]_{\beta\alpha} = 
	\langle\hat{0}|\mbox{T}[\psi_\beta(x)
     \bar{\psi}_\alpha (x')]|\hat{0}\rangle ,
\label{neutrino propagator}
\end{equation}
then ${\cal E}$ can be cast in the form
\begin{equation}
	{\cal E} = 
	-i \int d^3x \left\{ \partial_t 
	\mbox{tr}[\gamma_4 S_F(x,x^\prime)]
     \right\}_{x^\prime \rightarrow x}
\label{new E}
\end{equation}
where tr denotes the trace over Dirac indices.  Following Schwinger
we introduce the Fourier transform $S_F({\vec {x}},{ \vec{x}}^\prime,E)$
defined by
\begin{equation}
	S_F(x,x^\prime) = \int_{-\infty}^\infty \frac{dE}{2\pi}
     e^{-iE(t-t^\prime)} S_F({ \vec{x},\vec{x}^\prime}, E) .
\label{SF transformed}
\end{equation}
Combining Eqs.~(\ref{new E}) and (\ref{SF transformed}) we have
\begin{eqnarray}
	{\cal E}  & = & 
	-i\int d^3x \left\{ \partial_t 
	\mbox{tr} \left[\gamma_4 \int_{-\infty}^\infty
     \frac{dE}{2\pi} e^{-iE(t-t^\prime)} S_F({ \vec{x},\vec{x}^\prime},E)
      \right]\right\}_{x^\prime \rightarrow x}
	\nonumber \\
     		& = &  
	-\frac{1}{2\pi} \int d^3x \left\{ 
	\mbox{tr}\left[ \int_{-\infty}^\infty
      dE \,\, E\gamma_4 S_F ({ \vec{x},\vec{x}}^\prime,E)\right]
	\right\}_{{ \vec{x}^\prime} \rightarrow { \vec{x}}} ,
\end{eqnarray}
where the limit $t^\prime \rightarrow t$ has been taken following the
action of $\partial_t$.  If the order of integration with respect 
to ${ \vec{x}}$ and E is interchanged then
\begin{eqnarray}
	{\cal E} & = & 
	\frac{-1}{2\pi} \int_{-\infty}^\infty dE~E
    \int d^3x~\left\{\mbox{tr}[\gamma_4 S_F({ \vec{x},\vec{x}}^\prime,E)
      ]\right\}_{{ \vec{x}^\prime}\rightarrow {\vec{x}}}
	\nonumber \\
    	& = &  
	\frac{-1}{2\pi} \int_{-\infty}^\infty dE~E \left\{\mbox{Tr}
     [\gamma_4 S_F(E)] \right\}.
\label{intermediate E}
\end{eqnarray}
In the last step of Eq.~(\ref{intermediate E}) 
we have introduced the operator $S_F(E)$
whose matrix elements give the c-number function
$S_F({ \vec{x},\vec{x}^\prime}, E)$:
\begin{equation}
	\langle{\vec{x}}| S_F(E)|{ \vec{x}^\prime}\rangle
	 = S_F ({ \vec{x},\vec{x}^\prime}, E).
\label{SF}
\end{equation}
The limit ${ \vec{x}^\prime} \rightarrow { \vec{x}}$ followed by
$\int d^3 x$ can then be viewed as a formal trace 
in configuration space,
so that in the
notation of Eq.~(\ref{intermediate E}) we can write symbolically,
\begin{equation}
\mbox{Tr} \equiv \mbox{tr} \times \int d^3 x
\label{new trace}
\end{equation}

The expression for $\cal E$ in Eq.~(\ref{intermediate E}) 
can be recast into a more
useful form by expressing the interacting neutrino propagator $S_F(E)$
in terms of the free propagator $S_F^{(0)}(E)$.  Following the
discussion in Appendix A, we assume that
the
low-energy coupling of the neutrons and neutrinos can be expressed
in terms of the effective Lagrangian density
\begin{equation}
     {\cal L}_I (x)  =  
\frac{G_F}{\sqrt{2}} a_n N_\mu(x) \ell_\mu(x).
\label{effective Lagrangian}
\end{equation}
Here
$N_\mu(x)$ is the neutron current, $\ell_\mu(x)$ is the neutrino
current,
\begin{equation}
    \ell_\mu(x) = i\bar{\psi} (x) \gamma_\mu (1+\gamma_5)\psi(x),
\label{neutrino current}
\end{equation}
and $a_n = - 1/2$.  From Eqs.~(\ref{effective Lagrangian}) 
and (\ref{dirac equation}) the complete
Lagrangian density for neutrinos is
\begin{eqnarray}
	{\cal L} (x) & = & {\cal L}_0(x) + {\cal L}_I (x) 
	\nonumber \\
	& = &
    - \bar{\psi}(x)[\gamma \cdot \partial + m]\psi(x)
     + \frac{G_F}{\sqrt{2}} a_n N_\mu(x) \ell_\mu (x) ,
\label{neutrino Lagrangian}
\end{eqnarray}
where $m$ is the neutrino mass, which we will assume to be zero
at this stage.  The equation of motion for $\psi(x)$ can be obtained
from the Euler-Lagrange equation,
\begin{equation}
	\frac{\partial}{\partial x_\lambda} \left[ \frac{\partial {\cal L}}
      {\partial (\partial \bar{\psi}/\partial x_\lambda)}\right]
      - \frac{\partial {\cal L}}{\partial \bar{\psi}} = 0,
\label{E-L equation}
\end{equation}
and is given by
\begin{equation}
	\left[ \gamma \cdot \partial - \frac{iG_Fa_n}{\sqrt{2}}
       \gamma \cdot N (1+\gamma_5)\right] \psi(x) = 0.
\label{neutrino equation}
\end{equation}
It follows from Eq.~(\ref{E-L equation}) 
that $S_F ({ \vec{x},\vec{x}^\prime},E)$
is a solution of the equation
\begin{equation}
	\left[ \gamma \cdot \eta  - \gamma \cdot \tilde{N}\right] S_F
     ({ \vec{x},\vec{x}^\prime},E) = -i \delta^3 ({ \vec{x}}
      - { \vec{x}^\prime}),
\label{SF equation}
\end{equation}
where
\begin{equation}
	\tilde{N}_\mu \equiv \frac{iG_Fa_n}{\sqrt{2}} N_\mu (1+\gamma_5),
\end{equation}
\begin{equation}
	\gamma \cdot \eta \equiv { \vec{\gamma}} \cdot
    {\vec{\partial}} - \gamma_4 E,
\end{equation}
and ${ \vec{\partial}} \equiv \partial/\partial { \vec{x}}$.
If the state vectors are normalized such that
\begin{equation}
	\langle\vec{x}|\vec{x}'\rangle = \delta^3
    (\vec{x} -  \vec{x}') ,
\end{equation}
then Eqs.~(\ref{neutrino equation}) and (\ref{SF}) lead to
\begin{equation}
	[ \gamma \cdot \eta - \gamma \cdot \stackrel{\sim}{N}]
     S_F(E) = -i I,
\label{new SF equation}
\end{equation}
where $I$ is the identity operator.  Multiplying both sides of Eq.~(\ref{new
SF equation}) by $\gamma_4$ gives
\begin{equation}
   [E-H] S_F(E) = i\gamma_4 ,
\label{E-H equation}
\end{equation}
where
\begin{equation}
	H = - i{ \vec{\alpha}} \cdot { \vec{\partial}}
      - \gamma_4 \gamma \cdot \tilde{N} 
	= H_0 - \gamma_4 \gamma \cdot \tilde{N},
\end{equation}
using ${ \vec{\alpha}} = i\gamma_4 {\vec{\gamma}}$.
It follows that
\begin{mathletters}
\begin{eqnarray}
	S_F(E) & = & i [E-H]^{-1} \gamma_4 ,
\label{SF E-H equation} \\
	S_F^{(0)}(E)& = & i[E-H_0]^{-1} \gamma_4 .
\label{SF0 E-H equation} 
\end{eqnarray}
\end{mathletters}
Combining Eqs.~(\ref{intermediate E}) 
and (\ref{SF E-H equation}) we can express ${\cal E}$ in the form
\begin{equation}
	{\cal E} = \frac{-i}{2\pi} \int_{-\infty}^\infty
     dE ~E \left\{\mbox{Tr} \left(\frac{1}{E-H}\right)\right\} ,
\label{intermediate E 2}
\end{equation}
and carrying out a partial integration allows 
Eq.~(\ref{intermediate E 2}) to be written as
\begin{equation}
	{\cal E} = \frac{-i}{2\pi}\mbox{Tr} \left\{ E \ln(E-H)
     \Biggr|_{-\infty}^\infty  - \int_{-\infty}^\infty
      dE \ln(E-H) \right\} .
\label{intermediate E 3}
\end{equation}
Since ${\cal E}_0$ can be obtained from Eq.~(\ref{intermediate E
3}) by substituting
$H_0$ for H, we find for $W$
\begin{eqnarray}
	W & = & {\cal E} - {\cal E}_0 
	\nonumber \\
	& = & \frac{-i}{2\pi}
     	\mbox{Tr}
	\left\{ E \ln\left(\frac{E-H}{E-H_0}\right) \Biggr|_{-\infty}^\infty
      - \int_{-\infty}^\infty dE \ln\left(\frac{E-H}{E-H_0}\right) \right\}~~ .
	\nonumber \\
	& = & \frac{i}{2\pi}
     	\mbox{Tr}\left\{
      \int_{-\infty}^\infty dE \ln\left(\frac{E-H}{E-H_0}\right) \right\}~~ .
\label{general U}
\end{eqnarray}
Using Eqs.~(\ref{E-H equation}) and 
(\ref{SF0 E-H equation}) 
the remaining term in Eq.~(\ref{general U}) can be written as
\begin{equation}
	\frac{E-H}{E-H_0} = 
	1 + \frac{\gamma_4 \gamma\cdot \stackrel{\sim}{N}}{E-H_0}
      = 1 - i \gamma_4 \gamma \cdot \stackrel{\sim}{N} S_F^{(0)} (E) \gamma_4 .
\label{E-H ratio}
\end{equation}
Combining Eqs.~(\ref{general U}) and (\ref{E-H ratio}) gives
\begin{equation}
	W = \frac{i}{2\pi} 
	\mbox{Tr} \left\{ \int_{-\infty}^{\infty}
	dE\,
      \ln[1 - i \gamma_4 \gamma \cdot \stackrel{\sim}{N}
      S_F^{(0)} (E) \gamma_4]\right\}.
\label{general U 2}
\end{equation}
Since the integrand represents the infinite series
\begin{equation}
	\ln(1-\Delta) = -\sum_{k=1}^\infty \frac{\Delta^k}{k} ,
\label{ln Delta again}
\end{equation}
it follows that each term in the series will be of the form
\begin{equation}
	\mbox{tr}[\gamma_4 \gamma \cdot \tilde{N} 
	(-i) S_F^{(0)} (E)\cdots\gamma_4]
      = \mbox{tr}[\gamma \cdot \tilde{N} (-i) S_F^{(0)} (E)\cdots],
\end{equation}
using the cyclic property of the trace.  The Dirac matrices $\gamma_4$ can thus
be dropped from Eq.~(\ref{general U 2}) which then leads via Eq.~(\ref{SF
equation}) to the Schwinger formula for
$W$ \cite{HAR70,SCH54}:
\begin{equation}
	W = \frac{i}{2\pi} \mbox{Tr} \left\{ \int_{-\infty}^\infty dE~\ln
     [1 + \frac{G_F a_n}{\sqrt{2}}  N_\mu \gamma_\mu (1+\gamma_5) S_F^{(0)}
      (E)]\right\}.
\label{Schwinger W}
\end{equation}

Following Ref.~\cite{HARUNP}, we show in Sec. III that the Schwinger formula
leads directly to a finite expression for the 2-body potential
$V^{(2)}(r_{12})$, without having to resort to the 
regularization methods employed
by either FS \cite{FS68} or HS \cite{HS94}.  
Evidently the Schwinger formula must
also incorporate a regularization procedure (since the full 
2-body amplitude for
neutrino-exchange is divergent), but this regularization is built in at the
outset when $W$ is expressed as the difference $({\cal E}-{\cal E}_0)$ in
Eq.~(\ref{U difference}).  

	As noted at the beginning of this Appendix, $W$ is the
analog of the Coulomb energy $W_{C}$ in Eq.~(\ref{Z W_C}) in the
sense of incorporating both the integration over the charge
distribution and the combinatorics associated with these charges.
For example, when Eq.~(\ref{Schwinger W}) is expanded to
$\O(G_{F}^{4})$ as in Eq.~(\ref{W4a}), the integrations over all
space are explicitly indicated, and the combinatoric factors
arise from counting the number of ways that $N$ particles can be
assigned to the coordinates $x_{1},\ldots,x_{4}$ in
$T_{\mu\nu\lambda\sigma}(x_{1},x_{2},x_{3},x_{4})$.  In practice
we will not calculate $W$ directly from Eq.~(\ref{Schwinger W}),
but rather use the Schwinger formula as a generating functional
to obtain the $k$-body potentials
$V^{(k)}(r_{12},\ldots,r_{k1})$, as in Eqs.~(\ref{general V2}),
(\ref{V4}), (\ref{V6}), and (\ref{Vk}). These potentials will then be
integrated over the spherical volume in Section~IV to produce the
$U^{(k)}$, by adapting some formalism from geometric probability.
The final expression for $U$ is then obtained in Sec.~V by
supplying the combinatoric factor ${N \choose k}$, which allows
us to write 
\begin{equation}
	W = \sum_{k = 2\atop even}^{N}W^{(k)} = 
		\sum_{k=2 \atop even}^{N}U^{(k)}
		{N \choose k}.
\end{equation}

	In order to apply the Schwinger formula in Eq.~(\ref{Schwinger
W}) it is necessary to
exhibit the explicit functional form of the neutrino propagator
$S_F^{(0)} (E)$.  From Eq.~(\ref{neutrino propagator}) we see that
\begin{equation}
	S_F^{(0)} (x,x^\prime) = 
	\langle 0|\mbox{T}[{\psi}(x)\bar{\psi}(x^\prime)]|0\rangle,
\end{equation}
where our conventions for $S_F^{(0)}$ (including various factors of $i$)
follow those of Luri\'e \cite{LUR68}.  
Since $S_F^{(0)}(x,x^\prime)$ is translationally
invariant, we can set $x^\prime = 0$ without loss of generality in which
case \cite{LUR68}
\begin{equation}
	 S_F^{(0)} (x,0) = S_F^{(0)}(x) 
	= -(\gamma \cdot \partial - m)
      \Delta_F^{(0)} (x) \stackrel{m=0}{\longrightarrow} -
     \gamma \cdot \partial \Delta_F^{(0)} (x).
\end{equation}
Here $\Delta_F^{(0)} (x)$ is the free propagator for a (massless) scalar field,
which is given in configuration space by
\begin{equation}
	\Delta_F (x) = \frac{1}{4\pi^2(x^2 + i\epsilon)} .
\label{Simple Delta F}
\end{equation}
Inverting Eq.~(\ref{SF transformed}) we then have
\begin{eqnarray}
    	S_F^{(0)}(\vec{x},E) & = & 
	\int_{-\infty}^\infty dt~e^{iEt} S_F^{(0)}(x) 
	\nonumber \\
      	& = & 
	[\vec{\gamma}\cdot \vec{\partial} - \gamma_4E] \int_{-\infty}^\infty
      dt~e^{iEt}
     \frac{(-1)}{4\pi^2 (r^2 - t^2 + i\epsilon)} .
\label{inverted SF}
\end{eqnarray}
The integral in Eq.~(\ref{inverted SF}) 
can be evaluated using contour integration,  by closing
the contour in the upper half-plane for $E > 0$, and in the lower half-plane
for $E < 0$.  The result is
\begin{equation}
	S_F^{(0)} (\vec{x},E) = \gamma \cdot \eta \left[
     \frac{i}{4\pi} \frac{e^{i|E|(|\vec{x}|+i\epsilon)}}
                            {|\vec{x}|+i\epsilon} \right],
\label{inverted SF 2}
\end{equation}
where
\begin{equation}
	\gamma \cdot \eta = 
	\vec{\gamma} \cdot \vec{\partial} - \gamma_4E.
\end{equation}
To describe the propagation of a neutrino from $\vec{r}_{j}$ to
$\vec{r}_{i}$
where $\vec{r}_i$ and $\vec{r}_j$ are the coordinates of two neutrons,  
let $\vec{x} \rightarrow \vec{r}_{ij} = \vec{r}_i - \vec{r}_j$, 
with $r_{ij} = |\vec{r}_{ij}|$.
Thus the explicit
expression for $S_F^{(0)}(\vec{r}_{ij},E)$ to be used in evaluating the many-body
contributions is
\begin{eqnarray}
    S_F^{(0)} (\vec{r}_{ij},E) & = & 
	[\vec{\gamma}\cdot
      \vec{\partial}_{{ij}} - \gamma_4 E]
     \left[ \frac{i}{4\pi} \frac{e^{i|E|(r_{ij}+i\epsilon)}}
                                 {r_{ij} + i\epsilon} \right] 
	\nonumber \\
    	&\equiv & 
	\gamma \cdot \eta (ij) \Delta_F (\vec{r}_{ij},E) ,
\label{final inverted SF}
\end{eqnarray}
where $\vec{\partial}_{ij} \equiv \partial/\partial \vec{r}_{ij}$.
In the Tr notation of Eqs.~(\ref{new trace}) and (\ref{Schwinger
W}), the propagator
$S_F^{(0)} (\vec{r}_{ij},E)$ is
to be thought of as the ${ij}$ matrix element of the operator $S_F^{(0)}(E)$
in Eq.~(\ref{Schwinger W}).  We note that when $\ln[1+ \cdots]$ 
in Eq.~(\ref{Schwinger W}) is expanded using
Eq.~(\ref{ln Delta again}), 
the integrand in Eq.~(\ref{Schwinger W}) will contain a 
polynomial in $E$ in which
odd powers of $E$ can be dropped due to the symmetric integration limits.

\pagebreak

\section{Probability Distributions for Points in a Spherical
Volume}

	As noted in Sec.~II, the average value $\langle
g\rangle$ of a function $g(r)$ taken over a 3-dimensional
spherical volume of radius $R$ is
\begin{equation}
	\langle g\rangle = \int^{2R}_{0} dr\,
		{\cal P}_{3}(r)g(r),
\label{g average again}
\end{equation}
where $r = r_{12} = |\vec{r}_{1} - \vec{r}_{2}|$ is the
separation of two points, and ${\cal P}_{3}(r)$ is the
probability distribution.  The functional form of ${\cal
P}_{n}(r)$ for an $n$-dimensional ball has been discussed by a
number of authors whose work is summarized in Refs.~\cite{SAN76}
and \cite{KM63}.  It is convenient to introduce the scaled
dimensionless variable $s = r/2R$ which satisfies $1 \geq s
\geq 0$,
and to re-express ${\cal P}_{3}(r)$, $g(r)$, and $\langle
g\rangle$ in terms of ${\cal P}_{3}(s)$ and $g(s)$ so that
\begin{equation}
	\langle g\rangle = \int^{1}_{0} ds\,
		{\cal P}_{3}(s)g(s).
\label{g average s}
\end{equation}
${\cal P}_{n}(s)$, which denotes the probability that the scaled
distance between two points in an $n$-dimensional ball will be in the
interval ($s$, $s + ds$) is then given by \cite{SAN76,KM63}
\begin{equation}
	{\cal P}_{n}(s) = 2^{n}n s^{n - 1}
		I_{1 - s^{2}}[(n + 1)/2, 1/2].
\label{P_n(s)}
\end{equation}
$I_{x}(p,q)$ is the incomplete beta function defined by
\begin{equation}
	I_{x}(p,q) = \frac{\Gamma(p + q)}{\Gamma(p)\Gamma(q)}
		\int^{x}_{0} dt\,t^{p - 1}(1 - t)^{q - 1}.
\label{Ix}
\end{equation}
The results for $n = 1, 2, 3$ are of particular interest in
physics, and the corresponding probability distributions are
given by \cite{SAN76,KM63}
\begin{eqnarray}
	{\cal P}_{1}(s) & = & 2(1 - s),
\label{P_1(s)}
	\\
	{\cal P}_{2}(s) & = & \frac{16}{\pi}s
	\left[\cos^{-1}s - s(1 - s^{2})^{1/2}\right],
\label{P_2(s)}
	\\
	{\cal P}_{3}(s) & = & 12s^{2}
		\left(1 - s\right)^{2}(2 + s).
\label{P_3(s)}
\end{eqnarray}
The expression for ${\cal P}_{3}(s)$ in Eq.~(\ref{P_3(s)}) has
been obtained independently by Overhauser 
\cite{Overhauser}.

	The following properties of ${\cal P}_{n}(s)$ will be
useful in the ensuing discussion:
\begin{equation}
	\int^{1}_{0} ds\, {\cal P}_{n}(s) = 1,
\label{P_n normalization} 
\end{equation}
\begin{eqnarray}
	{\cal P}_{n}(1) & = & 0, \,\,n \geq 1;
	\phantom{space}
	{\cal P}_{n}(0)  =  0, \,\,n \geq 2;
\label{P_n 1,0}
	\\
	{\cal P}_{n}'(1) & = & 0, \,\,n \geq 2;
	\phantom{space}
	{\cal P}_{n}'(0)  =  0, \,\,n \geq 3,
\label{P_n' 1,0}
\end{eqnarray}
where the primes denote differentiation with respect $s$.
Another useful result is the mean value
$\overline{s}(n)$ of the separation of two points in an
$n$-dimensional ball which is given by \cite{SAN76}
\begin{equation}
	\overline{s}(n) = \frac{1}{2}
		\left(\frac{n}{n + 1}\right)^{2}
		\frac{\Gamma(n + 2)\Gamma(n/2)}
		{\Gamma(n + 3/2)\Gamma((n + 1)/2)}.
\label{mean sn}
\end{equation}

	For present purposes we are interested in ${\cal P}_{3}
\equiv {\cal P}(s)$, which is used in Sections II and IV to
evaluate the 2-body contributions arising from photon-exchange
and neutrino-exchange respectively.  As can be seen from
Eq.~(\ref{P_3(s)}), ${\cal P}(1) = {\cal P}(0) = 0$, as required by
Eq.~(\ref{P_n 1,0});  Eq.~(\ref{P_n' 1,0}) can be verified by 
noting that
\begin{equation}
	{\cal P}'(s) = 12s(1 - s)(4 - 5s - 5s^{2}).
\label{P's}
\end{equation}
Since both ${\cal P}(s)$ and ${\cal P}'(s)$ vanish at the
endpoints of the physical region ($s = 1$ and $s = 0$), it follows
that ${\cal P}(s)$ is a steeply falling function of $s$ at the
wings of the distribution.  The behavior of ${\cal P}(s)$ near $s
= 0$ and 1 can be verified from Fig.~\ref{P plot figure}
which exhibits
${\cal P}(s)$ and ${\cal P}'(s)$ in the interval 
$1 \geq s \geq 0$.
Two other quantities of
interest are the location $s_{0}$ of the maximum of ${\cal
P}(s)$, and the mean separation $\overline{s}(3)$ of two points
in a 3-dimensional sphere.  From Eq.~(\ref{P's}) ${\cal P}'(s) =
0$ gives local minima for ${\cal P}(s)$ (in the physical region $1
\geq s \geq 0$) at $s = 0,1$ and a local maximum for ${\cal
P}(s)$ when $(4 - 5s - 5s^{2}) = 0$.  The physical root is then
$s_{0} = [-1/2 + (21/20)^{1/2}] = 0.5247$
which is close to the middle of the physical region, $s =
0.5$.  To evaluate $\overline{s}(3)$ we use Eq.~(\ref{mean sn})
with $n = 3$ which gives
$\overline{s}(3) = 18/35 = 0.5143$.
The result $\overline{s}(3) \simeq 0.5$ is understandable given
that ${\cal P}(s)$ peaks near $s_{0} \simeq 0.5$, and the slight
difference between $\overline{s}(3)$ and $s_{0}$ is a reflection
of the fact that ${\cal P}(s)$ is not symmetric about $s = 0.5$.

	For purposes of averaging the $k$-body 0-derivative
contributions over a spherical volume it is in principle
necessary to know the $k$-body generalization of ${\cal P}(s)$ in
3-dimensions, and to be able to integrate this distribution over
the coordinates of $k \leq N = \O(10^{57})$ particles.  At present
the dependence of this function,
${\cal P}^{(k)}(s_{ij}) \equiv
{\cal P}^{(k)}(s_{12},s_{23},\ldots,s_{k1})$,
on the variables $s_{ij} =
|\vec{r}_{i} - \vec{r}_{j}|/2R$ is not known.
Moreover, even if
${\cal P}^{(k)}(s_{ij})$
were known, the task of evaluating the $k$-body generalization of
the integral in Eq.~(\ref{more complete U0 4}) would be beyond
present computational capabilities.  For these reasons we use the
preceding discussion of ${\cal P}(s)$ to approximate
$V^{(k)}(\vec{r}_{12},\vec{r}_{23},\ldots,\vec{r}_{k1})$ in
Eq.~(\ref{final Vk}) by replacing $r_{ij}$ with its mean value,
\begin{equation}
	r_{ij} \rightarrow \langle r_{ij} \rangle
		= 2R\,\overline{s}(3) \simeq R.
\label{MVA}
\end{equation}
The integrals in Eq.~(\ref{U0 k}) can then be carried out as in
Eq.~(\ref{U0 k condition}) and give
\begin{equation}
	U^{(k)}_{0} \simeq \frac{4}{\pi}
	\left(\frac{G_{F}a_{n}}{2\pi\sqrt{2}}\right)^{k}
	i^{k}k!\frac{1}{R^{k}(kR)^{k + 1}}
	\int^{2R}_{0}dr_{12} \cdots
	\int^{2R}_{0}dr_{k1}\,{\cal P}^{(k)}(r_{ij}),
\label{approximate U0 k 1}
\end{equation}
where $r_{ij} = 2Rs_{ij}$.
Using the normalization condition, Eq.~(\ref{k-prob
normalization}), we find, 
\begin{equation}
	U^{(k)}_{0} \simeq \frac{4}{\pi R}
	\left(\frac{G_{F}a_{n}}{2\pi\sqrt{2}R^{2}}\right)^{k}
	\frac{i^{k}k!}{k^{k + 1}},
\label{approximate U0 k}
\end{equation}
and this is the expression which will be used in Secs.~IV and
V.  The ``mean value approximation'' in
Eq.~(\ref{MVA}) can be justified by comparing Eq.~(\ref{MVA})
to the actual value of $\langle r_{ij}\rangle$ for the 2-body
electromagnetic,
3-body weak, and 4-body weak potentials which have been evaluated directly
\cite{Tu94}.  We find for these cases, respectively:
$\langle r_{ij} \rangle  =   0.83R$,
$\langle r_{ij} \rangle  =   0.48R$,
$\langle r_{ij} \rangle  =   0.62R$.
We note that the mean value
approximation in Eq.~(\ref{MVA}) overestimates $\langle r_{ij}
\rangle$, and hence it {\em underestimates} $|U^{(k)}_{0}|$ and
ultimately $|W|$ in Eq.~(\ref{W estimate}).  Moreover, the
uncertainties arising from the mean value approximation are no
worse than those inherent in estimating $R$ itself
\cite{ST83,AB77}.  The fact that all the values of $\langle r_{ij}
\rangle$ are reasonably close to $R$ can be understood
intuitively as follows:  The constraints in Eqs.~(\ref{P_n
1,0}) and (\ref{P_n' 1,0}) serve to suppress ${\cal P}(s)$ at
the wings of the distribution and, since ${\cal P}(s)$ is
normalized, the result is a peaking of ${\cal P}(s)$ near $s
\simeq 0.5$ which corresponds to $r_{ij} \simeq R$.

\pagebreak

\section{Evaluation of $\sum_{\lowercase{k}}W^{(\lowercase{k})}$}

	From Eq.~(\ref{WforN 2}) the weak energy $W$ is given by
\begin{equation}
	W \simeq   W^{(2)} +
	\sqrt{\frac{2}{\pi}}
	\frac{2}{R}
	\sum_{k = 4 \atop even}^{N}
	\frac{i^{k}k!}{k^{3/2}}
	\left(
	\frac{G_{F}a_{n}}{2\pi\sqrt{2}eR^{2}}
	\right)^{k}
	{N \choose k}.
\label{WforN 2 again}
\end{equation}
In Sec.~V we approximated $\sum_{k}$ by replacing $k^{3/2}$ in
the denominator of Eq.~(\ref{WforN 2 again}) by $N^{3/2}$, noting
that the sum is dominated by terms with $k \simeq N$.
In this Appendix
we show how $\sum_{k}$ can be evaluated with the factor of
$k^{3/2}$ included, should a more refined approximation to the
sum be desired \cite{HR94}.  Define 
\begin{equation}
	A_{k} = i^{k}k!{N \choose k}
	\phantom{this much space}	
	\Gamma_{R} =
	\frac{G_{F}a_{n}}{2\pi\sqrt{2}eR^{2}},
\label{A_k}
\end{equation}
and
\begin{equation}
	\sum_{k} = \sum_{k}\frac{A_{k}\Gamma_{R}^{k}}{k^{3/2}}.
\label{sum_k}
\end{equation}
In Eq.~(\ref{sum_k}) let $\Gamma_{R} = e^{y}$ so that
\begin{equation}
	\sum_{k} = \sum_{k}\frac{A_{k}e^{ky}}{k^{3/2}}
		= \sum_{k}A_{k}\frac{1}{\Gamma(3/2)}
		\int^{y}_{-\infty}dz\,e^{kz}(y - z)^{1/2}.
\label{sum_k 2}
\end{equation}
The last step in Eq.~(\ref{sum_k 2}) can be verified by
substituting $t = (y - z)$ which allows the integral in
Eq.~(\ref{sum_k 2}) to be expressed in terms of the gamma
function:
\begin{equation}
	\int^{y}_{-\infty}dz\,e^{kz}(y - z)^{1/2}
		= 
	   e^{ky}\int^{\infty}_{0}dt\,e^{-kt}t^{1/2}
		=
		\frac{e^{ky}}{k^{3/2}}\Gamma(3/2).
\end{equation}
Returning to Eq.~(\ref{sum_k 2}) we substitute $z = \ln\omega$ to
recast $\sum_{k}$ into the form
\begin{equation}
	\sum_{k}\frac{A_{k}e^{ky}}{k^{3/2}}
		= \frac{1}{\Gamma(3/2)}\sum_{k}A_{k}
		\int^{\Gamma_{R}}_{0}d\omega\,
		\omega^{k - 1}(\ln\Gamma_{R} - \ln\omega)^{1/2}.
\label{k_sum 3}
\end{equation}
Hence,
\begin{equation}
	\sum_{k}\frac{A_{k}\Gamma^{k}_{R}}{k^{3/2}}
		=
	\sum_{k}\frac{A_{k}e^{ky}}{k^{3/2}}
		= \frac{1}{\Gamma(3/2)}
		\int^{\Gamma_{R}}_{0}
		\frac{d\omega}{\omega}\,
		(\ln\Gamma_{R} - \ln\omega)^{1/2}
		\sum_{k}A_{k}\omega^{k}.
\label{k_sum 4}
\end{equation}
Using Eq.~(\ref{A_k}) we see that the expression for $\sum_{k}$
on the right-hand side of Eq.~(\ref{k_sum 4}) now has the same form
as the sum previously evaluated in Sec.~V starting from
Eq.~(\ref{extended sum}).  Hence by utilizing the result in
Eq.~(\ref{new e
sum}), the sum on the left-hand side of
Eq.~(\ref{k_sum 4}) can be expressed in terms of a
one-dimensional integral, which can be evaluated numerically if
necessary.

\pagebreak

\section{The Schwinger-Hartle Formalism for Massive Neutrinos}
	In this Appendix we generalize the Schwinger-Hartle
formalism presented in Appendix B to the case of massive
neutrinos.  Although the Schwinger formula itself retains the
same form as in Eq.~(\ref{Schwinger W}), the expression for
the free neutrino propagator $S^{(0)}_{F}(E)$ is modified when
the neutrino mass $m$ is different from zero.  The final
expression for the massive propagator
$S_{Fm}^{(0)}(E)$ is given in Eq.~(\ref{SF xE 2}), and the
application of the Schwinger-Hartle formalism to the present
problem is discussed in Sec.~VII.

	To establish our conventions when $m \neq 0$, we begin
with the expression for the propagator $\Delta_{Fm}^{(0)}$ of a
massive scalar field, which is given in our metric by
\cite{Lurie2,A/B2}
\begin{equation}
	\Delta_{Fm}^{(0)}(x) = -\frac{1}{(2\pi)^{3}}
		\int_{C_{F}} d^{3}k
		\int \frac{dk_{0}}{2\pi i}\,\,
		\frac{e^{i(\vec{k}\cdot\vec{x}
			- k_{0}x_{0})}}
		{(k_{0} - \omega_{k})(k_{0} + \omega_{k})}.
\label{DeltaF0 1}
\end{equation}
Here $\omega_{k} \equiv +(\vec{k}^{2} + m^{2})^{1/2}$ and $C_{F}$
is the usual Feynman contour in the complex $k_{0}$ plane.  The
integral over $k_{0}$ is straightforward and gives
\begin{equation}
	\Delta_{Fm}^{(0)}(x) = \frac{1}{2(2\pi)^{3}}
		\int d^{3}k\,e^{i\vec{k}\cdot\vec{x}}
		\left(
		\frac{e^{-i\omega_{k}|x_{0}|}}{\omega_{k}}
		\right).
\label{DeltaF0 2}
\end{equation}
Since $\Delta_{Fm}^{(0)}(x)$ is an invariant function of $x^{2} =
\vec{x}^{2} - x_{0}^{2}$, it will have the same form for
spacelike or timelike $x^{2}$.  In the former case
$\Delta_{Fm}^{(0)}(x)$ can be easily evaluated by setting $x_{0} =
0$ which gives \cite{A/B2},
\begin{equation}
	\Delta_{Fm}^{(0)}(x) = \frac{m}{8i\pi^{2}r}
		\int_{-\infty}^{\infty} dy\,
		\sinh y
		e^{imr\sinh y},
\label{DeltaF0 3}
\end{equation}
where $r = |\vec{x}|$ and $\sinh y = |\vec{k}|/m$.
$\Delta_{Fm}^{(0)}$ can be expressed in terms of the derivative of
the Hankel function $H_{0}^{(2)}$ which has the integral
representation \cite{A/B2}
\begin{equation}
	H_{0}^{(2)}(r) = \frac{i}{\pi}
		\int_{-\infty}^{\infty} dy\,
		e^{-ir\cosh y}.
\label{H0 2}
\end{equation}
It follows from Eq.~(\ref{H0 2}) that
\begin{equation}
	H_{0}^{(2)}(-ir) = \frac{i}{\pi}
		\int_{-\infty}^{\infty} dy\,
		e^{ir\sinh y},
\end{equation}
and hence \cite{A/B2}
\begin{equation}
	\Delta_{Fm}^{(0)}(x) 
		= \frac{i}{8\pi r}
		\frac{d}{dr} H_{0}^{(2)}(-imr) 
		= \frac{im^{2}}{8\pi r}
		\frac{H_{1}^{(2)}(-imr)}{-imr}.
\label{DeltaF0 4}
\end{equation}
The expression for $\Delta_{Fm}^{(0)}(x)$ can be written in
manifestly covariant form by letting $r \rightarrow
\sqrt{x^{2}}$, so that finally \cite{A/B2},
\begin{equation}
	\Delta_{Fm}^{(0)}(x) 
		= -\frac{m}{8\pi}
		\frac{H_{1}^{(2)}(-im\sqrt{x^{2}})}
		{\sqrt{x^{2}}}.
\label{DeltaF0 5}
\end{equation}
It is useful to check Eq.~(\ref{DeltaF0 5}) in the two limiting
cases of interest to us: a) For $m \rightarrow 0$ we use the
series expansion \cite{ARF2}
\begin{equation}
	H_{\nu}^{(2)}(z) \simeq \frac{i(\nu - 1)!}{\pi}
		\left(\frac{2}{z}\right)^{\nu},
\end{equation}
to obtain
\begin{equation}
	\Delta_{Fm}^{(0)}(x) 
		\stackrel{m \rightarrow 0}{\rightarrow}
		 -\frac{m}{8\pi \sqrt{x^{2}}}
		\left[\frac{i}{\pi}
		\left(\frac{2}{-im\sqrt{x^{2}}}\right)
		\right]
		= \frac{1}{4\pi x^{2}},
\label{DeltaF0 5 m-0}
\end{equation}
in agreement with Eq.~(\ref{Simple Delta F}).  b) When $|z|$ is large
$H_{\nu}^{(2)}(z)$ can be represented by the asymptotic expansion
\cite{ARF3}
\begin{equation}
	H_{\nu}^{(2)}(z) \simeq 
		\left(\frac{2}{\pi z}\right)^{1/2}
		\exp\left\{-i[z - (\nu + 1/2)\pi/2]\right\}.	
\end{equation}
Hence with $z = -im\sqrt{x^{2}}$ we have approximately,
\begin{equation}
	\Delta_{Fm}^{(0)}(x)  \simeq
		\frac{m^{2}}{4\sqrt{2}}
		\left(\frac{1}{\pi m\sqrt{x^{2}}}\right)^{3/2}
		e^{-m\sqrt{x^{2}}},
\label{DeltaF0 5 large z}
\end{equation}
which shows the characteristic exponential decrease of
$\Delta_{Fm}^{(0)}(x)$.  The analogous behavior for neutrinos
eventually leads to the ``saturation'' of the many-body
contribution to $W$, and thus to a resolution of the
neutrino-exchange energy-density catastrophe.

	The massive fermion propagator in configuration space can
be obtained from Eq.~(\ref{DeltaF0 5}) and, as in the massless case, we
are interested in calculating 
\begin{equation}
	S_{Fm}^{(0)}(\vec{x},E) =
	\int_{-\infty}^{\infty}dt \,
	e^{iEt}S_{Fm}^{(0)}(x)
	= (-\gamma\cdot\eta + m)
	\int_{-\infty}^{\infty}dt \,
	e^{iEt}\Delta_{Fm}^{(0)}(x),
\label{massive SF}
\end{equation}
where $\gamma\cdot\eta = \vec{\gamma}\cdot\vec{\partial} -
\gamma_{4}E$, and $t = x_{0}$.  We define
\begin{eqnarray}
	\Delta_{Fm}^{(0)}(\vec{x},E) 
		& = &
	\int_{-\infty}^{\infty}dt \,
	e^{iEt}
	\Delta_{Fm}^{(0)}(x)
		=
	\int_{-\infty}^{\infty}dt \,
	e^{iEt}
	\left[\left(\frac{im^{2}}{8\pi}\right)
	\frac{H_{1}^{(2)}(-im\sqrt{r^{2} - t^{2})}}
	{-im\sqrt{r^{2} - t^{2})}}
	\right]
	\nonumber \\
		& = &
		2\int_{-\infty}^{\infty}dt \,
		\cos(Et)
	\left[\left(\frac{im^{2}}{8\pi}\right)
	\frac{H_{1}^{(2)}(-im\sqrt{r^{2} - t^{2})}}
	{-im\sqrt{r^{2} - t^{2})}}
	\right],
\label{DeltaF xE}
\end{eqnarray}
where the last step follows by noting that the expression in
square brackets is an even function of $t$.  The integral in
Eq.~(\ref{DeltaF xE}) can be evaluated by making use of the
relations \cite{A/B2,G/R2}
\begin{equation}
	\int^{\infty}_{0}dt\,
		H_{0}^{(2)}(\alpha\sqrt{\beta^{2} - t^{2}})
		\cos(\gamma t) =
		\frac{i\exp(-i\beta\sqrt{\alpha^{2} + \gamma^{2}})}
		{\sqrt{\alpha^{2} + \gamma^{2}})},
\label{H02 1}
\end{equation}
\begin{equation}
	\frac{d}{dz}H_{0}^{(2)}(z) = -H_{1}^{(2)}(z).
\label{H02 2}
\end{equation}
Differentiating both sides of Eq.~(\ref{H02 1}) with respect to
$\beta$, and using Eq.~(\ref{H02 2}) we find 
\begin{equation}
	\int_{0}^{\infty}dt\,
	\frac{H_{1}^{(2)}(\alpha
		\sqrt{\beta^{2} - t^{2}})
	\cos(\gamma t)}
	{\alpha	\sqrt{\beta^{2} - t^{2}}}
	= -\frac{1}{\alpha^{2}\beta}
		\exp(-i\beta\sqrt{\alpha^{2} + \gamma^{2}}).
\end{equation}
If we identify $\alpha = -im$, $\beta = -r$, and $\gamma = E$,
then
\begin{equation}
	\Delta_{Fm}^{(0)}(\vec{x},E) =
		\left(\frac{im^{2}}{8\pi}\right)
		\left(\frac{-2}{m^{2}r}\right)
		\exp(ir\sqrt{E^{2} - m^{2}}),
\label{DeltaF xE 2}
\end{equation}
and
\begin{eqnarray}
	S_{Fm}^{(0)}(\vec{x},E) & = &
	(\vec{\gamma} \cdot \vec{\partial} -
		\gamma_{4}E - m)
		\frac{i}{4\pi r}
		\exp(ir\sqrt{E^{2} - m^{2}})
	\nonumber \\
	& \equiv & (\gamma\cdot\eta - m)
		\Delta_{Fm}(\vec{r},E).
\label{SF xE 2}
\end{eqnarray}
In the $m \rightarrow 0$ limit the expression in Eq.~(\ref{SF
xE 2}) reduces to the massless result given in Eq.~(\ref{final
inverted SF}).

	When the Schwinger formula in Eq.~(\ref{final inverted
SF}) is expanded
in a perturbation series, and the Dirac traces are evaluated,
integrals of the form
\begin{equation}
	\bar{F}_{n}(z) = 
		\int_{-\infty}^{\infty}dE\,E^{n}\,
		\exp(iz\sqrt{E^{2} - m^{2}})
\label{Fn}
\end{equation}
arise which are the $m \neq 0$ analogs of $\bar{I}_{n}(z)$ in
Eq.~(\ref{general In}).  
Since $\bar{F}_{n}(z) = 0$ when $n$ is odd
we can write
\begin{equation}
        \bar{F}_{n}(z) = \left\{ \begin{array}{ll}
                2\int^{\infty}_{0}dE\,E^{n}
                \exp(iz\sqrt{E^{2} - m^{2}})
                \equiv F_{n}(z) &\,\,\, \mbox{even $n$}
                        \\
                0 & \,\,\,\mbox{odd $n$}.
                \end{array} \right.
\label{general Fn}
\end{equation}
The integrals in Eqs.~(\ref{Fn}) and (\ref{general Fn})
 can be evaluated recursively
beginning with $F_{0}(z)$ which may be cast in the form
\begin{equation}
	F_{0}(z) = 
		2\int_{im}^{\infty}dp\,e^{izp}
		\frac{p}{\sqrt{p^{2} + m^{2}}},
\label{F0}
\end{equation}
where $p^{2} = E^{2} - m^{2}$.  $F_{0}(z)$ can be expressed in
terms of a modified Bessel function by analytically continuing
the integral \cite{G/R2 316}
\begin{equation}
		\int_{u}^{\infty}dp\,
		\frac{pe^{-\tau p}}{\sqrt{p^{2} - u^{2}}}
		= uK_{1}(\tau u).
\end{equation}
Here $K_{n}(z)$ is the modified Bessel function defined by
\begin{equation}
	K_{\nu}(x) = \frac{\pi}{2}i^{\nu + 1}
			H_{\nu}^{(1)}(ix),
\end{equation}
where $H_{\nu}^{(1)}$ is the Hankel function of the first kind.
Substituting $u = im$ and $\tau = iz$ then leads to 
\begin{equation}
	F_{0}(z) = 2imK_{1}(mz).
\label{final F0}
\end{equation}
For small $x$ the leading term in the series expansion of
$K_{\nu}(x)$ is 
\begin{equation}
	K_{\nu}(x) \simeq 2^{\nu - 1}(\nu - 1)! x^{-\nu},
\label{leading Kn}
\end{equation}
and hence for small $z$
\begin{equation}
	F_{0}(z) \simeq 2im \frac{1}{mz} = \frac{2i}{z},
\end{equation}
in agreement with Eq.~(\ref{I0}).  However, the regime of
interest when $m \neq 0$ is when $x = mz$ is large, in which case
$K_{\nu}(x)$ can be approximated by the asymptotic series
\cite{ARF3}
\begin{equation}
	K_{\nu}(x) \simeq \sqrt{\frac{\pi}{2x}} e^{-x}
		\left[1 + \frac{(4\nu^{2} - 1)}{1!8x}
		+ \frac{(4\nu^{2} - 1)(4\nu^{2} - 9)}
		   {2!(8x)^{2}} + \cdots \right].
\label{asymptotic Kn}
\end{equation}
Combining Eqs.~(\ref{final F0}) and (\ref{asymptotic Kn}) then
leads to
\begin{equation}
	F_{0}(z) \simeq  2im\sqrt{\frac{\pi}{2mz}} e^{-mz}
		\left[1 + \frac{3}{8mz} + \cdots \right],
\label{approximate F0}
\end{equation}
which exhibits the characteristic exponentially falling behavior
that leads to ``saturation'' of the neutrino-exchange forces.

	To evaluate $F_{n}(z)$ for $n \neq 0$ we 
differentiate $F_{0}(z)$ twice with respect to $z$: 
\begin{eqnarray}
	\frac{d^{2}F_{0}(z)}{dz^{2}} 
		& \equiv  &
		F_{0}^{(2)} =
		i^{2}2\int^{\infty}_{0}dE\,
		(E^{2} - m^{2})
		\exp(iz\sqrt{E^{2} - m^{2}})
	\nonumber \\
		& = & i^{2}F_{2}(z) + m^{2}F_{0}(z),
\label{Fbar 2}
\end{eqnarray}
and hence
\begin{equation}
	F_{2}(z) = (-i)^{2}F_{0}^{(2)}(z) + m^{2}
			F_{0}(z).
\label{F2}
\end{equation}
Proceeding in a similar fashion, we find upon differentiating
both sides of Eq.~(\ref{Fbar 2}) twice with respect to $z$,
\begin{equation}
	F_{4}(z) = F_{0}^{(4)}(z)
		+ 2m^{2}F_{2}(z) - m^{4}F_{0}(z).
\label{F4}
\end{equation}
Combining Eqs.~(\ref{F2}) and (\ref{F4}) then leads to
\begin{equation}
	F_{4}(z) = F_{0}^{(4)}(z)
			- 2m^{2}F_{0}^{(2)}(z)
			+ m^{4}F_{0}(z).
\label{F4 2}
\end{equation}
It follows from Eqs.(\ref{F2}) and (\ref{F4 2}) that
$F_{n}(z)$ can be evaluated recursively for any even $n$ in
terms of $F_{0}(z)$ and its derivatives 
$F_{0}^{(2)}(z)$,
$F_{0}^{(4)}(z)$,\ldots,
$F_{0}^{(n)}(z)$.

	The expressions for 
$F_{2}(z)$,
$F_{4}(z)$,\ldots
$F_{n}(z)$, can be further simplified by explicitly
evaluating the various derivatives 
$F_{0}^{(n)}(z)$ that arise in the expression for
$F_{n}(z)$.  From Eq.~(\ref{final F0}),
\begin{equation}
	F_{0}^{(n)}(z) = 2im^{n + 1}K_{1}^{(n)}(mz),
\label{F0n}
\end{equation}
where the superscript $(n)$ denotes the $n$th derivative on both
sides of Eq.~(\ref{F0n}).  The $n$th derivative of the modified
Bessel function $K_{1}(x)$ can be expressed in terms of
$K_{0}(x)$ and $K_{1}(x)$ by utilizing the relations \cite{A/S}
\begin{equation}
	\frac{d}{dz} K_{0}(z) \equiv K_{0}^{(1)}(z)
		= -K_{1}(z),
\label{K1'a}
\end{equation}
\begin{equation}
	K_{1}^{(1)}(z) = -K_{0}(z) - \frac{1}{z}K_{1}(z).
\label{K1'b}
\end{equation}
Differentiating Eq.~(\ref{K1'b}) and using Eq.~(\ref{K1'a}) leads
to
\begin{equation}
	K_{1}^{(2)}(z) = \frac{1}{z}K_{0}(z) +
		\left(1 + \frac{2!}{z^{2}}\right)K_{1}(z).
\end{equation}
Proceeding in this way we can express the $n$th derivative
$K_{1}^{(n)}(z)$ in terms of $K_{0}(z)$ and $K_{1}(z)$.  For
purposes of calculating the 4-body contribution in Sec.~VII the
explicit expressions for $K_{1}^{(3)}(z)$ and $K_{1}^{(4)}(z)$
are needed, and these are given by
\begin{equation}
	K_{1}^{(3)}(z) =
		-\left(1 + \frac{3}{z^{2}}\right)K_{0}(z)
		- \left(
		\frac{2}{z} + \frac{3!}{z^{3}}\right)K_{1}(z),
\label{K1(3)}
\end{equation}
\begin{equation}
	K_{1}^{(4)}(z) =
		\left(
		\frac{2}{z} + \frac{12}{z^{3}}\right)K_{0}(z)
		+ \left(1 +
		\frac{7}{z^{2}} + \frac{4!}{z^{4}}\right)K_{1}(z).
\label{K1(4)}
\end{equation}
We note from Eq.~(\ref{massive V04 c}) that when the trace of the 4-body matrix
element is calculated the result will depend on the functions
$F_{0}^{(4)}(z)$,
$m^{2}F_{0}^{(2)}(z)$, 
and $m^{4}F_{0}(z)$  which are given by
\begin{mathletters}
\begin{eqnarray}
	m^{4}F_{0}(z) & = &
		2im^{5}K_{1}(mz)
	\\	
	m^{2}F_{0}^{(2)}(z)  & = &
	2i\left[\frac{m^{4}}{z}K_{0}(mz) +
		\left(m^{5} + m^{3}\frac{2!}{z^{2}}\right)
		K_{1}(mz)\right],
	\\
	F_{0}^{(4)}(z) & = &
	2i\left[
		\left(\frac{2m^{4}}{z} + \frac{12m^{2}}{z^{3}}
		\right)K_{0}(mz)
		+
		\left(m^{5} + \frac{7m^{3}}{z^{2}} +
			m\frac{4!}{z^{4}}\right)
		K_{1}(mz)\right].
\end{eqnarray}
\end{mathletters}
In the $m = 0$ limit, the only term which survives is the
contribution proportional to $m/z^{4}$, and since this term is
also the source of the bound on $m$ in Eq.~(\ref{final m limit}) we examine it
in more detail.

	As $m \rightarrow 0$ we have from Eq.~(\ref{leading Kn})
\begin{equation}
	2im\frac{4!}{z^{4}}K_{1}(mz) \simeq 2im\frac{4!}{z^{4}}
		\,\frac{1}{mz} =
		2i\frac{4!}{z^{5}},
\end{equation}
in agreement with Eq.~(\ref{In}).  As before, we are interested
in the behavior of this term when $m \neq 0$ and $z$ is large.
Using Eq.~(\ref{asymptotic Kn}) we find
\begin{equation}
	F_{0}^{(4)}(z) \simeq 2i\frac{4!}{z^{5}}
			e^{-mz}
		\left[\sqrt{\frac{\pi mz}{2}}
		\left(1 + \frac{3}{8mz} + \cdots \right)
		\right].
\end{equation}
Since the expression in square brackets is slowly varying (in
$z$) compared to the remaining $z$-dependent factors, the net
effect of having $m \neq 0$ is to replace $1/z^{5}$ by
$\exp(-mz)/z^{5}$ in $F_{0}^{(4) }$.  This result is the
basis for the bound on $m$ derived in Sec.~VIII.
	


\input{ref-aol.tex}     


\begin{table}
\vspace{12pt}
\caption{Summary of the necessary conditions for the existence of
a large many-body effect.  For each interaction, ``Yes''
indicates that the condition is met, and ``No'' indicates that
it is not.  Among known interactions only neutrino-exchange meets
all three conditions.  See text for further details.}
\vspace{12pt}
\begin{tabular}{lccccc}
\multicolumn{1}{c}{Condition} &
	Strong &
	Electomagnetism &
	Weak \tablenote[1]{$Z^{0}$-exchange.} &
	Gravity &
	Neutrino Exchange \\
	\hline 
Long-Range & No & Yes & No & Yes & Yes \\
Bulk Matter ``Charge'' & Yes & No & Yes & Yes & Yes \\
Large Coupling Strength &  Yes & Yes & Yes & No & Yes 
\label{many-body table}
\end{tabular}
\end{table}

\begin{table}
\vspace{12pt}
\caption{Parameters for two typical white dwarfs.}
\label{white dwarf table}
\vspace{12pt}
\begin{tabular}{lcc}
	&
	Sirius B &
	40 Eri B \\
	\hline 
$M$ \tablenote[1]{From Ref.~\cite{ST83b}} 
& $1.053(28)M_{\odot}$ & $0.48(2)M_{\odot}$  \\
$R$ $^{\rm a}$ 
& $0.0074(6)R_{\odot}$  & $0.0124(5)R_{\odot}$ \\
$N_{e}$ &  $6.3 \times 10^{56}$ & $2.9 \times 10^{56}$ \\
$\Lambda_{R}$ \tablenote[2]{Defined by Eq.~(\ref{Lambda
R}).} 
& $4.1 \times 10^{4}$ & $6.6 \times 10^{3}$ \\
$W/M_{U}c^{2}$ \tablenote[3]{$M_{U}$ is the mass of the
universe given by Eq.~(\ref{mass of universe}).}
& $10^{(3 \times 10^{57} + 28 - 13 - 88)}$ 
	& $10^{(1\times 10^{57} + 28 - 13 - 88)}$
\end{tabular}
\end{table}


\begin{figure}
\caption{The 2-body neutron-neutron potential arising from
neutrino-exchange.  The solid (dashed) lines denote neutrons
(neutrinos).}
\label{2-body neutrino exchange figure}
\end{figure}

\begin{figure}
\caption{Contributions to the 4-body potential energy arising
from neutrino exchange.  As before, solid (dashed) lines denote
neutrons (neutrinos).  Each of the diagrams (a), (b), and (c) is
topologically different from the others, as can be seen by
redrawing the graphs as shown.  For each of these diagrams there
is another that is obtained by reversing the sense of the
neutrino loop momentum, as we show explicitly in Fig.~4.}
\label{First 4-body figure}
\end{figure}

\begin{figure}
\caption{a) Plot of the function ${\cal P}(s) = {\cal P}_{3}(s)$ 
in Eqs.~(2.12) and (C7). b) Plot of ${\cal P}'(s)$ in Eq.~(C12).}
\label{P plot figure}
\end{figure}

\begin{figure}
\caption{Combinatorics for the 4-body diagrams.
As before, the solid (dashed) lines denote neutrons
(neutrinos).  For $k =4$ there are 3! = 6 topologically distinct
diagrams that can be drawn,  corresponding to the 6 possible
permutations of the integers 1, 2, 3, and 4 arranged at the
vertices of the neutrino loop.  However, diagrams (a'), (b'),
and (c') are obtained from diagrams (a), (b), and (c)
respectively by reversing the sense of the neutrino loop
momentum.  Hence there are $(k - 1)!/2$ pairs of diagrams, [(a) +
(a')], [(b) + (b')], [(c) + (c')], etc.}
\label{Second 4-body figure}
\end{figure}

\begin{figure}
\caption{The 3-body contributions arising from neutrino exchange.
The solid (dashed) lines represent neutrons (neutrinos).  As noted in
the text, both diagrams must be included to reproduce Furry's
theorem for the vector contribution.}
\label{3-body figure}
\end{figure}

\begin{figure}
\caption{Constraints on the masses of $\nu_{e}$, $\nu_{\mu}$, and
$\nu_{\tau}$.  For each neutrino (or antineutrino) the shaded
regions are excluded either by the lower bound in Eq.~(8.12) or
by the upper bounds in Eq.~(1.1).}
\label{neutrino bound figure}
\end{figure}

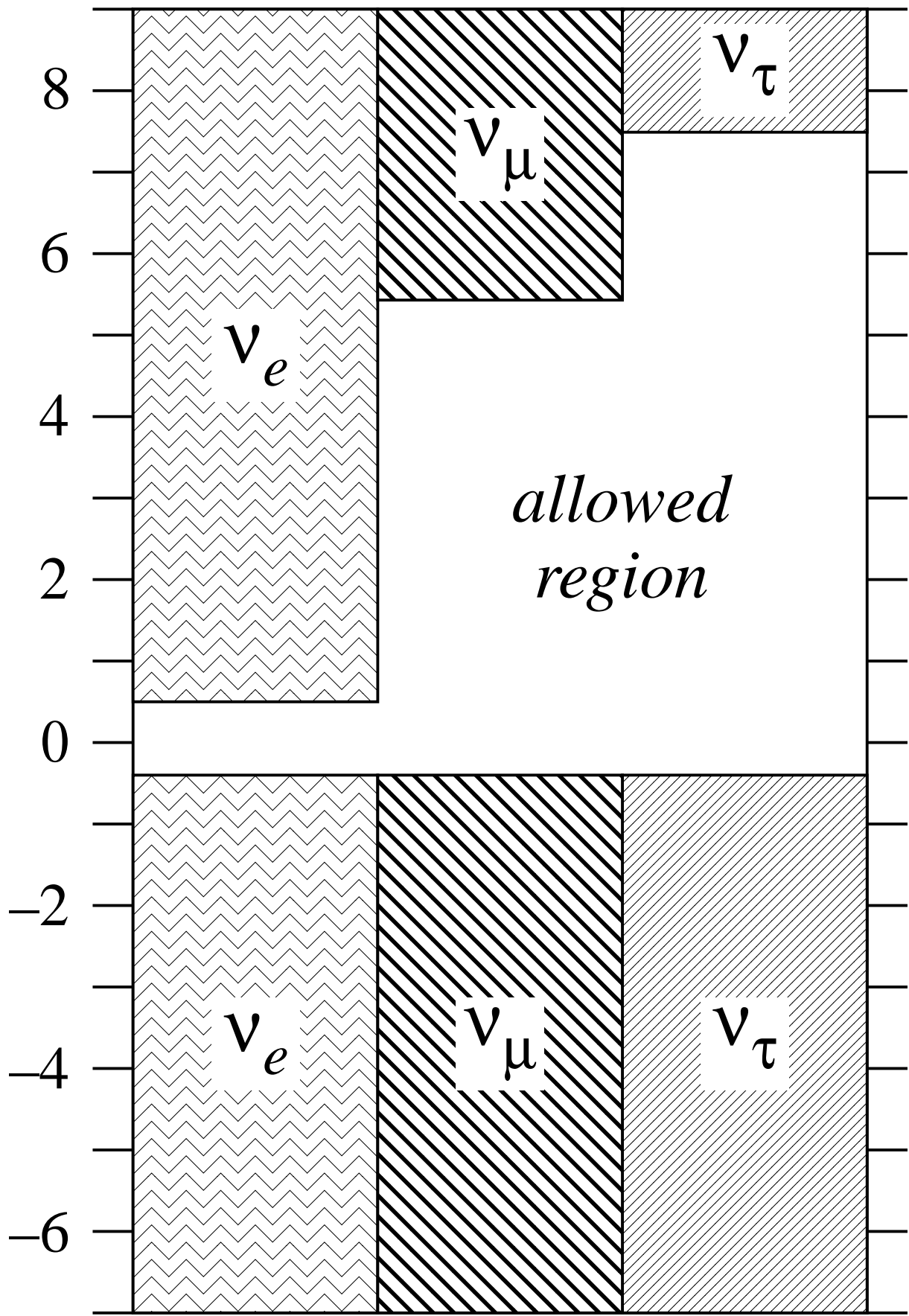
\end{document}

%% file: beginning-aol.tex
\title{Long-Range Forces and Neutrino Mass}

\author{Ephraim Fischbach}

\address{Physics Department, Purdue University, West
Lafayette, IN  47907-1396}

\maketitle
\begin{abstract}

	We explore the limits on neutrino mass which follow from a
study of the long-range forces that arise from the exchange of
massless or ultra-light neutrinos.  Although the 2-body
neutrino-exchange force is unobservably small, the many-body
force can generate a very large energy density in neutron stars
and white dwarfs.  We discuss the novel features of
neutrino-exchange forces which lead to large many-body effects,
and present the formalism that allows these effects to be
calculated explicitly in the Standard Model.  After considering,
and excluding, several possibilities for avoiding the
unphysically large contributions from the exchange of massless
neutrinos, we develop a formalism to describe the exchange of
massive neutrinos.  It is shown that the stability of
both neutrons stars and white dwarfs in the presence of many-body
neutrino-exchange forces implies a lower bound, $m \gtrsim 0.4$
eV/$c^{2}$ on the mass $m$ of any neutrino.
\\
\\
\noindent {\small Accepted for publication in Annals of Physics}
\end{abstract}

\vspace{.5in}

\noindent
Number of Manuscript Pages: 96 \\
Number of Tables: 2 \\
Number of Figures: 6 \\

\pagebreak

\noindent
Proposed Running Head: {\em Long-Range Forces and Neutrinos}

\vspace{.5in}

\noindent
Direct all correspondence:

\vspace{12pt}

{\singlespace
Dr. Ephraim Fischbach 

Purdue University 

1396 Physics Building 

West Lafayette, IN 47907-1396 

Phone: (317) 494-5506

Fax: (317) 494-0706

E-Mail: {\tt ephraim@physics.purdue.edu}

}

\pagebreak